\documentclass[3p,twocolumn]{elsarticle}
\usepackage{amsmath,bm,epsfig,subfigure}
\usepackage{epstopdf}
\usepackage{indentfirst}
\usepackage{amssymb}
\usepackage{graphicx}
\usepackage{float}
\usepackage{romanbar}

\usepackage{etoolbox}
\makeatletter

\renewcommand{\fnum@figure}{Fig. \thefigure}

\patchcmd{\ps@pprintTitle}
  {Preprint submitted}
  {Revised to comply with the accepted version proofread and sent for publication}
  {}{}
\makeatother

\usepackage{color}
\usepackage[normalem]{ulem}
\usepackage{lineno}
\modulolinenumbers[5]
\usepackage[toc,page]{appendix}
\usepackage{natbib}
\usepackage{ mathrsfs }

\DeclareMathAlphabet{\mathpzc}{OT1}{pzc}{m}{it}

\journal{Physical Review E}

\begin{document}

\begin{frontmatter}

\title{Flat bands and compactons in mechanical lattices}


\author[]{Nathan Perchikov\corref{mycorrespondingauthor}}
\cortext[mycorrespondingauthor]{Corresponding author}
\ead{perchico@gmail.com}

\author{O.V. Gendelman}

\address{Faculty of Mechanical Engineering, Technion, Haifa 32000, Israel}

\begin{abstract}
Local configurational symmetry in lattice structures may give rise to stationary, compact solutions, even in the absence of disorder and nonlinearity. These compact solutions are related to the existence of flat dispersion curves (bands). Nonlinearity can destabilize such compactons. One common flat-band-generating system is the one-dimensional cross-stitch model, in which compactons were shown to exist for the photonic lattice with Kerr nonlinearity.The compactons exist there already in the linear regime and are not generally destructed by that nonlinearity. Smooth nonlinearity of this kind does not permit performing complete stability analysis for this chain. We consider a discrete mechanical system with flat dispersion bands, in which the nonlinearity exists due to impact constraints. In this case, one can use the concept of the saltation matrix for the analytic construction of the monodromy matrix. Besides, we consider a smooth nonlinear lattice with linearly connected massless boxes, each containing two symmetric anharmonic oscillators. In this model, the flat bands and discrete compactons also readily emerge. This system also permits performing comprehensive stability analysis, at least in the anti-continuum limit, due to the reduced number of degrees-of-freedom. In both systems, there exist two types of localization. The first one is the complete localization, and the second one is the more common exponential localization. The latter type is associated with discrete breathers (DBs). Two principal mechanisms for the loss of stability are revealed. The first one is the possible internal instability of the symmetric and/or antisymmetric solution in the individual unit cell of the chain. One can interpret this instability pattern as internal resonance between the compacton and the DB. The other mechanism is global instability related to resonance of the stationary solution with the propagation frequencies. Different instability mechanisms lead to different bifurcations at the stability threshold.
\end{abstract}


\end{frontmatter}


\section{Introduction}
\label{sect1}

The aim of the present research is to investigate the nonlinear dynamical phenomena related to the presence of flat bands (FBs) in the oscillatory spectra of mechanical and physical systems with special structural symmetry. Such structures are known to possess a number of unique and appealing physical and dynamical properties -- in particular, completely localized, or compact, solutions as well as ``hidden modes'', which cannot be excited externally in the linear setting. Flat bands in physical systems have been observed and broadly discussed in the recent literature. The effect of interactions and disorder on wave transport in periodic potentials, such as electrons in crystals, is strongly amplified if the bandwidth (kinetic energy) is small. A particularly interesting situation arises when some of the dispersion bands become strictly flat with macroscopically degenerate eigenstates. In this limit, any relevant perturbation will lift the degeneracy and determine the emerging highly correlated and nontrivial eigenstates. In these systems it may be possible to control the interactions, the potential energy associated with which successfully competes with kinetic energy, such that interesting wave transport phenomena are produced. Engineering FB lattice models have been extended to three-dimensional (3D), two-dimensional (2D), and even one dimensional (1D) settings. A number of FB construction pathways using graph theory were suggested in the literature. One example is the case where compact states that are fully localized on several lattice sites are used. 

The origin of the compact flat band states (FBSs) is the destructive interference effectively decoupling the FBSs from the rest of the lattice, similar to the one in the known case where antisymmetric bound states embedded in and decoupled from the continuum are observed, as in \cite{Flach2014}, where the context is Fano lattices, the mechanism of resonance in which the notion of flat bands helps to explain. Fano lattices appear, for example, in the context of plasmonics, which is relevant for laser technology, and also in the broader context of meta-materials \cite{Lukyanchuk2010}. An interesting feature of the flat bands is that they are (still hypothetically) characterized by fractional power-law dependence between correlation length and level of disorder \cite{Flach2013}. Furthermore, flat bands have extremely singular densities of states. Therefore, such systems are prone to form correlated states with broken symmetries \cite{Heikkila}. To elaborate on the aforementioned  feature of destructive interference, it may be argued that flat (dispersion) bands associated with the propagation of waves in latticed materials occur when perfect destructive interference allows for compact localized eigenstates (CLSs), or, modes with nonzero amplitude only at a finite number of lattice sites \cite{Flach_arXiv2014}. This characterization of the flat bands turns out to be closely related to the notion of discrete breathers.

Discrete breathers (DBs) have long been a subject of both theoretical analysis and experimental studies \cite{Ovchinnikov, Flach1, Flach2}. Recently, it was demonstrated that one can derive exact solutions for DBs in vibro-impact chain models. Such lattices have been investigated analytically both for the Hamiltonian case \cite{GendelmanManevitch2008,Grinberg2016} and for the forced-damped case \cite{Gendelman2013,Perchikov2014,Grinberg2016}. Noteworthy in the perspective of the present research are the recent achievements in the analysis of FB structures, especially such as the ones in \cite{Flach2013} and \cite{Flach_arXiv2014}, and in what concerns the effects produced by the nonlinearity, also the recent success in the analysis of the dynamics and stability of discrete breathers, in specific, the analytic results presented in \cite{Gendelman2013,Perchikov2014}.

In a spatially-extended system comprised of complex elements, the internal symmetry that produces the flat dispersion bands, leads to local dynamics that can exist with no long-range interactions. This means that a single element can be excited locally, which corresponds to the ideal compacton in the discrete lattice.

In continuous systems, compacton solutions with non-analytical profiles were first given in explicit form by Rosenau and Hyman in \cite{RosenauHyman1993}. They emerge as solutions of a special class of nonlinear partial differential equations, the $K(m,n)$ equations, which possess essential gradient nonlinearity. Although physically motivated, these continuous equations were not derived as a rigorous limit of a specific discrete system \cite{RosenauHyman1993}. In fact, in an earlier work by Nesterenko, in \cite{VFNesterenkoP}, a partial differential equation with gradient nonlinearity related to nonlinear dispersion, was derived consistently as a continuous limit of a granular mechanical lattice. That work presented a specific solution suggesting the interpretation of being arbitrarily close to having a compact support \cite{VFNesterenkoB}. An important feature of the solution was that it was characterized by amplitude-independent width. This characteristic, as well as the high level of spatial localization of the obtained soliton solution were validated by experimental observation in a one-dimensional granular medium \cite{Lazaridi1985}. Shortly thereafter, in \cite{Wright1985}, a solitary wave with exactly compact support was shown to exist as a one-parameter class of functions for a specific scenario in the case of nonlinear waves in an elastic rod. In \cite{Kivshar1994}, a one-dimensional lattice of atoms with purely anharmonic interactions was considered, a compacton solution was found analytically, and as in \cite{VFNesterenkoP} was found to be characterized by amplitude-independent width. The compacton solution was also considered in the context of dynamic friction following the Frenkel-Kontorova model \cite{BrownKivshar1998}. In \cite{Dusuel1998} the mechanical analog of the double-well nonlinear interaction potential that produces compacton-like solutions, was constructed in the form of a lattice of pendula on an elastic foundation with strong coupling. It was found there that static compactons are stable, whereas their mobile counterparts are unstable. In \cite{Kevrekidis2002}, exact results were obtained for linearly stable compactons on discrete 1D lattices with interaction of the Klein-Gordon type. In \cite{Gaeta2007}, a discrete chain with linear elastic inter-site interactions with the addition of non-smooth on-site potentials was examined in the context of emergence of compactons.

In the recent work in \cite{RosenauZilburg2015}, discrete systems were examined, both in one dimension (a chain) and in two dimensions (a rectangular array), using the Klein-Gordon model with cubic nonlinearities on-site and between the sites. Compact breather solutions were obtained, which are stable and non-traveling, in the continuum limit. In \cite{DAmbroise2015} the discrete compact breathers were discussed in the context of Bose-Einstein condensates and nonlinear optics using various interaction models, such as the discrete nonlinear Shr\"{o}dinger equation, as well as the Klein-Gordon and the Fermi-Pasta-Ulam models. The question of mobility of a discrete soliton in the context of a practical application is addressed in \cite{Johansson2015} for a chain with a Peierls-Nabarro potential barrier typically applied for a description of the dislocation movement in imperfect crystals. In \cite{James2012} a discrete mechanical chain with Hertzian interaction is analyzed and traveling compactons with amplitude-independent wave-speed are revealed. In \cite{Ndjoko2012}, DNA chains are studied using the extended discrete nonlinear Shr\"{o}dinger equation and solitary waves with compact support are obtained. Some of the obtained compactons are robust and some quickly decompose, depending on the DNA molecule model stacking parameters. A recent work, Ref. \cite{Johansson2015b}, deals with the problem of compactification by tuning parameters, such as the measure of interaction nonlinearity (the exponent in a power-law dependence), on-site energies and anisotropic coupling. Compact solitary waves are then obtained for the three lattice-sites support, in the context of the discrete nonlinear Shr\"{o}dinger equation. Stability analysis of the obtained compact solutions is performed numerically, by approximate application of Floquet theory.

It should be clarified that the compactons in the FBSs differ strongly from the compactons in the systems with the nonlinear gradient coupling, mentioned above. In the latter models, the solutions with the compact support appear only in the continuum limit;  the discrete counterparts are characterized by hyper-exponential localization, but are, strictly speaking, not compact. The FBS compactons are strictly compact even in the discrete models, due to the symmetry-related destructive interference.

In \cite{Zong2016}, the flat bands and the compact states were observed in 2D photonic kagome lattices, with possible application to distortion-free image transmission. In \cite{Muller2016}, 1D and 2D diamond lattices were studied in the context of kinetic-energy-driven ferromagnetic transition in the Hubbard model, using perturbation theory. The transition was found to be related to perturbed flat bands acquiring small dispersion allowing originally localized electrons to correlate. In \cite{Ramezani_arXiv2017}, a photonic lattice is analyzed, where a flat band emerges due to PT (parity time) symmetry for specific gain and loss values. In \cite{Ge2015}, a similar system is considered.  It is shown there that both randomly positioned states and pinned states at the symmetry plane in the flat band can undergo thresholdless PT symmetry breaking. Furthermore, the flat band is shown to be unaffected by weak disorder, contrary to the modes in the dispersive bands. 

In \cite{Maimistov_arXiv2016}, a rhombic 1D optical waveguide lattice was analyzed taking into account cubic nonlinearity, and FB solutions were found to exist yet stay unstable due to mode-coupling until a power-threshold is reached. In \cite{Poli2017}, a dimerized Lieb lattice with next-to-nearest-neighbor couplings was studied. A flat band was found, corresponding to a point-defect zero-mode, spectrally isolated and spatially localized, producing sub-lattice polarization, which provides a route to mode selection via loss-imbalance. The revealed mechanism may be useful for photonic mode-shaping and guiding, and can, possibly, be extended to laser settings. In artificial 2D materials, the symmetry-breaking-originated flat-band structure provides a mechanism for the design of robust defect-states. In \cite{Valiente2017}, scattering theory was applied to collisions in highly degenerate (flat) bands. It was shown how to construct localized states in flat bands that are the scattering states of impurity-potentials in one dimension, particularized to the sawtooth lattice.
In \cite{Huang2016} it was found that a planar composition of 2D square photonic lattices with Kerr nonlinearity (nonlinear dielectrics), with rotations corresponding to Pythagorean triples or, alternatively, co-rotated hexagonal lattices, produces localization-delocalization-transition with stable thresholdless solitons.

In \cite{Mukherjee_arXiv2017} a stable flat band in a rhombic quasi-1D lattice optical waveguide is observed in the presence of external driving. Robust localization is found to originate from destructive interference associated with the symmetry in the lattice geometry. Excitation of dispersive bands then leads to Bloch oscillations and coherent destruction of tunneling. In \cite{Morales_Vicencio_PRA2016}, a method is presented for the construction of different single and multiple flat bands using 1D optical lattices with configurational symmetry, such as the sawtooth and rhombic chains. Cubic Kerr-effect-related nonlinearities and the continuum-limit are addressed. Next-to-nearest-neighbor and other extra-interactions are used to create localization for specific power and coupling values. This is an example of existence of compactons in discrete nonlinear quantum lattices, established numerically. 

One additional result is relevant for the perspective of the present work. In \cite{Kevrekidis2002}, perfectly compact solutions are obtained analytically in discrete lattices for artificial on-site and inter-site potentials. That is, the potentials are not inspired by physical reasoning but rather are designed to produce compact solutions, the so-called compactlets. This result is noteworthy since it is a unique example, disregarding the present work, where compact solutions are obtained analytically for discrete nonlinear lattices, albeit artificial ones. Moreover, linear stability analysis is performed numerically, in an approximate fashion. A survey of different results related to local-support solutions in the aforementioned physical systems is given in  \cite{Jason2016}.

The objective of the present work, in the discussed context, is to obtain analytic results for the existence and stability of perfectly compact solutions in physically reasonable discrete mechanical lattices, possible asymptotically.

In \cite{FlachOVG2017}, FB generators were constructed for one-dimensional settings. It is shown there that there exist certain classes of lattices having configurational symmetry that lead to FBs. A typical one is the so-called cross-stitch lattice.

As mentioned, the compacton modes exist in FB systems already in completely linear settings, and persist if the added nonlinearities respect the symmetry. In the nonlinear setting, however, the compactons can become unstable. The stability analysis of such periodic solutions is rather complicated, as it involves direct computation of the monodromy matrix. This paper addresses certain special types of mechanical lattices with FBs. Nonlinear interactions in the considered models lead to possible instability of the compactons, and the exploration of this instability is one of our main research goals. We circumvent, to some extent, the difficulties related to the stability analysis, by taking, in the first instance, the nonlinearity to be introduced into the considered system in the form of impact constraints, which can easily be realized in mechanical lattices. We use an augmented version of the 1D cross-stitch lattice, and analyze the stability of the compacton solutions. The principal idea is to use the methodology of Floquet analysis employing analytically constructed saltation matrices, the way it was done for the case of discrete breathers, as in \cite{Gendelman2013, Perchikov2014}. Both numerical and asymptotic methods of analysis, available for models with vibro-impact potential, are employed herein for the compactons and the counterpart DB modes, to examine stability. The two basic instability mechanisms revealed in the vibro-impact cross-stitch lattice, namely the local and global instabilities, appear also in a chain of massless boxes with internal nonlinear oscillators, which was examined in the second part of this paper, following the insight gained in \cite{Perchikov2016}.

It was recently shown in \cite{Perchikov2016}, that in a discrete mechanical system comprised of a massless box in harmonic on-site potential, having two identical oscillators inside, characterized by two-term (linear-cubic) interaction forces with the box, there exists a flat band corresponding to antisymmetric dynamics of the  internal masses. The compact dynamics remains stable in a finite parameter range in the nonlinear regime. A (more or less) rigorous stability analysis of the compact solution is enabled, in part, due to the choice of a massless box and the type of coupling. The single-element result can be naturally interpreted to apply to the anti-continuum limit (weak normalized inter-site potential) in a chain of such elements. The present work shows also, by numerical integration, that the compacton solution remains stable also for a chain of such elements, for finite inter-site stiffness values.

The structure of the paper is as follows: Section \ref{Sect3} introduces the augmented version of the cross-stitch model with non-smooth nonlinearities, Secs. \ref{Sect4}-\ref{Sect5} present the complete stability analysis and numerical verification of the results for the non-smooth model; Sec. \ref{Sect2} addresses the chain of massless boxes with internal nonlinear oscillators, related to the system analyzed in \cite{Perchikov2016}. Section \ref{Sect2b} establishes the existence of stable compacton solutions for the chain by direct numerical integration, and is followed by Sec. \ref{Sect6} with concluding remarks.

\section{Description of the model-- (augmented) cross-stitch lattice with vibro-impact on-site potential}
\label{Sect3}

As a model system, we employ the one-dimensional cross-stitch lattice discussed in \cite{Morales_Vicencio_PRA2016} and \cite{FlachOVG2017}, with the addition of  identical shear springs of generically different stiffness coefficient, connecting the two chains site-to-site (rather than diagonally, as the cross-stitching). Furthermore, the on-site vibro-impact potential with unit restitution coefficient (elastic impacts) is added to each mass in the lattice, similarly to the case in \cite{Perchikov2014}. Geometric nonlinearities are neglected, and only the motion in the axis direction is considered. The system is sketched in Fig. \ref{Fig9a} below.

\begin{figure}[H]
\begin{center}
{\includegraphics[scale = 0.5]{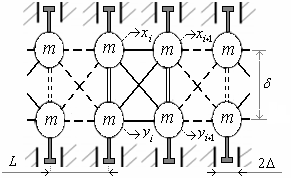}}
\end{center}
\caption{\small Sketch of the 1D horizontal `cross-stitch' chain. Lines: linear springs ($k$); double lines: shear springs ($k_{\text{shear}}$); (gray) anvils: rigid impactors; $\delta \ll L$}
\label{Fig9a}
\end{figure}

The equations of motion for the cross-stitch lattice given in Fig. \ref{Fig9a} are as follows (where $x_n(t),y_n(t)$ are the horizontal displacements of the lower and upper chain elements, respectively):
\begin{equation}
\label{CSeqmot1}
\begin{split}
\forall \ n \in \mathbb{N} \le N_s: \ m\ddot{x}_n+k_{\text{shear}}(x_n-y_n)\\+k(4x_n-x_{n-1}-x_{n+1}-y_{n-1}-y_{n+1})=0,\\
m\ddot{y}_n+k_{\text{shear}}(y_n-x_n)+k(4y_n-x_{n-1}-\\ x_{n+1}-y_{n-1}-y_{n+1})=0, \ \forall \
|x_n|,|y_n|  < \Delta; \\ \frac{\dot{y}_n|_{t|_{|y_n|\to \Delta}+\delta}}{\dot{y}_n|_{t|_{|y_n|\to \Delta}-\delta}} \underset{\delta \to 0}\to\frac{\dot{x}_n|_{t|_{|x_n|\to \Delta}+\delta}}{\dot{x}_n|_{t|_{|x_n|\to \Delta}-\delta}}\underset{\delta \to 0}\to -1
\end{split}
\end{equation}

\subsection{Linear dynamics}

Taking the difference of the first two equations in Eq.  (\ref{CSeqmot1}), one obtains the equation of motion for the detached mode corresponding to the flat band:
\begin{equation}
\label{CSeqmot2}
\begin{split}
m\ddot{v}_n+(4k+2k_{\text{shear}})v_n=0
\end{split}
\end{equation}
\begin{equation}
\label{Defvn}
v_n \triangleq x_n-y_n, \ \forall \ n \in \mathbb{N} \le N_s
\end{equation}

The flat band frequency is given by ($k_s\triangleq k_{\text{shear}}$):
\begin{equation}
\label{LamCom3}
\omega_{\text{FB}}  \triangleq \sqrt{\frac{4k+2k_s}{m}}
\end{equation}

Taking the sum of the two first equations in Eq.  (\ref{CSeqmot1}), one obtains the known equation of motion for a linear chain (with $2k$ for link stiffness):
\begin{equation}
\label{CSeqmot3}
\begin{split}
m\ddot{u}_n+2k (2u_n-u_{n-1}-u_{n+1})=0
\end{split}
\end{equation}
where $u_n \triangleq x_n+y_n, \ \forall \ n \in \mathbb{N} \le N_s$. The following dispersion relation stems from Eq. (\ref{CSeqmot3}):
\begin{equation}
\label{CSD}
\begin{split}
\omega_1(q)=2\sqrt{\frac{k}{m}}(1-\cos{q})^{1/2}
\end{split}
\end{equation}

\begin{figure}[H]
\begin{center}
{\includegraphics[scale = 0.5]{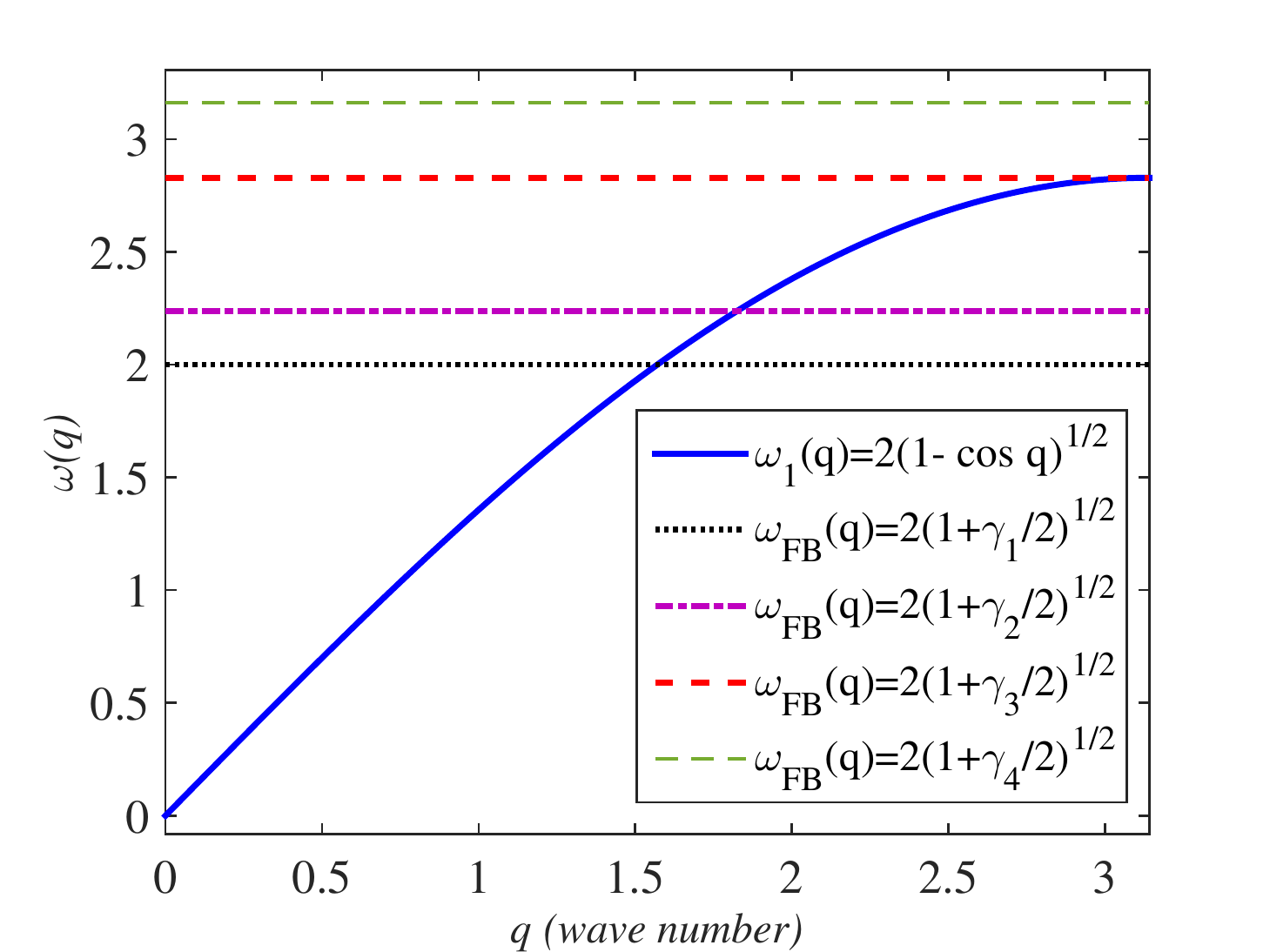}}
\end{center}
\caption{\small Dispersion relations (frequencies given in units of $\sqrt{k/m}$), $\gamma_1=0,\gamma_2=1/2,\gamma_3=2,\gamma_4=3, \gamma\triangleq k_s/k$}
\label{Fig9b}
\end{figure}

Dispersion curves are presented in Fig. \ref{Fig9b}, where one can observe intersection between the flat and the dispersive bands (no self-avoidance, as there is no linear coupling, unlike in \cite{Merhasin2005}).

To specify linear dynamics beyond dispersion analysis, the system (in the linear regime) can be represented in the following dimensionless dynamical systems' vector form:
\begin{equation}
\label{FL1}
{\textbf{x}}'=\textbf{A}\textbf{x}
\end{equation}
where $()'$ denotes differentiation with respect to time that is scaled by $\sqrt{m/k}$, and:
\begin{equation}
\label{FL2}
\textbf{A} \triangleq\begin{bmatrix}
\textbf{0}&\textbf{I}\\
-\hat{\textbf{A}}&\textbf{0}\\
\end{bmatrix}
\end{equation}
in which  $\textbf{0}$ is a matrix of zeros of size $N\times N$, $\textbf{I}$ is the identity matrix of rank $N$, and:
\begin{equation}
\label{FL3}
\begin{split}
\hat{A}_{ij}=(4+\gamma)\delta_{ij}-\gamma\delta_{i,j+1}\langle N-1-j\rangle^0 \\
-\gamma\delta_{i+1,j}\langle N-1-i\rangle^0-\delta_{i,j+2}\langle N-2-j\rangle^0 \\
-\delta_{i+2,j}\langle N-2-i\rangle^0-\delta_{i,j+3}\langle N-3-j\rangle^0 \\
-\delta_{i+3,j}\langle N-3-i\rangle^0
\end{split}
\end{equation}

As aforementioned, $\gamma \triangleq k_s/k$, and:
\begin{equation}
\label{Eqheav}
\langle{x}\rangle^0 \triangleq \begin{cases} 1 \ , \ \forall \ x \ge 0 \\ 0 \ , \ \forall \ x<0  \end{cases}
\end{equation}

Also, $N=2N_s$, $N_s$ being the number of sites.

\subsection{Nonlinear dynamics (regime of impacts)}

The nonlinearity enters into the play when some of the lattice elements are engaged in conservative impacts, described by the following relation:

\begin{equation}
\label{FN1}
\textbf{x}(t_i^+)=\textbf{C}_j\textbf{x}(t_i^-)
\end{equation}
with $t_i$ being the temporal series of impact instances (measured in units of $\sqrt{m/k}$), and where:
\begin{equation}
\label{FN2}
\textbf{C}_j \triangleq\begin{bmatrix}
\textbf{I}&\textbf{0}\\
\textbf{0}&\hat{\textbf{C}}^{(j)}
\end{bmatrix} \ , \  \hat{C}^{(j)}_{ik}\triangleq \delta_{ik}-2\delta_{ik}\delta_{ij}
\end{equation}
with $j=1,2$ denoting the index of the particle at the localization site.

\section{Linear stability analysis}
\label{Sect4}
\subsection{General framework}
\label{sectGF}
To analyze the stability of periodic (modal) solutions, the monodromy matrix can be represented as follows \cite{Champneys} (see \ref{AppendixA} for details):
\begin{equation}
\label{FS1}
\textbf{M}=(\textbf{L}\textbf{S}\textbf{L})^2 \ , \ \textbf{L}=\exp{\left(\frac{\pi}{2\hat\omega}\textbf{A}\right)} \ , \ \textbf{S}=\textbf{S}_2 \textbf{S}_1
\end{equation}
where $\hat\omega=\omega/\sqrt{k/m}$ is the normalized frequency of the periodic solution ($\omega$ being the frequency) and:
\begin{equation}
\label{FS2}
\textbf{S}_j=\textbf{C}_j+(\textbf{A}\textbf{C}_j-\textbf{C}_j\textbf{A})\textbf{D}_j \ , \  j=1,2
\end{equation}
\begin{equation}
\label{FS3}
\textbf{D}_1 \triangleq \frac{\textbf{L}\textbf{x}^0[\textbf{n}^{(1)}]^\top}{[\textbf{n}^{(1)}]^\top\textbf{A}\textbf{L}\textbf{x}^0} , \textbf{D}_2 \triangleq \frac{\textbf{C}_1\textbf{L}\textbf{x}^0[\textbf{n}^{(2)}]^\top}{[\textbf{n}^{(2)}]^\top\textbf{A}\textbf{C}_1\textbf{L}\textbf{x}^0}
\end{equation}
where $n_i^{(j)}=\delta_{ij}$ for $j=1,2$ and $i=1,2,..,N$. 

With no loss of generality, impulsive initial conditions are assumed:
\begin{equation}
\label{FS4}
x^0_i=v^0_i \langle i-N_s -1/2\rangle^0
\end{equation}
where Eq. (\ref{Eqheav}) is used and where $x^0_i$ is a phase-space state vector, in which displacements are normalized by $\Delta$ and velocities --  by $\Delta\sqrt{k/m}$.

\subsection{Analytical integration of the compacton mode}

For the compacton mode, using \cite{Perchikov2014}, one gets:
\begin{equation}
\label{FI1}
v_i(t)=2\bar{v}_1(t)\delta_{1i}
\end{equation}
where $v_i(t)$ is as defined in Eq. (\ref{Defvn}) normalized by $\Delta$ (thus $,y_i(t)=-x_i(t)=-\frac{1}{2}v_i(t)$), and where:
\begin{equation}
\label{FI2}
\begin{split}
\bar{v}_1(t)=\frac{\sqrt{\kappa}}{\tan\left(\frac{\pi\sqrt{\kappa}}{2}\right)}\left \lbrace \left \lvert \pi-2\pi \left\lbrace \frac{\omega t}{2\pi}\right\rbrace _{\text{frac}} \right \rvert -\frac{\pi}{2} \right. \\
+\left. \frac{4\kappa}{\pi}\left [ \Xi \left(\chi,\sqrt{\kappa}\right)-\frac{1}{16}\Xi \left(2\chi,\frac{\sqrt{\kappa}}{2}\right)\right ]  \right \rbrace
\end{split}
\end{equation}
\begin{equation}
\label{FI3}
\begin{split}
\Xi (\zeta,a) \triangleq \frac{6\pi^2\lbrace \frac{\zeta}{2\pi} \rbrace _{\text{frac}}\left(1-\lbrace \frac{\zeta}{2\pi} \rbrace _{\text{frac}}\right)-\pi^2}{6a^2} \\
-\frac{\pi\cos\left[\pi a\left(1- 2\lbrace \frac{\zeta}{2\pi} \rbrace _{\text{frac}}\right)\right]}{2a^3\sin(\pi a)}
\end{split}
\end{equation}
with: $\lbrace z \rbrace _{\text{frac}}\triangleq z-\lfloor z \rfloor, \chi\triangleq \omega t-\frac{3\pi}{2}$, $\kappa \triangleq \omega_{\text{FB}}^2/\omega^2$.

The eigenvalues of the monodromy matrix constructed as described in Sec. \ref{sectGF} for the compacton solution, cannot be computed analytically for the complete set of the parameter values (at least, we have not succeeded in doing that). However, it is possible to obtain asymptotic expansions of these eigenvalues for the asymptotic limits considered below.

\subsection{Stability boundaries of the compacton mode -- asymptotic results}

\emph{The case of $\gamma \ll 1$.}
It is possible to obtain analytic approximations for the stability bounds of the compact solution by taking advantage of the compactness. For this sake, a chain of three two-mass elements is considered. The central element corresponds to the excited part of the lattice. The two neighbors are the minimum required to represent periodic boundary conditions symmetrically. The monodromy matrix for this three-element system, containing six masses and having the dimension of $12\times 12$, can be constructed analytically. Furthermore, the matrix can be expanded for small values of $\gamma$ -- a third-order Taylor series proves sufficient. Order-balancing then requires that the near-critical dimensionless quarter-period of oscillations, $\hat{q}(\gamma)$,  be expanded in powers of the stiffness ratio,$\gamma$, as well (it cannot be large since a finite upper existence bound holds from linear dynamics).

Next, exact analytical calculation of the spectral norm of the asymptotic expansion of the monodromy matrix in radicals is performed (the degeneracy renders the 12th-order polynomial characteristic equation exactly solvable). In the unstable case (identified \emph{a posteriori}) two eigenvalues are equal to unity, there are four conjugate pairs on the unit circle and a pair of real reciprocals. Expanding the smooth portions of the radicals for the real larger-than-unity eigenvalue as four-term positive-power expansions in $\gamma$ (less proves insufficient) and calculating the limit, yields the following (critically-unstable-case) asymptotic expression:
\begin{equation}
\label{LamCom1}
|\lambda|_{\text{max}} \underset{\gamma \to 0} \to 1+4\sqrt{\frac{2}{3}}q_0\sqrt{3-4q_0^2}\gamma+\mathcal{O}\left(\gamma^2\right)
\end{equation}
where $q_0$ emerges from the orders-balancing dimensionless quarter-period expansion, namely:
\begin{equation}
\label{LamCom2}
\hat{q}(\gamma) \triangleq \frac{\pi}{2\hat{\omega}}\underset{\gamma \ll 1 }\to \frac{\pi\omega_{\text{FB}}}{4\omega}\to q_0 \gamma^{1/2}+\mathcal{O}(\gamma)
\end{equation}

The critical value rendering the spectral norm of the monodromy bounded by unity for $\gamma \ll 1$ is thus:
\begin{equation}
\label{LamCom4}
q_0^{\text{cr}} = \frac{\sqrt{3}}{2}
\end{equation}
which corresponds to the following stability-bounding asymptotic relation in the frequency--stiffness-ratio plane:
\begin{equation}
\label{LamCom5}
\frac{\omega_{\text{cr}}}{\omega_{\text{FB}}}\underset{\gamma \to 0} \to\frac{\pi}{2\sqrt{3}}\gamma^{-1/2}
\end{equation}

Since this critical value corresponds to a transition from a pair of complex conjugate eigenvalues on the unit circle to a pair of real reciprocal eigenvalues, one would expect a pitchfork bifurcation in the underlying chain at this stability bound.

Once the critical value is obtained for the three-site construction, we verify that it remains the critical value for stability also for arbitrary (longer) closed chains. In other words, it holds that the obtained radical remains a root also for the higher-order characteristic equation emerging for the larger chain. Consequently, the obtained critical bound should be regarded not as a feature of the three-site chain, but rather as an asymptotic result true for an arbitrary chain (including an infinitely long one, where closed and open chains are equivalent). 

For the case of vanishing shear stiffness, the classical cross-stitch lattice (with the addition of vibro-impact on-site nonlinearity) is reproduced, and for this case, the above result serves as a  proof of linear stability of a (one-site) compacton for an arbitrarily long closed chain for all frequencies.

Figure \ref{Fig10a} shows the agreement between the numerical calculation of the stability map for the compacton solution, and the asymptotic estimate.

\emph{The case of $\gamma \gg 1$.}
The $\gamma \gg 1$ case is the anti-continuum limit in which the chain decouples and one effectively has a single elementary cell including the two masses in vibro-impact potential, connected by a linear shear spring. The stability of the antisymmetric mode of vibration can be investigated analytically. It turns out that two eigenvalues of the corresponding monodromy matrix are equal to unity and two additional eigenvalues are real positive and have a product of unity. The smallest of these two is treated here for the sake of better presentation. 

It appears that this eigenvalue depends only on a certain combination of the two parameters of the problem, namely $\gamma$ and $\hat\omega$, and not on both of them independently. A plot of the dependence of the critical eigenvalue on that combination reveals that the system is always unstable in this limit, as can be learned from Fig. \ref{Fig10b}.
Moreover, calculation of the eigenvector of the monodromy matrix of the full chain (in the middle of the feasible range-of-change of the aforementioned combination of parameters) reveals strict localization (as one might expect by na\"ive understanding of the considered limit). This is shown in Fig. \ref{Fig10c} .

\subsection{Numeric stability analysis for a long cross-stitch lattice}

Numerical spectral analysis of analytically constructed monodromy matrices for different stiffness-ratio/frequency pairs, reveals the existence of a growing stable stripe, with upper and lower bounds, for decreasing shear-spring stiffness. The upper (frequency) stability limit corresponds to the pitchfork bifurcation, and physically is related to the internal instability of the antisymmetric mode in the unit cell. 

\begin{figure}[H]
\begin{center}
{{\includegraphics[scale = 0.4]{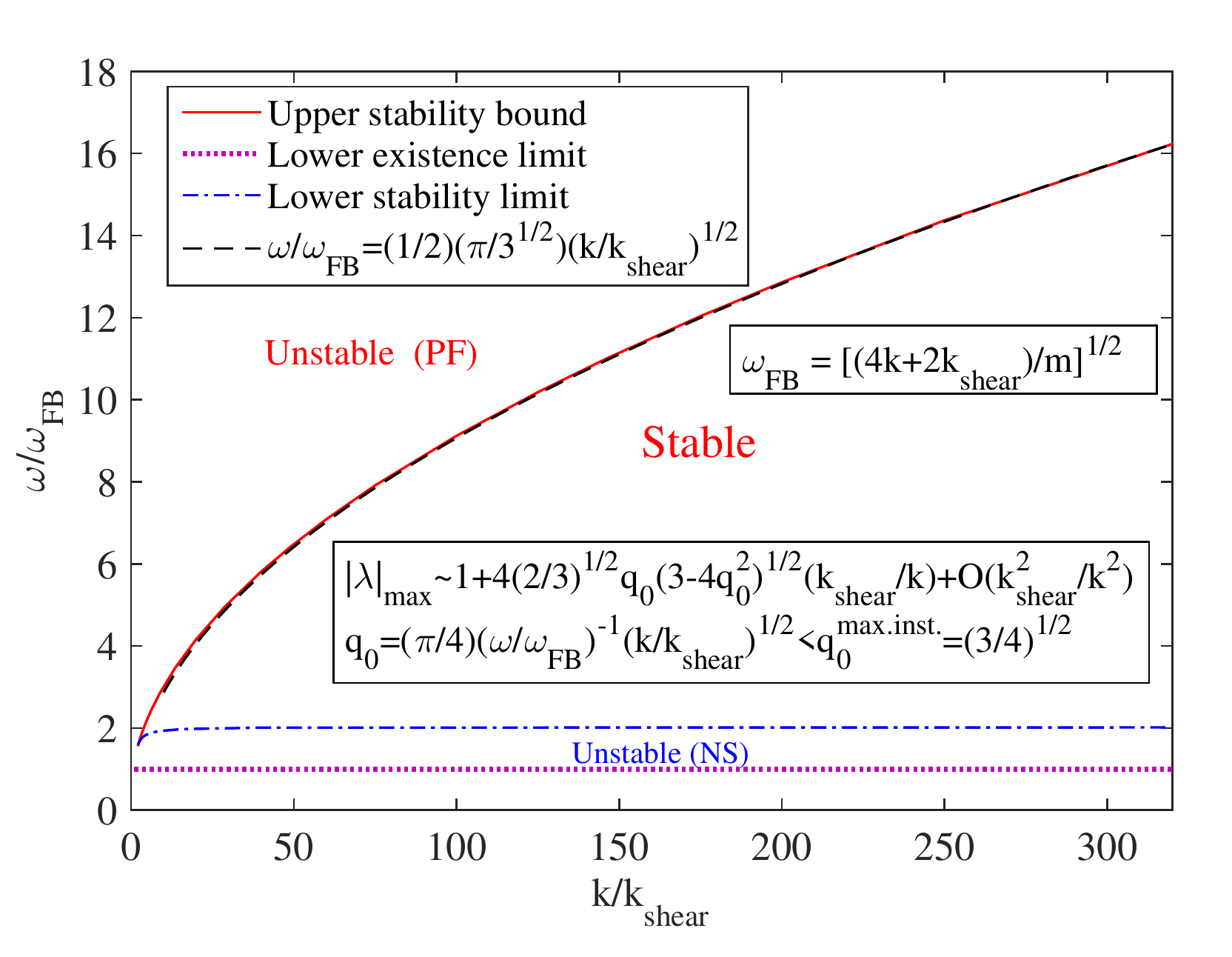}}}
\end{center}
\caption{\small Stability bounds of the compacton mode for $N \gg1$ -- numeric results in solid line (red online) and dash-dotted line (blue online) and analytic results in dashed black line. The labels `PF' and `NS' denote the pitchfork and Neimark-Sacker bifurcations, respectively}
\label{Fig10a}
\end{figure}

This conclusion is confirmed by almost perfect coincidence between the numeric simulation of the complete system and the analytic results for the single unit cell. The lower stability limit corresponds to the Neimark-Sacker bifurcation; this instability scenario is absent in the unit cell and should be attributed to interaction with the propagation zone of the lattice. The bounds, along with the corresponding asymptotic estimates and the flat band as a reference, are shown in Fig. \ref{Fig10a}.

\begin{figure}[H]
\begin{center}
{{\includegraphics[scale = 0.42]{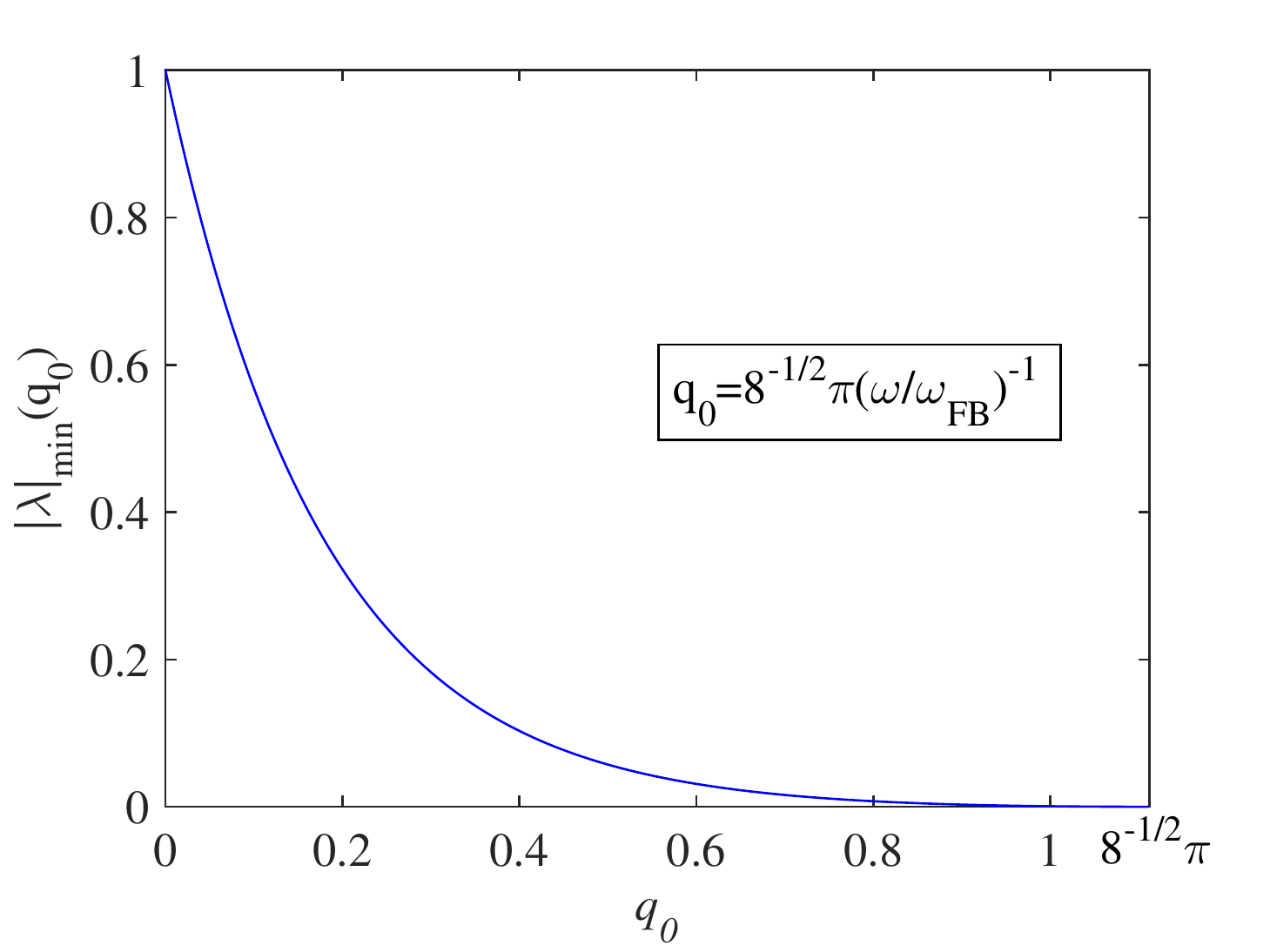}}}
\end{center}
\caption{\small Critical Floquet multiplier variation ($\gamma \gg 1$), compacton mode, $N=1$}
\label{Fig10b}
\end{figure}

\begin{figure}[H]
\begin{center}
{{\includegraphics[scale = 0.42]{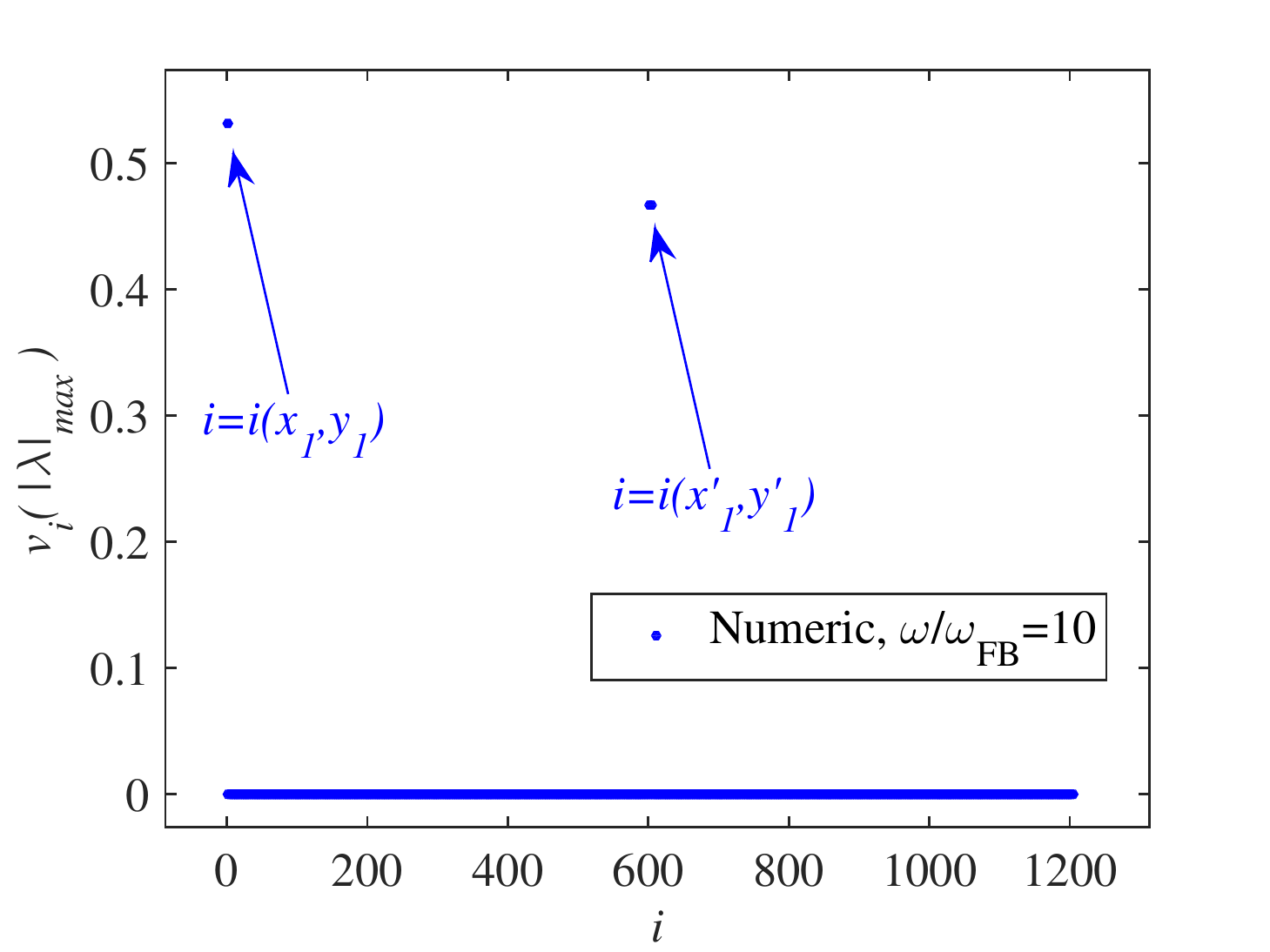}}}
\end{center}
\caption{\small Typical $\gamma \gg 1$ full chain monodromy eigenvector, showing localization in compacton mode ($N=301$)}
\label{Fig10c}
\end{figure}

\subsection{The alternative localization regime -- discrete breather (analytical DB-mode integration)}

As was established above, each unit-cell in the lattice can admit the antisymmetric mode, corresponding to the compacton solution. However, it can also oscillate in the symmetric mode. In that case, no destructive interference occurs, and thus one expects to observe the conventional discrete breather. Similar breathers were treated, for instance, in \cite{Grinberg2016}. Here, a more complete form is derived. The effective breather displacement, namely, $u_n(t) \triangleq x_n(t)+y_n(t)$ is given by a symmetric Fourier series (with the associated auxiliary quantities):

\begin{equation}
\label{FBI1}
u_n(t)=2\sum_{l=0}^{\infty}{\hat{V}_{nl}\cos{[(2l+1)\omega t}]}
\end{equation}
\begin{equation}
\label{FBI2}
\begin{split}
\hat{V}_{1,l} \triangleq \left({\bar \alpha_l}{\hat\alpha_l^{2}}\hat{W}_{1l} \right.
\left.+\hat\alpha_l{\bar \alpha_l^{2}}\hat{W}_{2l}\right) p
\end{split}
\end{equation}
\begin{equation}
\label{FBI3}
\begin{split}
\hat{V}_{n\neq 1,l} \triangleq \left(\hat\alpha_l^{n}\hat{W}_{1l} \right.
\left.+\bar \alpha_l^{n}\hat{W}_{2l}\right)p
\end{split}
\end{equation}
\begin{equation}
\label{FBI4}
\begin{split}
p \triangleq \Delta \left[\sum_{l=0}^{\infty}{\left({\bar \alpha_l}{\hat\alpha_l^{2}}\hat{W}_{1l} \right.
\left.+{\hat\alpha_l}{\bar \alpha_l^{2}}\hat{W}_{2l}\right)}\right]^{-1}
\end{split}
\end{equation}
where $\Delta$ is the amplitude of the breather, or half the distance between the limiters. In addition,
\begin{equation}
\label{FBI5}
\begin{split}
A_l \triangleq \frac{\hat{\alpha}_l^3}{\bar{\alpha}_l^3}\frac{2\alpha_l\hat{\alpha}_l^{N_s-3}-\hat{\alpha}_l^{N_s-4}-\bar{\alpha}_l/\hat{\alpha}_l}{2\alpha_l\bar{\alpha}_l^{N_s-3}-\bar{\alpha}_l^{N_s-4}-\hat{\alpha}_l/\bar{\alpha}_l}
\end{split}
\end{equation}
\begin{equation}
\label{FBI6}
\begin{split}
B_l \triangleq A_l\left[\left(4\alpha_l^2-2\frac{\alpha_l}{\bar{\alpha}_l}-1\right)\bar{\alpha}_l^{N_s}-\bar{\alpha}_l^2\right ] \\
-\left[\left(4\alpha_l^2-2\frac{\alpha_l}{\hat{\alpha}_l}-1\right)\hat{\alpha}_l^{N_s}-\hat{\alpha}_l^2\right ]
\end{split}
\end{equation}
\begin{equation}
\label{FBI7}
\begin{split}
\hat{W}_{1l} \triangleq \frac{2}{\pi}B_l^{-1} \ , \ \hat{W}_{2l} \triangleq -A_l \hat{W}_{1l}
\end{split}
\end{equation}
\begin{equation}
\label{FBI8}
\begin{split}
\hat\alpha_l \triangleq \alpha_l+\sqrt{\alpha_l^2-1} \ ,\bar \alpha_l \triangleq \alpha_l-\sqrt{\alpha_l^2-1}
\end{split}
\end{equation}
\begin{equation}
\label{FBI9}
\begin{split}
\alpha_l \triangleq 1-\hat{\omega}^2(l+1/2)^2
\end{split}
\end{equation}
(recall that $\hat{\omega}\triangleq\sqrt{\frac{m}{k}}\omega$). The initial conditions for the stability analysis of the breather mode are zero displacements with consistent  initial velocities:
\begin{equation}
\label{FBI10}
\begin{split}
u_n(t=t_0)=0,v_n(t=t_0)=0,\dot{v}_n(t=t_0)=0
\end{split}
\end{equation}
and the following consistent (inhomogeneous) initial condition, with $t_0=\pi/(2\omega)$:
\begin{equation}
\label{FBI11}
\begin{split}
\dot{u}_n(t=t_0)=-2\omega\sum_{l=0}^{\infty}{(-1)^{l}(2l+1)\hat{V}_{nl}}
\end{split}
\end{equation}

\subsection{Stability boundaries of the DB mode -- asymptotic results}

\emph{The case of $\gamma \ll1$}.
Analytic estimation of the stability bounds of the breather mode in the $\omega$--$\gamma$ plane can be obtained asymptotically for $\gamma \ll 1$. To this end, once again, the previously discussed triple of elementary cells is considered. This is the smallest chain needed to represent a symmetric system with periodic boundary conditions. As in the case of the compacton mode, the stability criterion here prescribes an order of unity critical value on a combination of the parameters. One of these parameters is assumed asymptotically small, hence it is expected that the other parameter (here the frequency) would be of order different than unity in the critical case. Clearly, the frequency cannot be much smaller than the linear-regime limit, and thus a large critical value is expected. Unlike in the case of the compacton mode, where there is only one velocity value in the analysis, which (thus) cancels out, in the case of the breather, the exact initial velocity profile is required. For the asymptotic analysis considered here the high-frequency limit of Eq. (\ref{FBI11}) for $N_s=3$, as obtained by rigorous asymptotic treatment, is sufficient and reads, ($\hat{q}$ from  Eq. (\ref{LamCom2})):
\begin{equation}
\label{FBSA1}
\begin{split}
\dot{u}_1|_{t_0}=-\hat{q}^{-1}\sqrt{k/m}\Delta \ , \ \dot{u}_{2,3}|_{t_0}=\hat{q} \ \sqrt{k/m}\Delta
\end{split}
\end{equation}

Having now all the quantities required for the construction of the monodromy matrix (as functions of $\gamma$ and $\omega$), one can obtain it analytically. Taking $N_s=3$, just as for the compacton case, constructing the monodromy matrix and taking its two-variable second-order Taylor-series approximation with respect to the small parameters $\gamma$ and $\hat{q}$ (as an equivalent but alternative route to the one taken in the compacton case), and then computing the spectral norm exactly and further expanding the smooth parts of the radicals in three leading terms, and order-balancing, one obtains the real maximal eigenvalue of $\textbf{M}$ (for the critically unstable case) as:
\begin{equation}
\label{FBSA2}
\lambda_{\text{max}} \underset{\gamma \to 0} \to 1+4\sqrt{\frac{2}{3}}q_1\sqrt{4q_1^2-3}\gamma-16q_1^2\gamma^2
 \end{equation}
where $q_1$ is given by:
\begin{equation}
\label{FBSA3}
q_1 \triangleq \frac{\pi}{\sqrt{32}}\left(\frac{\omega}{\omega_{\text{prop}}}\right)^{-1}\gamma^{-1/2}
\end{equation}
and
\begin{equation}
\label{wprop}
\omega_{\text{prop}}\triangleq \underset{q}\max[{\omega_1(q)}]=\sqrt{8k/m}
\end{equation}

The critical value rendering the spectral norm of the monodromy matrix bounded by unity is the same as for the compacton:
\begin{equation}
\label{FBSA5}
q_1^{\text{cr}} = \frac{\sqrt{3}}{2}
\end{equation}

This corresponds to the following stability-bounding asymptotic relation in the frequency--stiffness-ratio plane:
\begin{equation}
\label{FBSA6}
\frac{\omega_{\text{cr}}}{\omega_{\text{prop}}}\underset{\gamma \to 0} \to\frac{\pi}{2\sqrt{6}}\gamma^{-1/2}
\end{equation}

One notes that in the $\gamma\to 0$ limit, the same curve in the $\omega$--$\gamma$ plane is obtained as the stability limit for either the compacton or the breather, with the compacton mode being stable below that line and the breather mode being stable above it, as arises from Eqs.  (\ref{LamCom3},\ref{CSD},\ref{LamCom1},\ref{LamCom5},\ref{FBSA2},\ref{FBSA6}).

\emph{The case of $\gamma \gg 1$}.
In the limit of large $\gamma$, the cross-stitch chain degenerates to the single two-mass "molecule". The stability of the symmetric mode of vibration of such a molecule can be examined analytically. It appears that once the assumption of a large $\gamma$ is introduced, allowing one to consider a single two-mass element, no further asymptotics is required, and the eigenvalues of the monodromy matrix can be obtained in closed-form. Two eigenvalues are identically unity, which represents the degeneracy in position and velocity arising from the symmetry. Two additional eigenvalues are generally complex conjugates with absolute values of unity. Those eigenvalues are the critical ones, and they are given by:
\begin{equation}
\label{FBSA7}
\lambda_{\text{cr}}=e^{\pm\mathrm{i}\pi\sqrt{\gamma}\left(\omega/\omega_{\text{prop}}\right)^{-1}}
\end{equation}

This result is a first-order approximation in $\gamma^{-1}$. Clearly, higher terms would perturb those eigenvalues. For eigenvalues with nonzero imaginary parts, the perturbation could only change the arguments of the eigenvalues in the complex plane, since the symmetry would keep them complex conjugates and energy conservation would keep the product of their radii at the value of unity. Hence, they would both remain on the unit circle, and no instability would arise. However, for the special case where all four eigenvalues are equal to unity, a small perturbation can drive two of the eigenvalues apart on the real axis, keeping their product unitary. In this case, one eigenvalue would be real and greater than unity, corresponding to instability. Therefore, an infinite (countable) set of narrow instability tongues (turning to lines for $\gamma\to \infty$) would emerge on the $\omega$--$\gamma$ plane in the perturbed (chain) case, obeying the following asymptotic relationship:
\begin{equation}
\label{FBSA8}
\frac{\omega_{\text{cr}}}{\omega_{\text{prop}}}\underset{\gamma\to\infty}\to\frac{\sqrt{\gamma}}{2\mathcal{N}} \ , \ \forall \ \mathcal{N} \in \mathbb{N}
\end{equation}

\begin{figure}[H]
\begin{center}
{{\includegraphics[scale = 0.44]{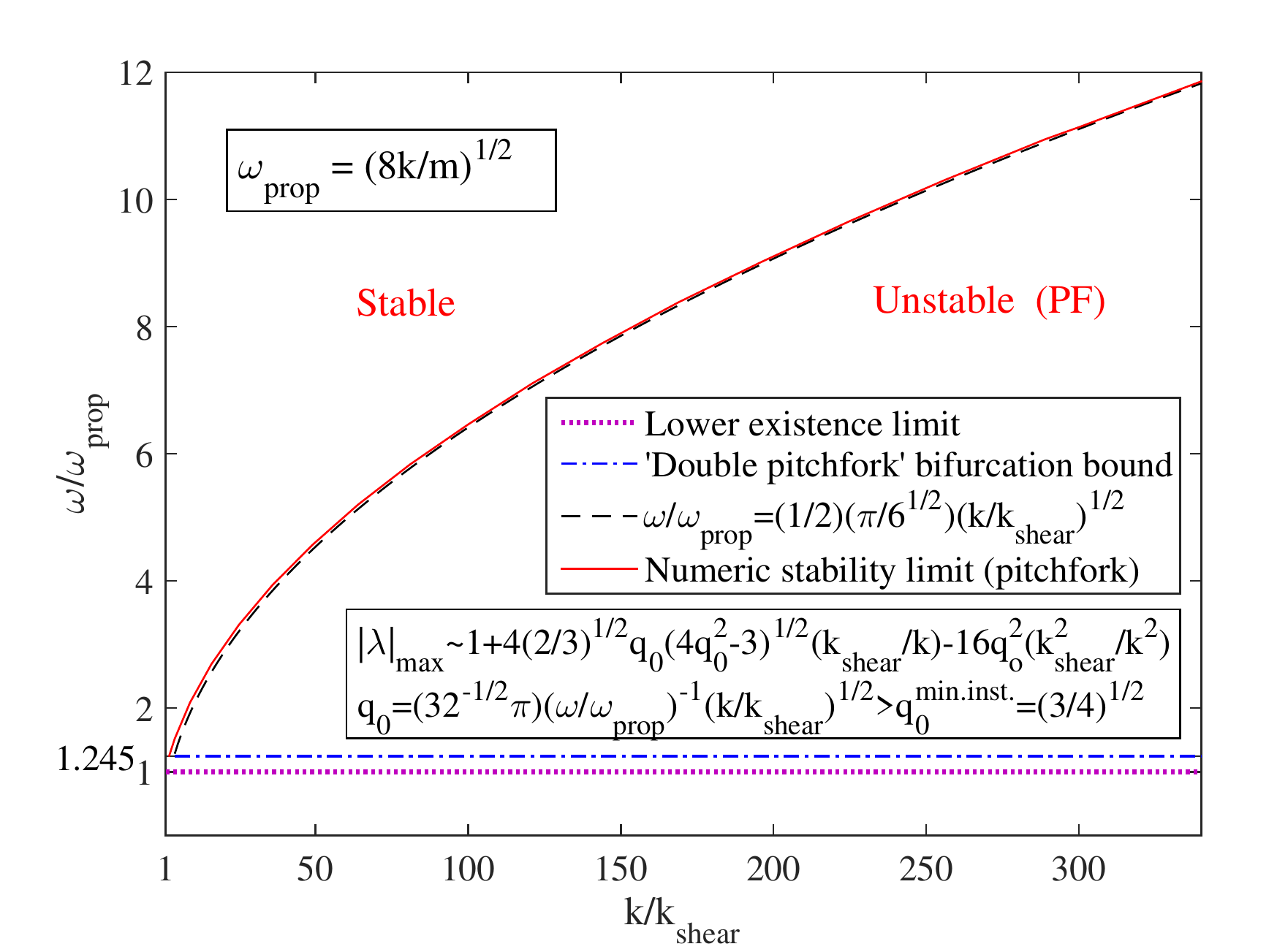}}}
\end{center}
\caption{\small Breather mode stability bounds for $\gamma \ll 1, N=301$ (asymptotic estimate shown as dashed line)}
\label{Fig11a}
\end{figure}

\begin{figure}[H]
\begin{center}
{{\includegraphics[scale = 0.44]{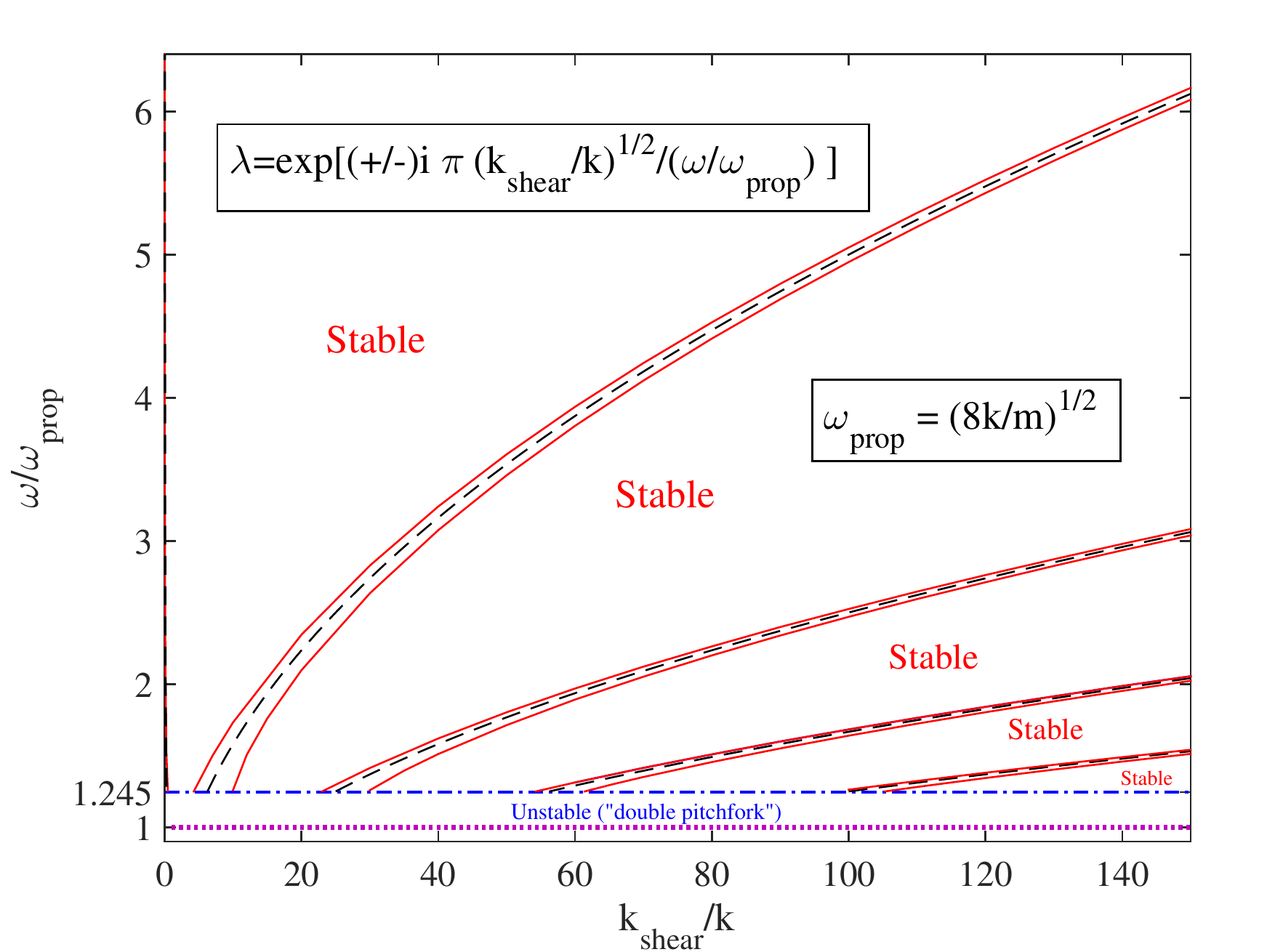}}}
\end{center}
\caption{\small Breather mode stability bounds for $\gamma \gg 1, N=301$ (asymptotic estimates shown as dashed lines)}
\label{Fig11b}
\end{figure}

The quality of the asymptotic estimates for the stability bounds is shown in Figs. \ref{Fig11a} and \ref{Fig11b}. Figure \ref{Fig11b} also features a lower stability bound, corresponding to something that can be called double pitchfork bifurcation. In this case, both chains experience the pitchfork-type loss of stability simultaneously, which implies the destruction of periodicity for the breather. This differs from the regular pitchfork bifurcation emerging at the upper bound, where stability is lost due to internal resonance, where part of the energy is transmitted from the symmetric to the antisymmetric mode.

\subsection{Stability bounds for the DB mode -- numeric results}

\section{Numerical integration}
\label{Sect5}

To validate the results of the stability analysis numerically, we employ an algorithmically stable integration scheme, in conjunction with a smooth model representing the vibro-impact potential. The widely-known high-odd-power force term is used for modeling impact. The solver employed here is a one-step method based on the trapezoidal rule and a backward differentiation formula of order 2 with a free interpolant \cite{Hosea1996}.

\subsection{Simulation of the compacton in and below the stability region}

\begin{figure}[H]
\begin{center}
{{\includegraphics[scale = 0.45]{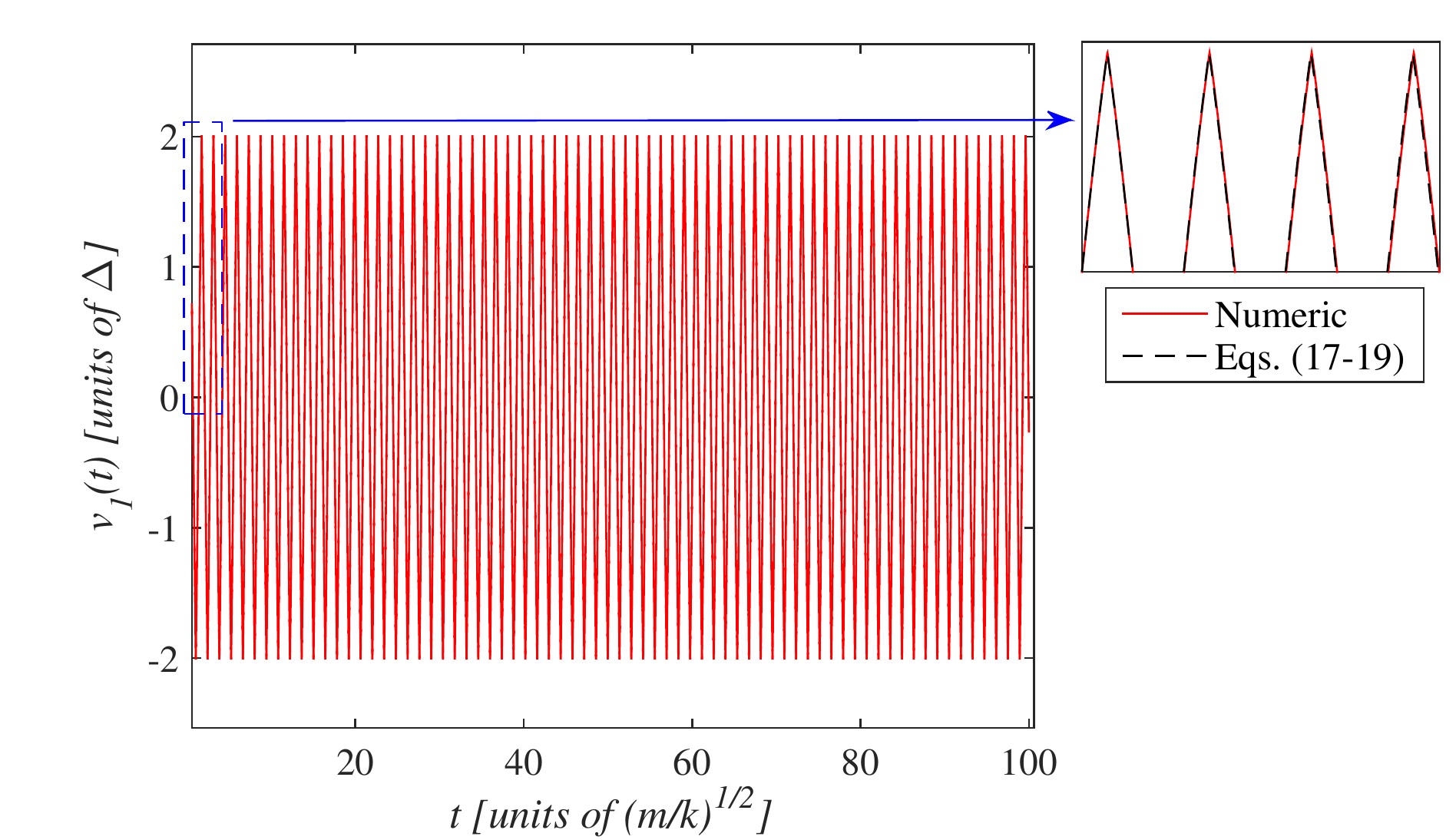}} \\
{\includegraphics[scale = 0.45]{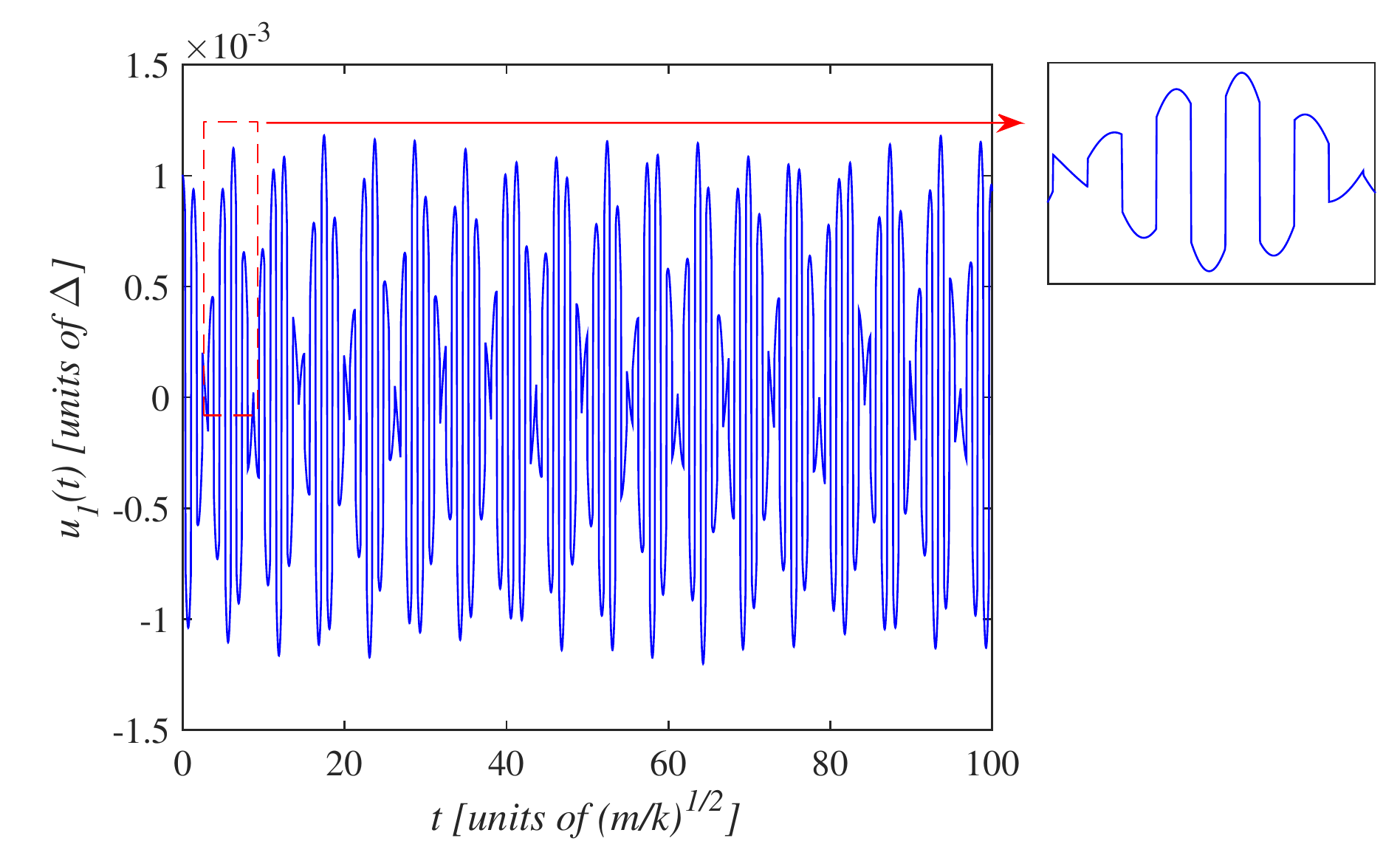}} \\
{\includegraphics[scale = 0.43]{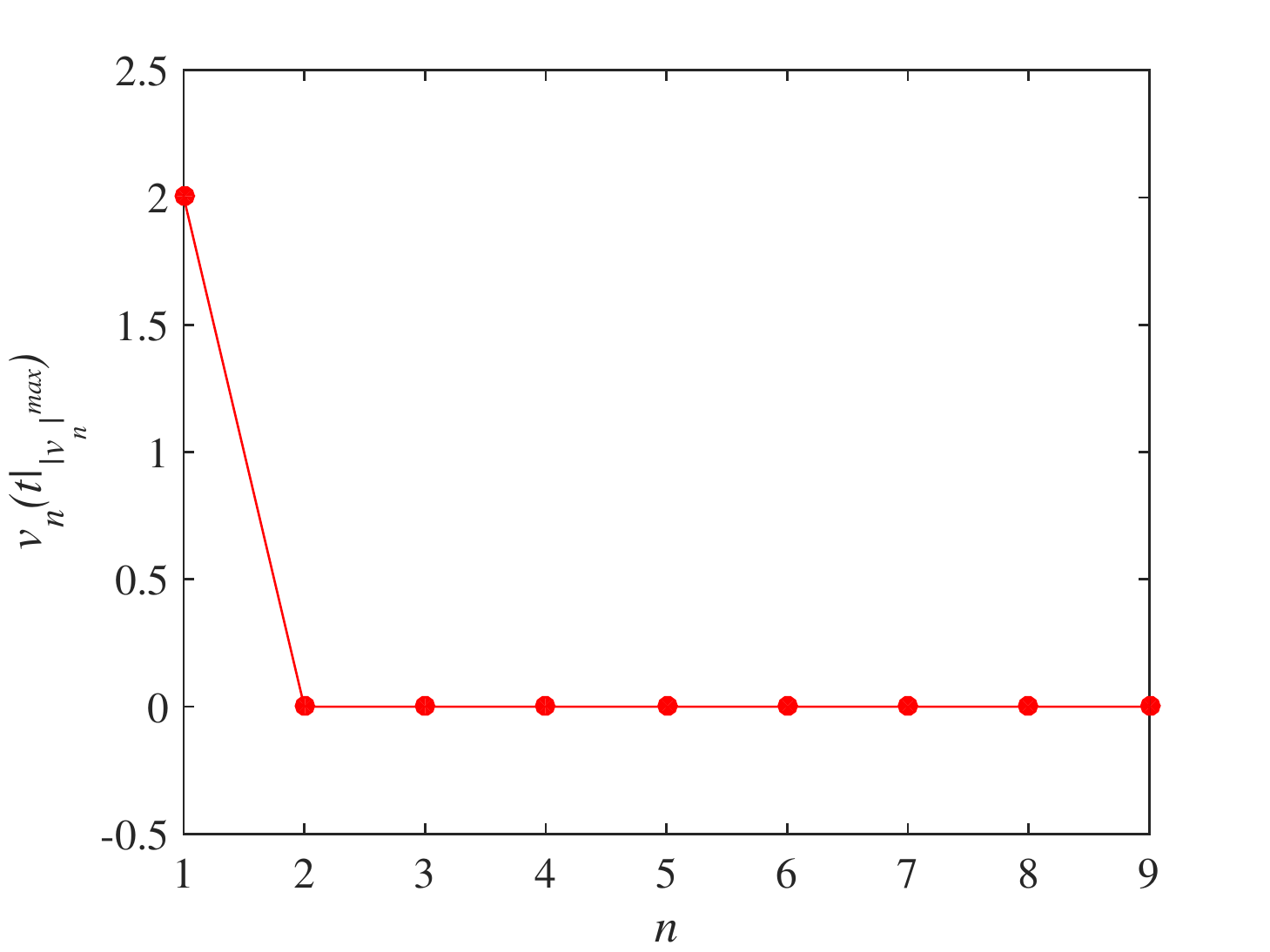}} \\
{\includegraphics[scale =0.43]{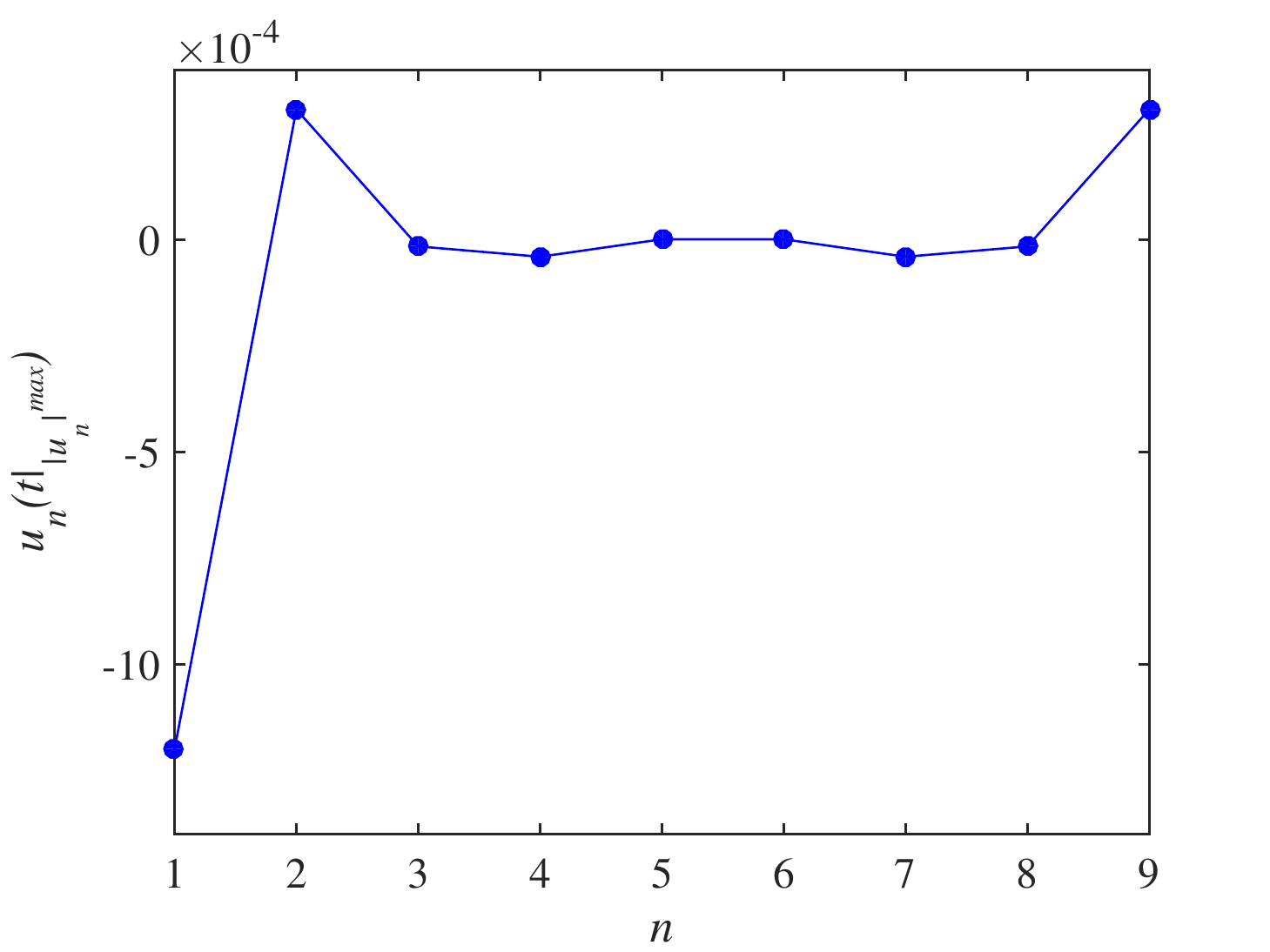}}}
\end{center}
\caption{\small Simulation results for the compacton -- stable case: $N=9,k/k_{\text{shear}}=10,\omega/\omega_{\text{FB}}=2.2,\dot{u}_1(0)=10^{-3}\sqrt{k/m}\Delta$}
\label{Fig12}
\end{figure}

For numerical integration, the system is presented in the following form:
\begin{equation}
\label{FBNI1}
\begin{split}
\ddot{\textbf{w}}=-\hat{\textbf{A}}\textbf{w}-\tilde{\textbf{f}}
\end{split}
\end{equation}
where $\hat{\textbf{A}}$ is defined in Eq. (\ref{FL3}). The displacements vector is: $\textbf{w}=(x_1,y_1,x_2,y_2,...,x_{N_s},y_{N_s})^{\top}$, and the nonlinear impact-representing forcing vector is given by:
\begin{equation}
\label{FBNI2}
\begin{split}
\tilde{f}_i\triangleq (2\xi+1)w_i^{4\xi+1}
\end{split}
\end{equation}
where $\xi$ is the force exponent, reproducing impact when $\xi\to\infty$. As in \cite{Perchikov2014}, $\xi=300$ was chosen here.
Figure \ref{Fig12} comprises, along with Fig. \ref{Fig10a}, perhaps the most important result in the present paper. It confirms the existence of a stable compacton in a nonlinear 1D lattice, by showing the dynamic boundedness of a perturbation to a perfectly compact solution. This is illustrated both by the time histories of the principal-site displacements and by the chain mode-amplitude profiles.

One can argue that the dashed line in Fig. \ref{Fig10a} gives a complete stability picture for the asymptotic limit, the solid (red online) curve therein confirms that one can rely on this limit, when the parameter values are finite, and Fig. \ref{Fig12} confirms that linear stability analysis provides reliable results for actual small yet finite perturbations.

\begin{figure}[H]
\begin{center}
{{\includegraphics[scale = 0.45]{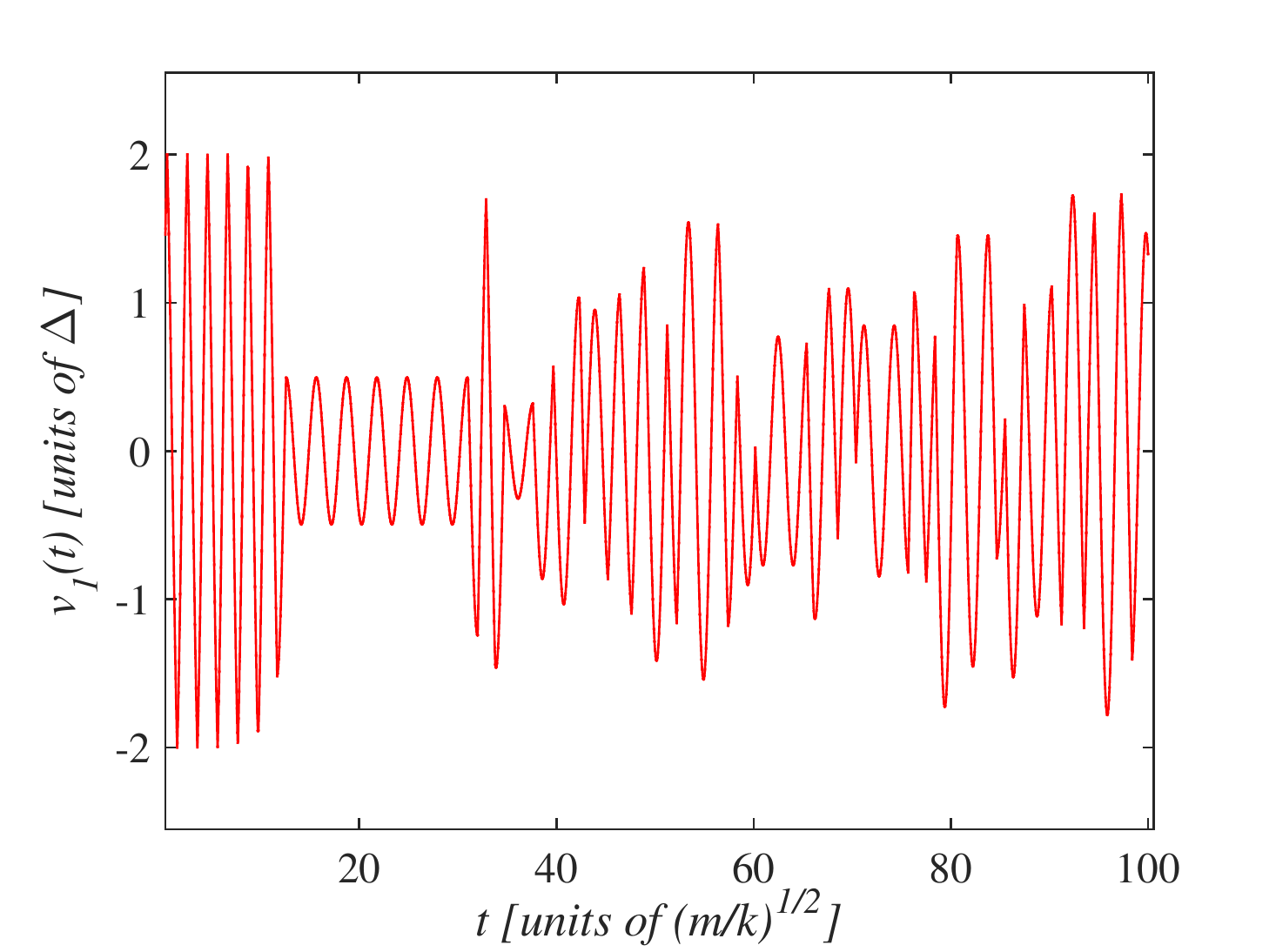}} \\
{\includegraphics[scale = 0.45]{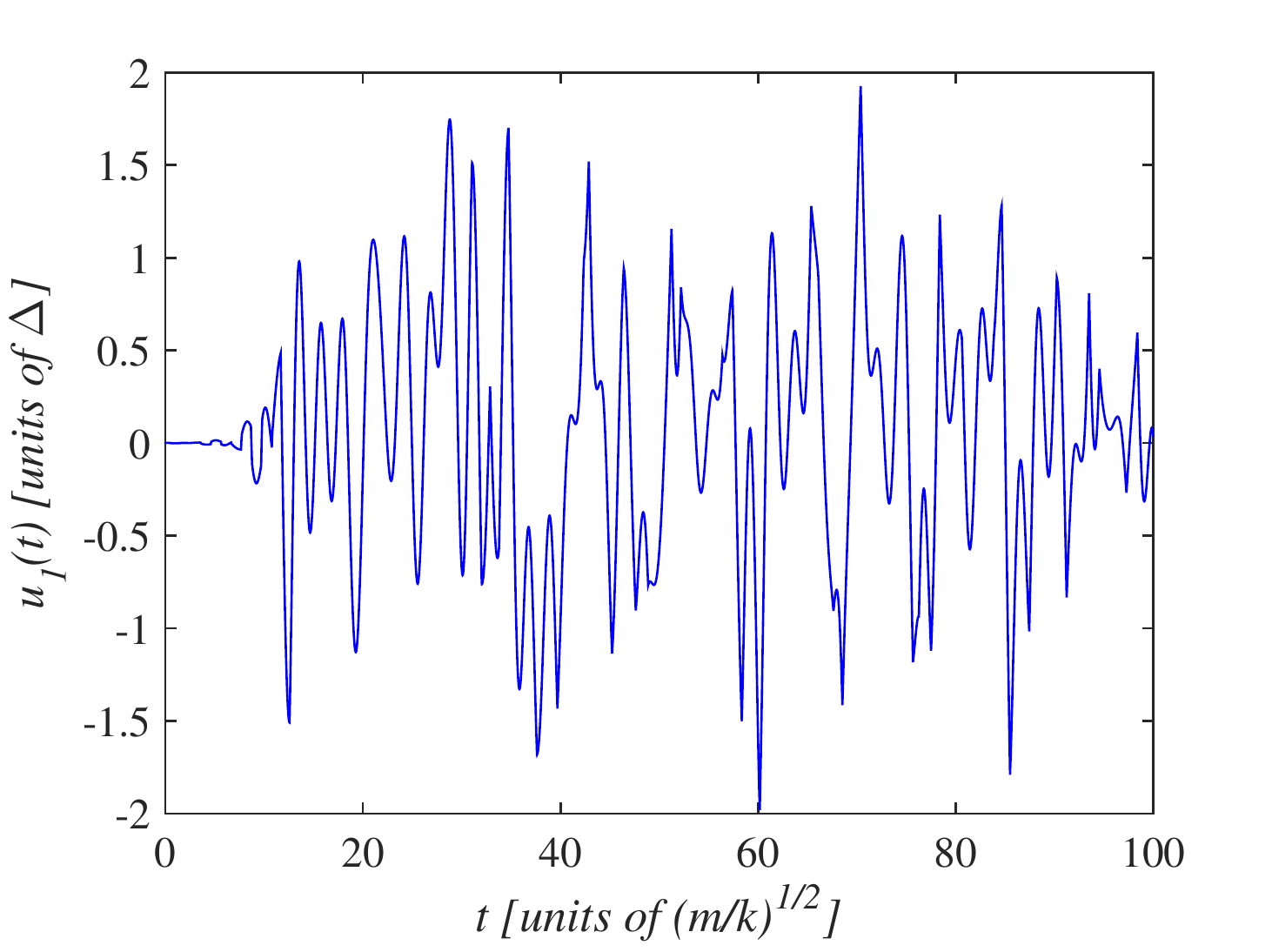}} \\
{\includegraphics[scale = 0.43]{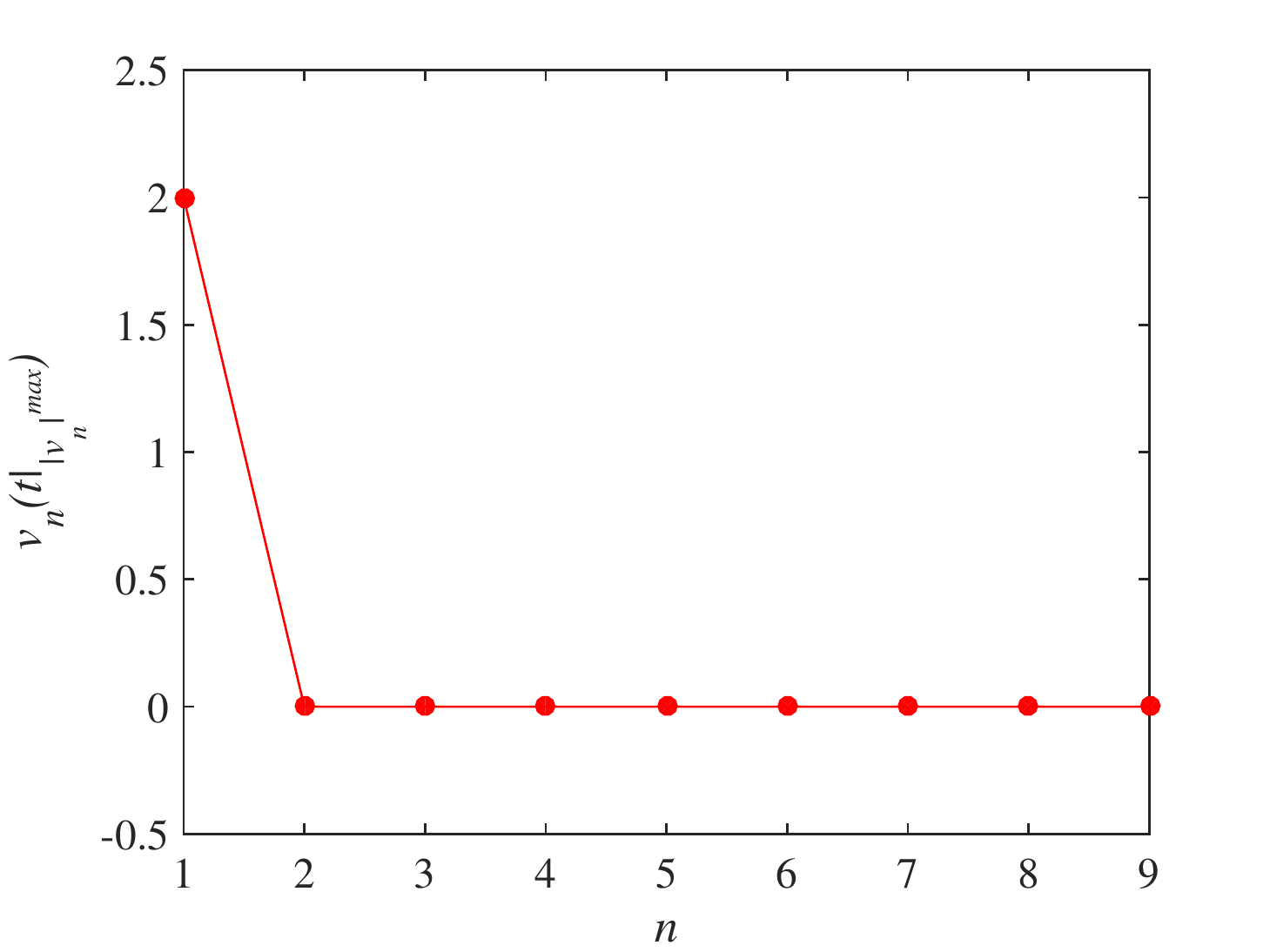}} \\
{\includegraphics[scale = 0.43]{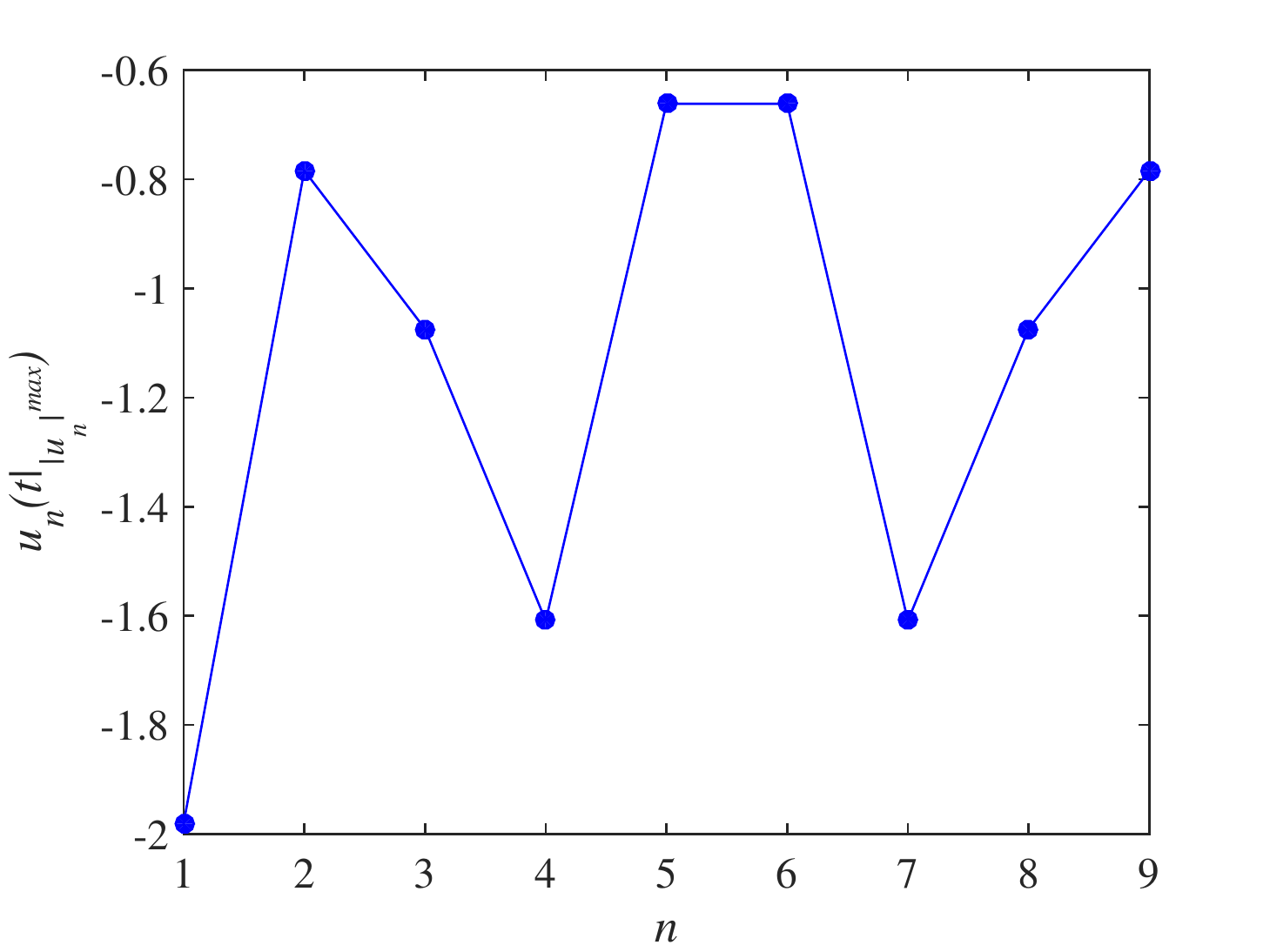}}}
\end{center}
\caption{\small Simulation results for the compacton  -- 'NS'-unstable case: $N=9,k/k_{\text{shear}}=10,\omega/\omega_{\text{FB}}=1.5,\dot{u}_1(0)=10^{-3}\sqrt{k/m}\Delta$}
\label{Fig13}
\end{figure}

Figure \ref{Fig13} shows that when the frequency is smaller than the critical (lower) bound, delocalization occurs, through the Neimark-Sacker (NS) bifurcation, leading to, arguably, chaotic-like trajectories. The initial state there is the same slightly perturbed compacton as in the case exhibited in Fig. \ref{Fig12}.

In both cases, the shear spring is taken to be an order of magnitude less rigid than the other springs. This is required for a stable solution to be feasible according to linear stability analysis.

\subsection{Simulation of the discrete breather in and below the stability region}
\begin{figure}[H]
\begin{center}
{{\includegraphics[scale =0.41]{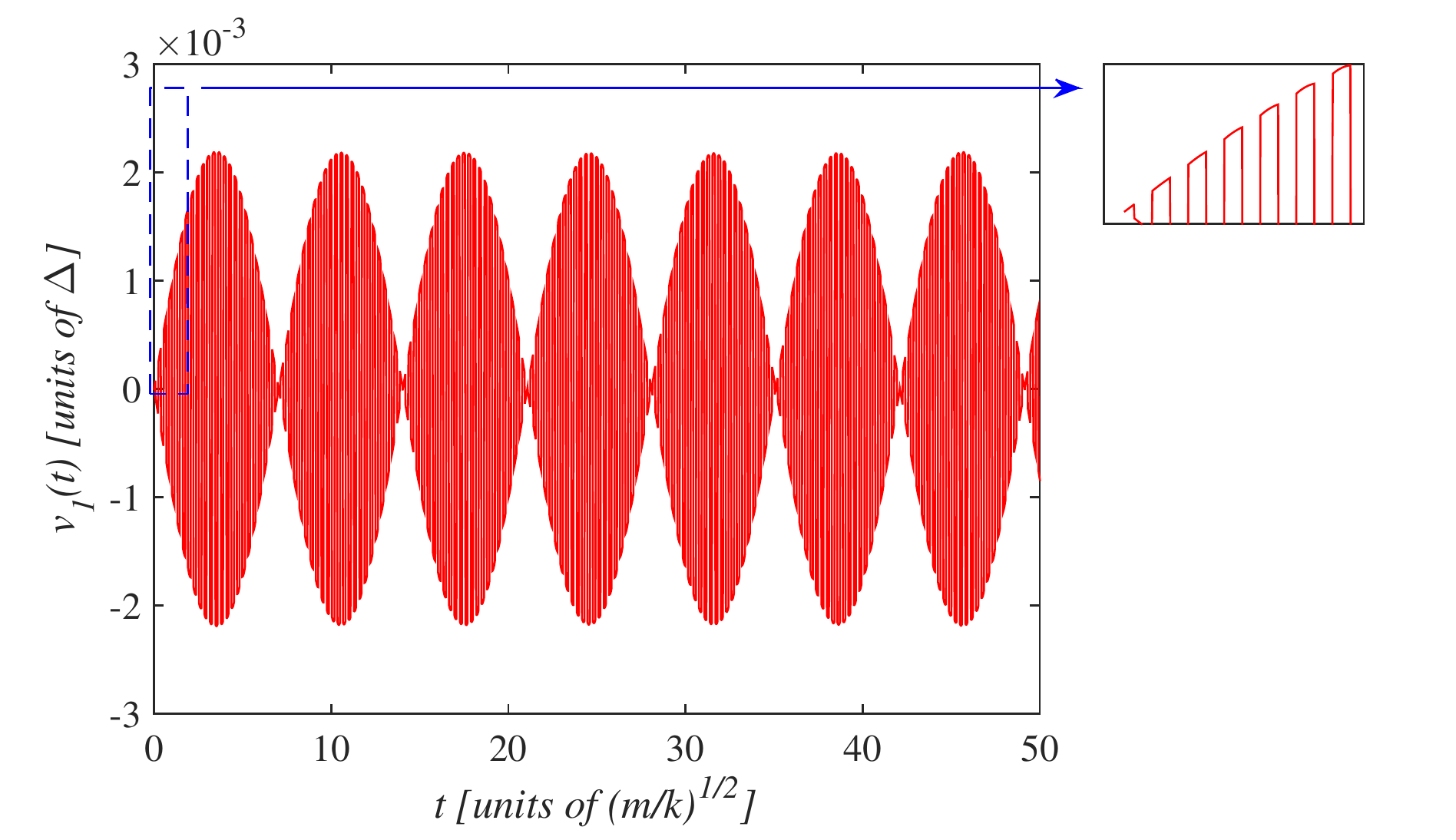}} \\
{\includegraphics[scale = 0.41]{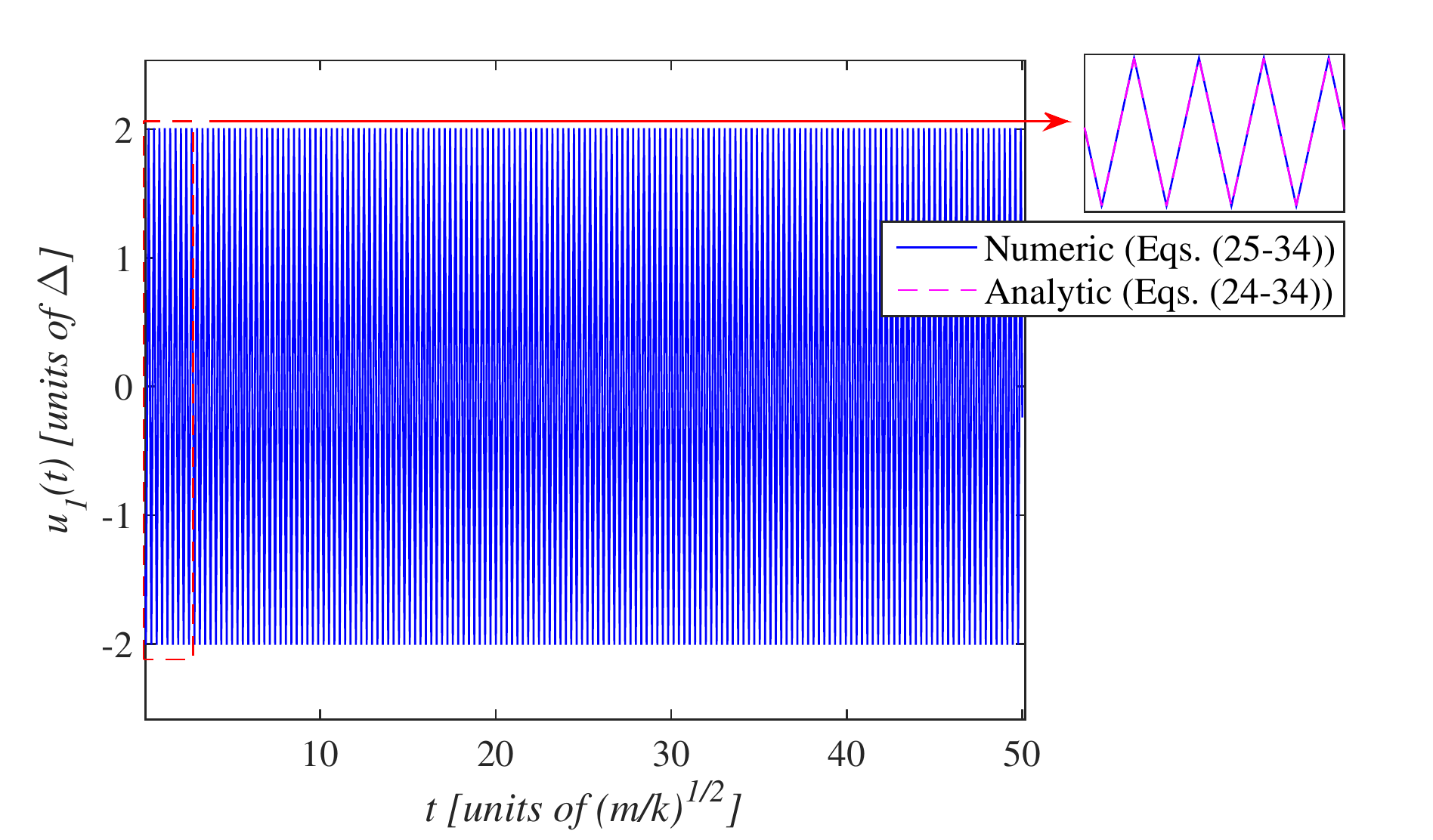}} \\
{\includegraphics[scale = 0.4]{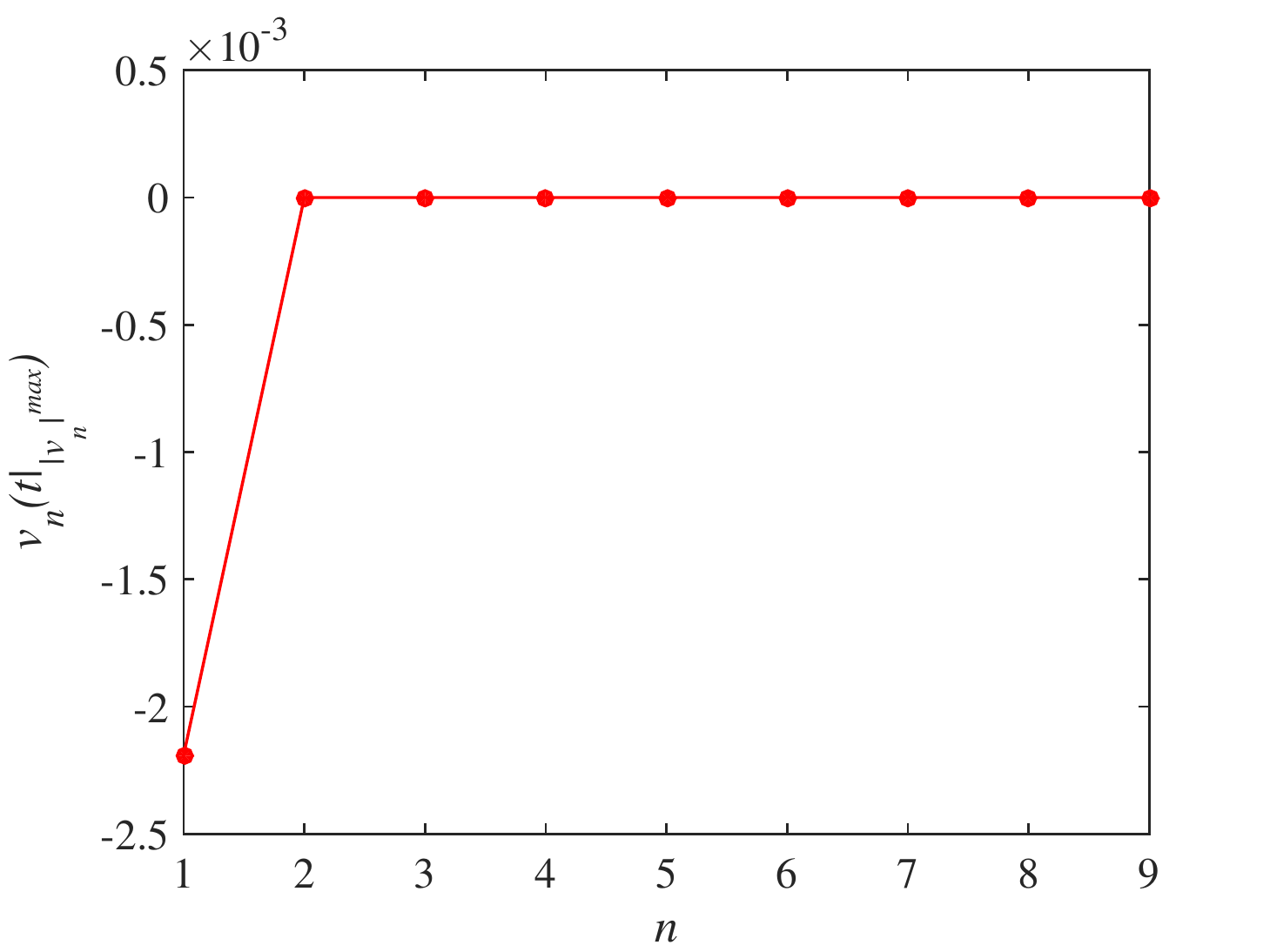}} \\
{\includegraphics[scale = 0.4]{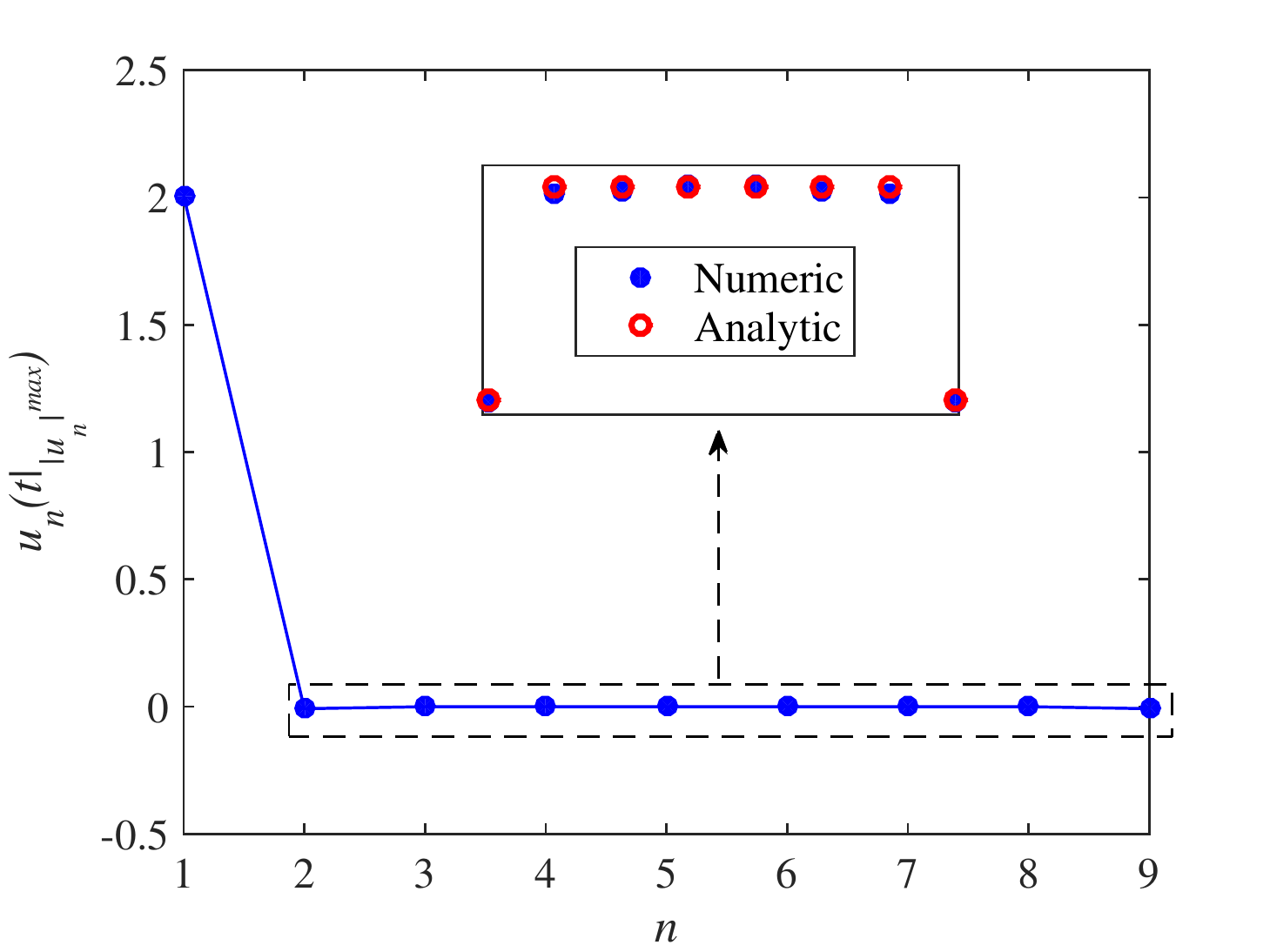}}}
\end{center}
\caption{\small Simulation results for the ``DB'' -- stable case: $N=9,k/k_{\text{shear}}=\omega/\omega_{\text{FB}}=10,\dot{v}_1(0)=10^{-3}\sqrt{k/m}\Delta$}
\label{Fig14}
\end{figure}

\begin{figure}[H]
\begin{center}
{{\includegraphics[scale =0.41]{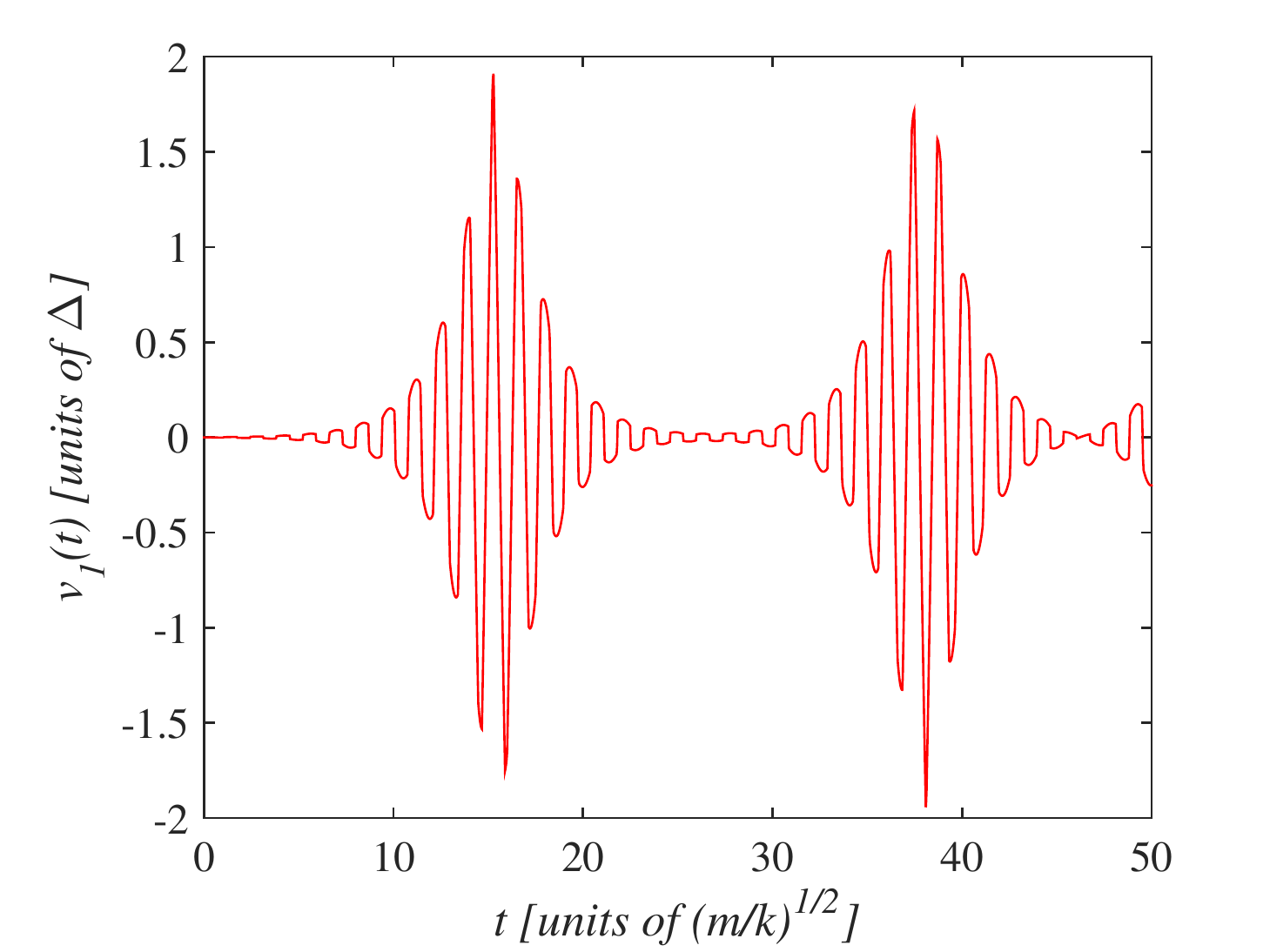}} \\
{\includegraphics[scale = 0.41]{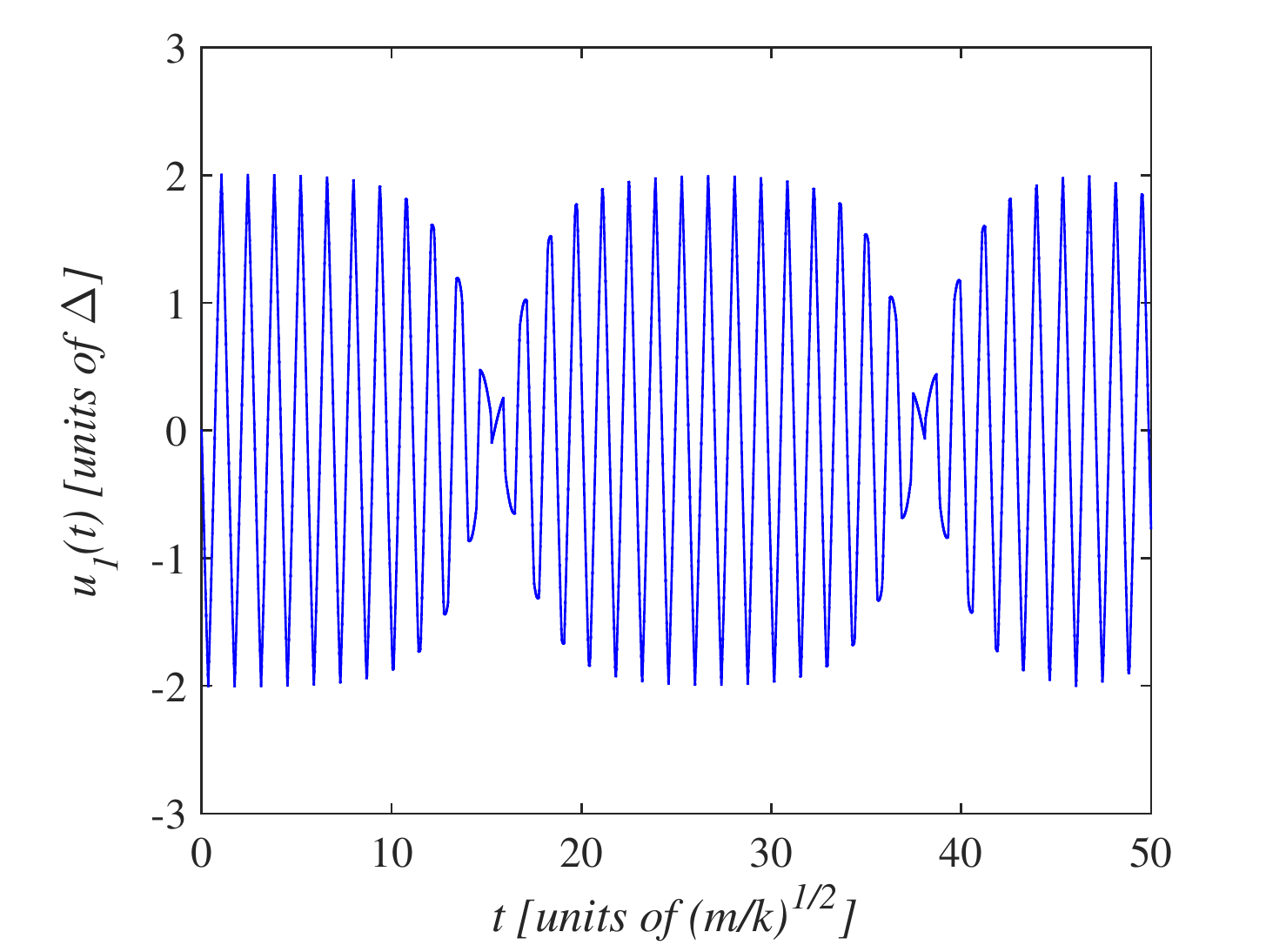}} \\
{\includegraphics[scale = 0.4]{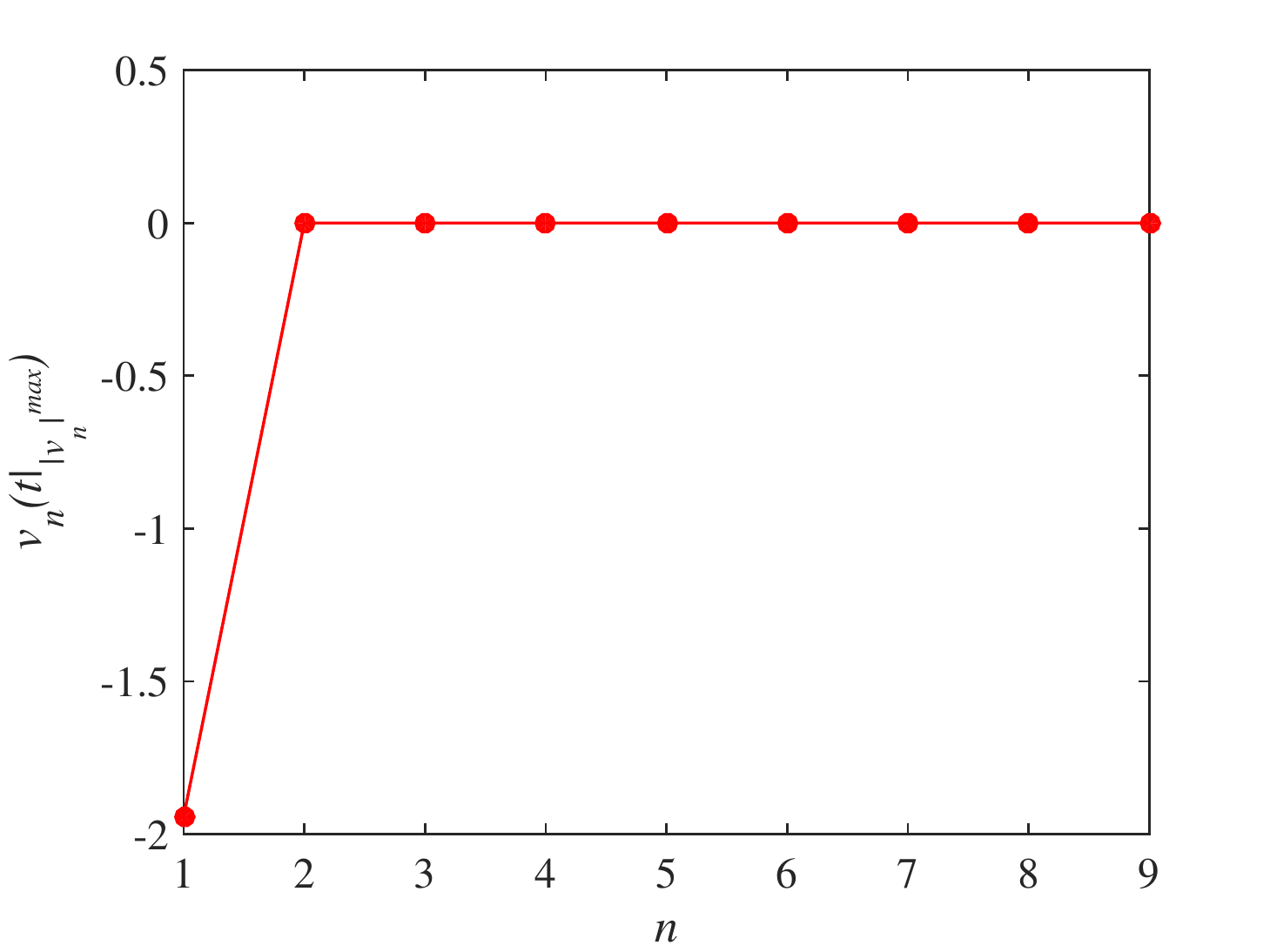}} \\
{\includegraphics[scale = 0.4]{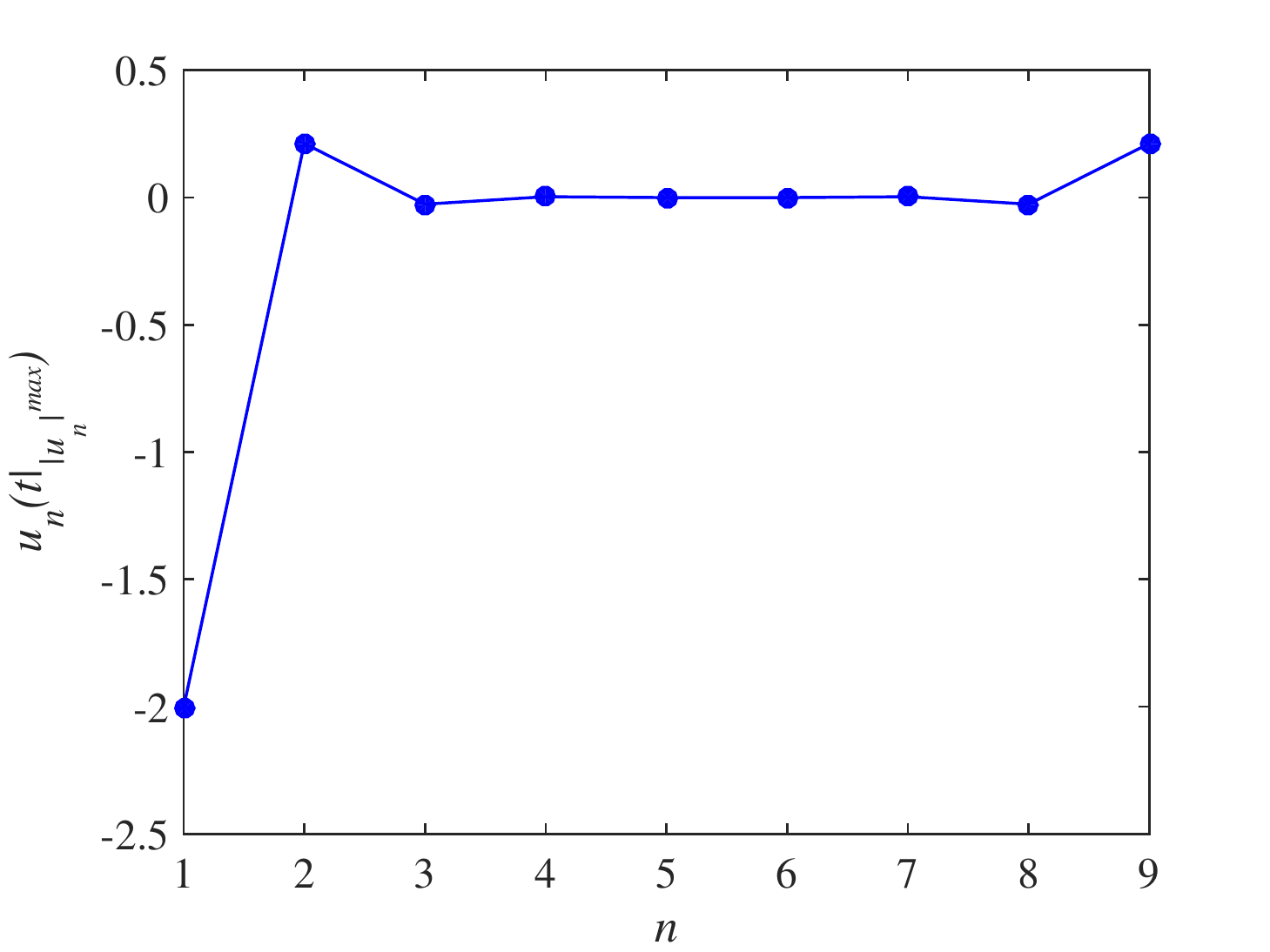}}}
\end{center}
\caption{\small Simulation results, 'DB' -- 'PF'-unstable case: $N=9,\frac{k}{k_{\text{shear}}}=10,\frac{\omega}{\omega_{\text{FB}}}=2.2,\dot{v}_1(0)=10^{-3}\sqrt{k/m}\Delta$}
\label{Fig15}
\end{figure}

The results for the numerical integration of the discrete breather are given in Figs. \ref{Fig14} and \ref{Fig15}.

Figure \ref{Fig14} confirms the stability of the breather for frequency above the lower threshold (noteworthy is the absence of an upper stability bound). Figure \ref{Fig15} illustrates the loss of stability of the breather mode for a frequency lower than the critical value. The long-time dynamics of the perturbed symmetric mode is not ``chaotic'' as the one emerging from the 'NS' bifurcation shown in Fig. \ref{Fig13}, but rather reminds (quasi) beating typical for local energy exchange between two nonlinear normal modes (NNMs). A similar picture is observed below, when crossing the 'PF' instability boundary to the unstable side for the compacton mode.

\subsection{Integration with compact initial conditions above the stable region -- global dynamics}
\label{sectGD}

Asymptotic analysis reveals that for the breather:
\begin{equation}
\label{FGD1}
\begin{split}
|\dot{u}_n(t)| \underset{q\to 0} \le \hat{q}^2 |\dot{u}_1(t)|
\end{split}
\end{equation}
(where $\hat{q}$ is as defined in Eq. (\ref{LamCom2})), which implies the asymptotic compactness of the breather mode. Moreover, the compacton mode is strictly compact anyway for all frequencies. Consequently, for high frequencies both modes are compact. Therefore, the entire dynamics is compact and is thus expected to be equivalent to the dynamics of a single element for high kinetic-to-potential energy ratios.

\begin{figure}[H]
\begin{center}
{{\includegraphics[scale = 0.39]{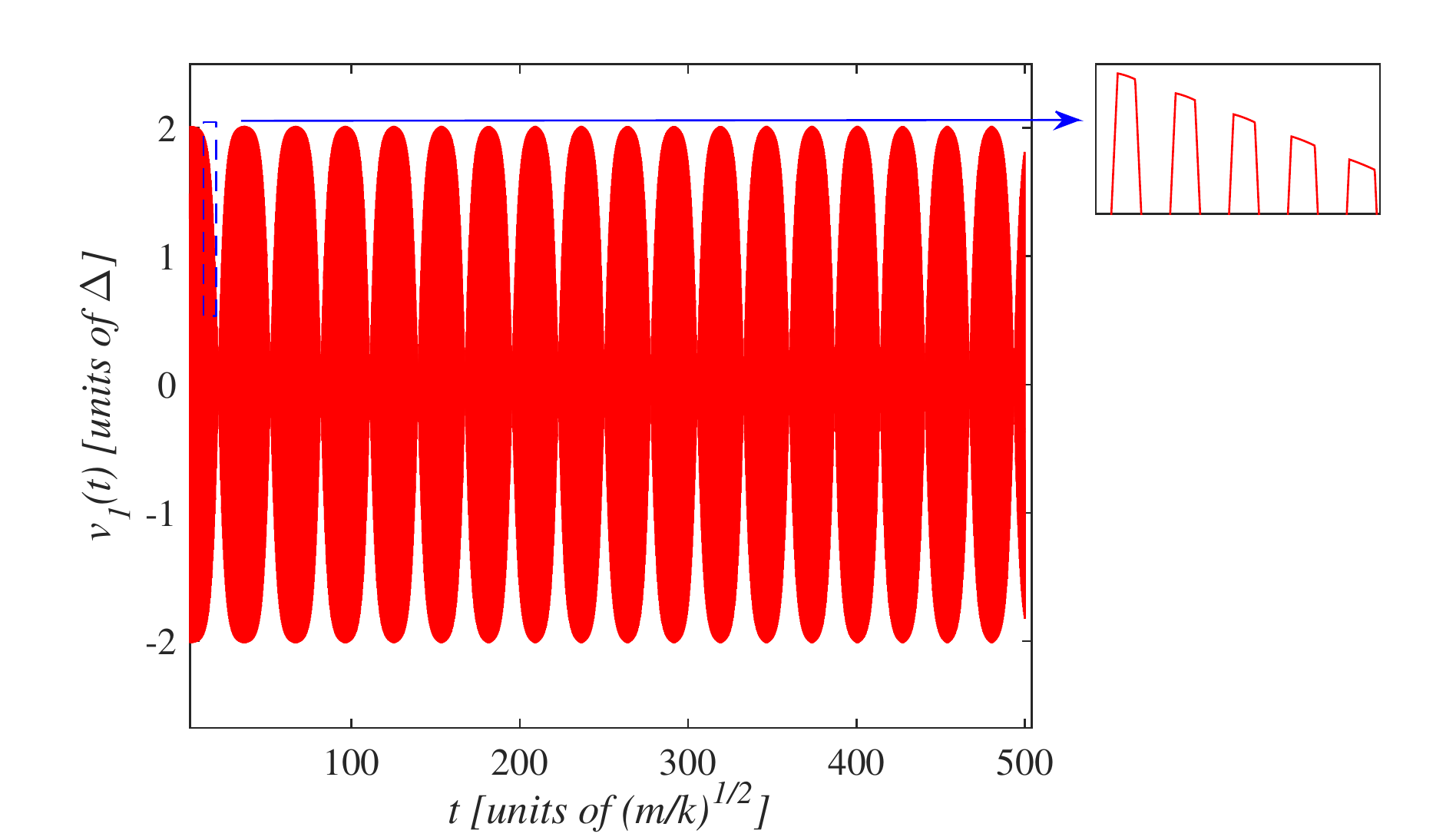}} \\
{\includegraphics[scale = 0.39]{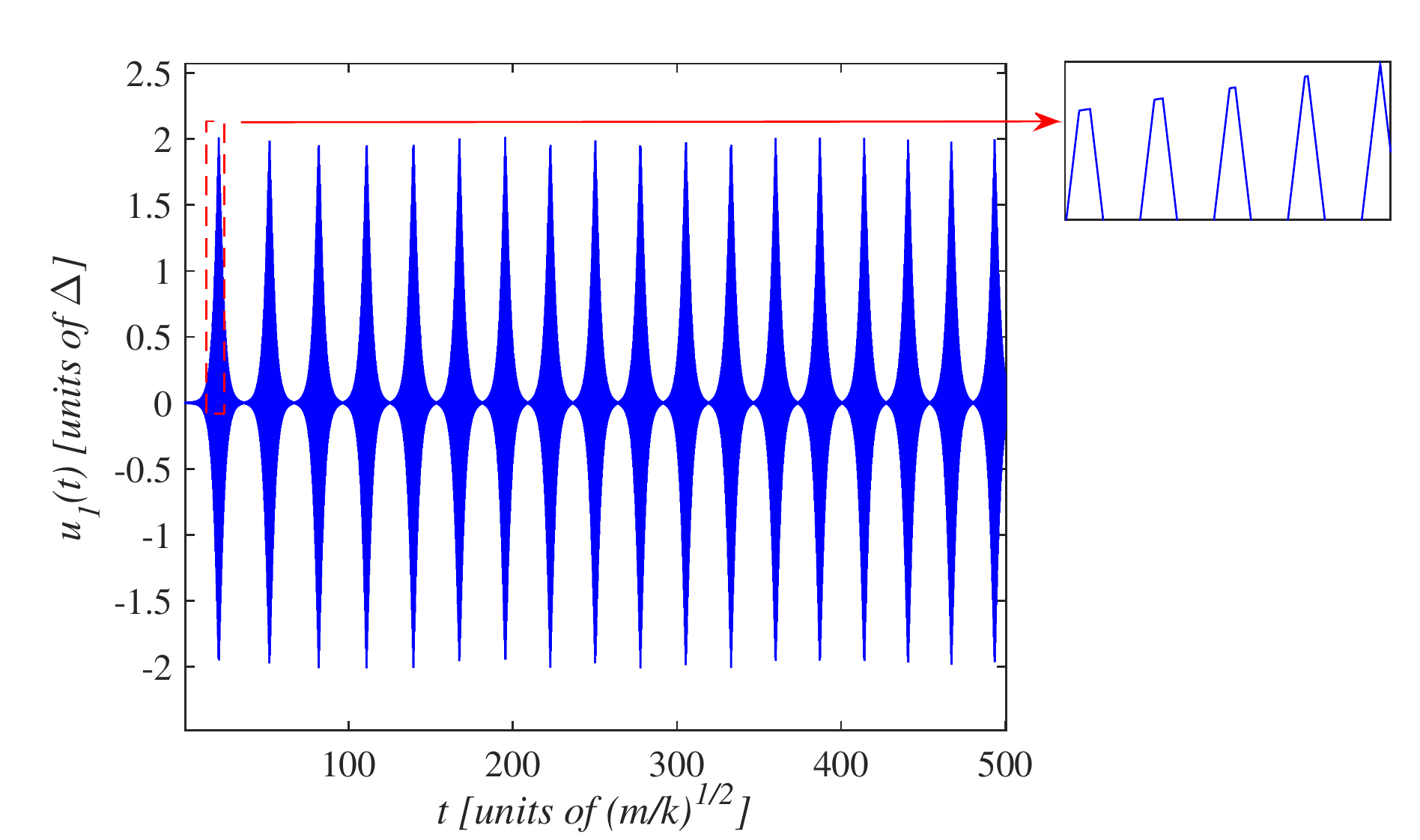}}}
\end{center}
\caption{\small Simulation results for (perturbed) compacton initial conditions -- 'PF' unstable case (principal-site histories): $N=9,k/k_{\text{shear}}=10,\omega/\omega_{\text{FB}}=10,\dot{u}_1(0)=10^{-3}\sqrt{k/m}\Delta$}
\label{Fig16}
\end{figure}

\begin{figure}[H]
\begin{center}
{{\includegraphics[scale = 0.39]{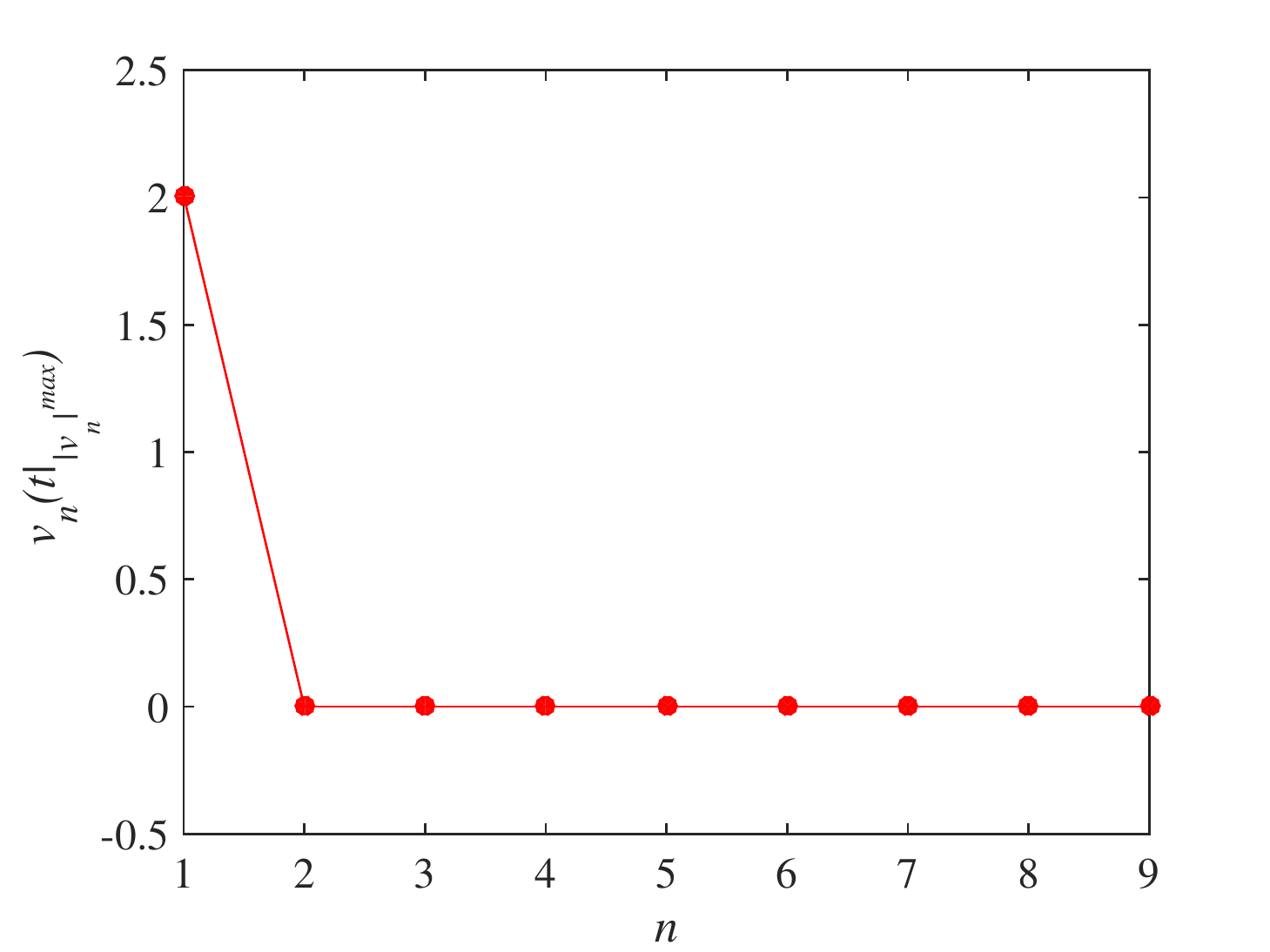}} \\
{\includegraphics[scale = 0.39]{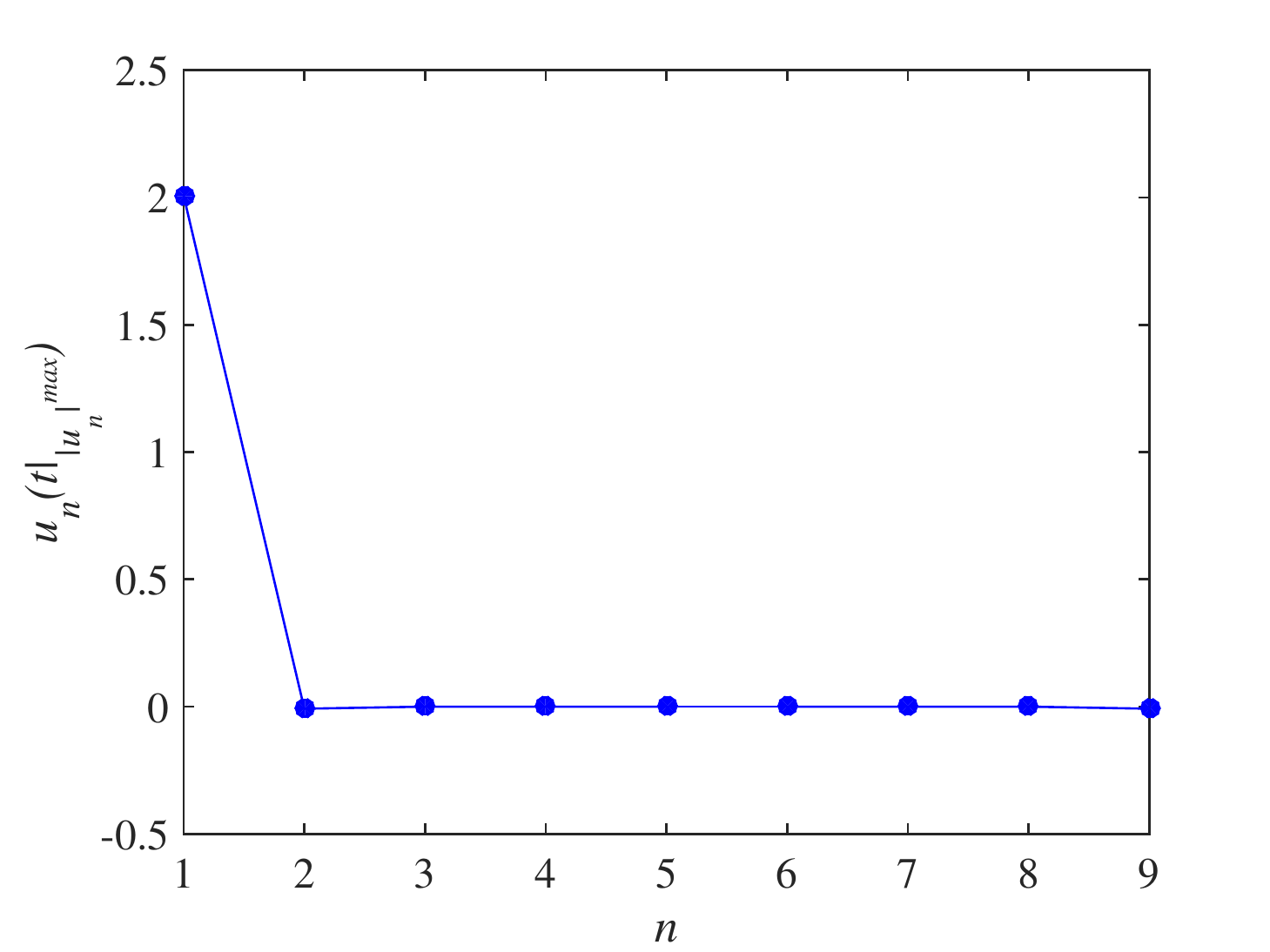}}}
\end{center}
\caption{\small Simulation results for (perturbed) compacton initial conditions -- 'PF' unstable case (maximal-displacement-time profiles)}
\label{Fig16b}
\end{figure}

In \cite{Sapsis2017}, a similar single element is analyzed and global dynamics is revealed through transformation to canonic variables (action--angle formalism). One result arising in \cite{Sapsis2017} is that the one-element cross-stitch system has a saddle point corresponding to the antisymmetric mode. This means that for high energies and finite stiffness-ratio values, the compacton solution should be unstable. Furthermore, the aforementioned saddle point is a global feature, and the nearest minimum corresponds to the symmetric mode. However, this does not yet imply that starting at the compacton initial conditions for high enough energies, one would stabilize the system in the symmetric mode. The reason is that, as shown in \cite{Sapsis2017}, there is no equal-energy contour line passing through both the compacton-related saddle point and the breather-related minimum. Such stable global dynamic a route is hence inexistent. What does exist is a contour line passing through the saddle point, reaching an intermediate location and going back to the saddle point. The intermediate location in the canonic phase-space is close enough to the symmetric mode. This dynamics correlates well-enough with the results of the numerical integration of the chain, given in Figs. \ref{Fig16} and \ref{Fig16b}.

\section {The second model system (1D chain) -- two-masses in a massless box with smooth internal potential and linear links}
\label{Sect2}

\begin{figure}[H]
\begin{center}
{\includegraphics[scale = 0.3]{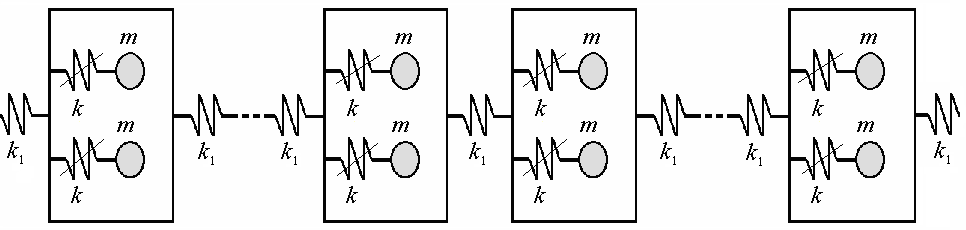}}
\end{center}
\caption{\small  Sketch of a chain of massless boxes with internal oscillators (line across internal spring symbol denotes anharmonic addition to the potential)}
\label{Fig1a}
\end{figure}

Figure \ref{Fig1a} presents the model setting for a chain of massless boxes with internal oscillators. For the required purposes, the interesting case would be that in which in every box the masses and spring characteristics are identical for the two oscillators (the reasoning is explained below). For such a case, the equations of motion for the system are as follows:
\begin{equation}
\begin{split}
m \ddot{\bar{u}}_n+k(\bar{u}_n-u_n)=f^{NL}_{(\bar{u}_n-u_n)} \\
m \ddot{\tilde{u}}_n+k(\tilde{u}_n-u_n)=f^{NL}_{(\tilde{u}_n-u_n)}  \\
m \ddot{\bar{u}}_n+m\ddot{\tilde{u}}_n+k_1(2u_n-u_{n+1}-u_{n-1})=0
\end{split}
\label{eq0}
\end{equation}
where $\bar{u}_n$ and $\tilde{u}_n$ denote the displacements of the first and the second oscillator, respectively, for each box, and $u_n$ stands for the displacement of the $n^{\text{th}}$ box itself. The conservative case (no external forcing and no damping) is assumed. The parameters $k$ and $k_1$ stand for spring rigidities inside and outside each box,  respectively, and $f^{NL}_{(x)}$ is an interaction force derived from a smooth anharmonic correction to on-site potential, representing local nonlinearity.

\subsection{Linear dynamics}

The set of equations given above can be written in normalized form, more convenient for mathematical treatment, which in the linear case becomes:
\begin{equation}
\begin{split}
\ddot{\bar{u}}_n+\kappa(\bar{u}_n-u_n)=0 \\
\ddot{\tilde{u}}_n+\kappa(\tilde{u}_n-u_n)=0 \\
\ddot{\bar{u}}_n+\ddot{\tilde{u}}_n+\gamma(2u_n-u_{n+1}-u_{n-1})=0
\end{split}
\label{eq1}
\end{equation}
(here $\kappa$ and $\gamma$ stand for spring rigidities normalized by mass, inside and outside each box, respectively).

As in the previous model system, due to the symmetry between the first and second equations in Eqs. (\ref{eq1}), linear transformation of variables is employed, as follows:

\begin{equation}
v_n \triangleq \bar{u}_n-\tilde{u}_n \ , \ x_n \triangleq \bar{u}_n+\tilde{u}_n
\label{eq1b}
\end{equation}

Combining Eqs. (\ref{eq1}) and (\ref{eq1b}), the following equivalent system of equations is obtained:
\begin{equation}
\begin{split}
\ddot{v}_n+\kappa v_n=0 \\
\ddot{x}_n+\kappa x_n=2\kappa u_n \\
\ddot{x}_n+\gamma(2u_n-u_{n+1}-u_{n-1})=0
\end{split}
\label{eq1c}
\end{equation}

The first equation in (\ref{eq1c}) decouples from the rest of the equations. Moreover, one observes that the first equation in (\ref{eq1c}) is a local relation, and indeed oscillatory spectral analysis results in a dispersion relation for $v_n(t)$ that has the following form:

\begin{equation}
\omega_0(q)=\sqrt{\kappa}
\label{eq5}
\end{equation}
in which $\omega_0$ stands for the frequency and $q$ denotes the wave number. Once again, due to the local configurational symmetry, a flat band emerges.

Next, proceeding to the two remaining equations in system (\ref{eq1c}), one finds that a plane wave oscillatory solution of this system is only possible in the case of synchronous motion of $u_n$ and $x_n$ (which is reasonable, since oscillatory motion requires mass inertia which is supplied to $u_n$ by $x_n$). For such motion, the corresponding dispersion relation takes the form:

\begin{equation}
\hat{\omega}_1(q)=\sqrt{\frac{1-\cos{q}}{1-\cos{q}+\delta}}
\label{eq2}
\end{equation}
where $\delta \triangleq \kappa/\gamma$ and $\omega_1=\sqrt{\kappa}\hat{\omega}_1<\omega_0$ (implying no-crossing, unlike in Fig. \ref{Fig9b}), $\omega_1$ being the frequency. One can observe that in the limit $\delta \gg 1$, the dispersion relation of a simple linear chain is recovered, whereas in the limit $\delta \ll 1$, this dispersion relation becomes (asymptotically) flat too.

\begin{figure}[H]
\begin{center}
{\includegraphics[scale = 0.45]{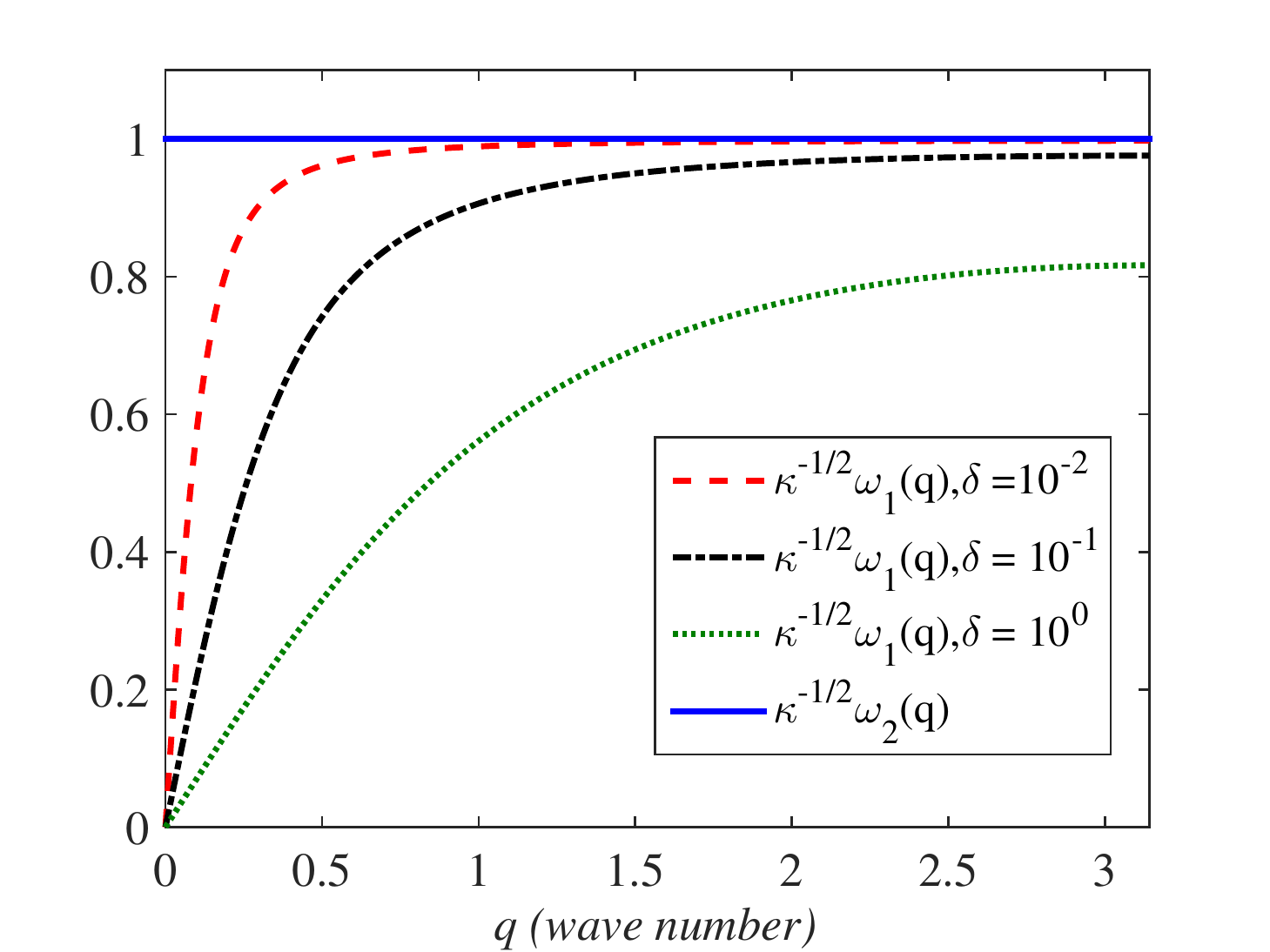}}
\end{center}
\caption{\small Dispersion relations (flat band in solid), $\delta=k/k_1, \kappa = k/m$}
\label{Fig2}
\end{figure}

Another simple linear mechanical system that has a flat band structure has therefore been demonstrated. To conclude this part of the description, a plot of the oscillatory spectra of the system is presented in Fig. \ref{Fig2}, illustrating the flat band in more detail.

\subsection{Nonlinear dynamics: analysis of a single (elementary) cell}

\begin{figure}[H]
\begin{center}
{\includegraphics[scale = 0.3]{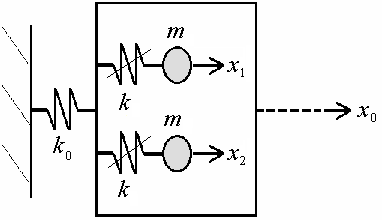}}
\end{center}
\caption{\small  Schematic depiction of the elementary cell subjected to nonlinear analysis}
\label{Fig1b}
\end{figure}

The basic system subjected to nonlinear analysis  is presented in Fig. \ref{Fig1b}.
\\

\subsubsection{Analytical integration of the compact mode}
\label{sect10B0}

In line with the well-known Fermi-Pasta-Ulam model, cubic nonlinearity is added to the internal interaction forces, producing the following equations of motion (where $p$ measures anharmonicity):
\begin{equation}
\begin{split}
m\ddot{x}_1+k(x_1-x_0)+p(x_1-x_0)^3=0 \\
m\ddot{x}_2+k(x_2-x_0)+p(x_2-x_0)^3=0 \\
k_0 x_0-k(x_1-x_0)-p(x_1-x_0)^3\\-k(x_2-x_0)-p(x_2-x_0)^3=0
\label{eq6.2}
\end{split}
\end{equation}

Since the conservative case is considered, with no forcing or damping, the general expression for the energy of the model system is given by:
\begin{equation}
\label{eq7.1}
\begin{split}
E=E_0=\frac{1}{2}m\dot{x}_1^2+\frac{1}{2}m\dot{x}_2^2+\frac{1}{2}k_0x_0^2 \\
+\frac{1}{2}k(x_1-x_0)^2+\frac{1}{4}p(x_1-x_0)^4 \\ +\frac{1}{2}k(x_2-x_0)^2+\frac{1}{4}p(x_2-x_0)^4
\end{split}
\end{equation}

Next, appropriate notation is introduced:
\begin{equation}
\bar{x} \triangleq \frac{x_1+x_2}{2}, v \triangleq x_1-x_2, y \triangleq \bar{x}-x_0, \tilde{p} \triangleq \frac{p}{2k}
\label{eq6.3}
\end{equation}
\begin{equation}
\label{eq7.9}
\begin{split}
 C \triangleq 4\hat{\eta}\hat{\epsilon}^2+(1+4\hat{\eta})\hat{\epsilon}+1+\hat{\eta}
\end{split}
\end{equation}
where the auxiliary quantities are given by:
\begin{equation}
\label{eq7.4}
\omega\triangleq\sqrt{k/m} \ , \ \hat{\eta} \triangleq 2k/k_0 \ , \ \hat{\epsilon} \triangleq \tilde{p}Y_0^2 \
\end{equation}

This can be used to represent the energy in the antisymmetric mode in terms of some amplitude $Y_0$ (later used, for convenience, in parametric analysis), which can be interpreted as the amplitude of oscillation in the \emph{symmetric} mode with the same energy:

\begin{equation}
\label{eq7.5}
\begin{split}
\frac{E_0}{k}=(1+\hat{\eta})Y_0^2+(1+4\hat{\eta})\tilde{p}Y_0^4+4\hat{\eta}\tilde{p}^2Y_0^6
\end{split}
\end{equation}

The antisymmetric mode corresponds to zero motion of the center-of-mass of the system, which leads, following the definitions in Eq. (\ref{eq6.3}), to the next relations:
\begin{equation}
\label{eq10A.1}
\begin{split}
\bar{x}(t) \equiv 0 \underset{(\ref{eq6.2},\ref{eq6.3})}\Rightarrow  x_0(t), y(t) \equiv 0, x_{1,2} = \pm v/2
\end{split}
\end{equation}

Taking the difference of the first two equations in Eqs. (\ref{eq6.2}), expressed in terms of $y,v$ and $x_0$, after eliminating $x_0$ using the third, algebraic equation in Eqs. (\ref{eq6.2}), and its first and second time derivatives, expressed in terms of the three aforementioned variables, leads to the following equation (required for stability analysis at a later instance):
\begin{equation}
\ddot{v}+\omega^2 (1+6 \tilde{p} y^2) v+\frac{1}{2}\tilde{p} \ \omega^2 v^3=0
\label{eq6B.1}
\end{equation}

Substituting Eq. (\ref{eq10A.1}) into the expression for the total energy of the system given in Eq. (\ref{eq7.1}), and using Eqs.  (\ref{eq6.3}), (\ref{eq7.4}) and (\ref{eq7.5}), one obtains the first integral of the antisymmetric  mode of the system, in the following form:
\begin{equation}
\label{eq10A.2}
\begin{split}
[\hat{v}']^2+\hat{v}^2+\frac{1}{4}\hat{\epsilon}\hat{v}^4=4C; \ \hat{v} \triangleq v/Y_0 \ , \\ \hat{v}' \triangleq d\hat{v}/d\tau \ , \ \tau \triangleq \omega t
\end{split}
\end{equation}
(here $Y_0(E_0,k_0,k,p)$ is perceived merely as convenient parametrization of the total energy in the system).

Further integration of Eq. (\ref{eq10A.2}) with a centered impulsive boundary condition ($\hat{v}(0)=0$), results in a closed form expression for the normalized antisymmetric mode displacement as a function of the dimensionless time, using the fifth Jacobi elliptic function 'sd', as follows:
\begin{equation}
\label{eq10A.4}
\hat{v}(\tau)=\frac{\sqrt{4C}\text{sd}\left[(1+4\hat{\epsilon}C)^{1/4}\tau\left| \frac{\sqrt{1+4\hat{\epsilon}C}-1}{2\sqrt{1+4\hat{\epsilon}C}} \right.\right]}{(1+4\hat{\epsilon}C)^{1/4}}
\end{equation}

One observes that for small amplitudes, the simple form $\hat{v}(t)=\sin(\omega t)$ is reproduced.

Equation (\ref{eq10A.4}) provides the explicit solution for the antisymmetic mode. In order to study its stability, Floquet-Hill method is applied, and Fourier-series representation of the solution is called for. The sine series representation of sd$(z|m)$ is given in \cite{AbramowitzStegun1964}. In the employed parametrization it gives rise to the following series for the antisymmetric mode:
\begin{equation}
\label{eq10A.5}
\hat{v}(\hat{\tau})=\sum\limits_{n=1}^{\infty}V_n\sin(n\hat{\tau}) \ , \ \hat{\tau} \triangleq \Omega_v \tau
\end{equation}
where
\begin{equation}
\label{eq10A.6}
\begin{split}
 \  \hat{m} \triangleq \frac{\sqrt{1+4\hat{\epsilon}C}-1}{2\sqrt{1+4\hat{\epsilon}C}}, \
\Omega_v \triangleq \frac{\pi}{2}\frac{(1+4\hat{\epsilon}C)^{1/4}}{K(\hat{m})}, \\
V_n \triangleq \frac{2\pi\sqrt{C}(1+4\hat{\epsilon}C)^{-1/4}}{\sqrt{\hat{m}(1-\hat{m})}K(\hat{m})}  \frac{(-1)^{\frac{n-1}{2}}\delta_{n,2\mathbb{N}-1}}{\cosh{\left[ \frac{n \pi K(1-\hat{m})}{2K(\hat{m})}\right]}}
\end{split}
\end{equation}
and $K(m)$ is the complete elliptic integral of the first kind (and $\delta_{a,b}$ is Kronecker's delta).

\subsubsection{Linear stability analysis of the compact mode}
\label{sect10B}

In order to examine the linear stability of the obtained solution, the equation of motion for the symmetric mode is required.

Finding from definition (\ref{eq6.3}) that:
\begin{equation}
\label{eq10B.1}
x_{1,2}-x_0=y \pm v/2
\end{equation}
and substituting this into the third equation in Eqs. (\ref{eq6.2}), one obtains the (algebraic) equation of motion for the displacement of the box, as follows:
\begin{equation}
\label{eq10B.2}
x_0=y\left(\hat{\eta}+2\hat{\eta}\tilde{p}y^2+\frac{3}{2}\hat{\eta}\tilde{p} v^2\right)
\end{equation}

Next, taking half of the sum of the first two equations in Eqs. (\ref{eq6.2}) and substituting Eq. (\ref{eq10B.1}), the third of definitions (\ref{eq6.3}) and the second time derivative of Eq. (\ref{eq10B.2}) into the result, eliminating the second time derivative of $v$ by isolating it from Eq. (\ref{eq6B.1}), one obtains the equation of motion for the symmetric mode (for mixed-mode dynamics), containing only $v$, $y$, their first time derivatives and the second time derivative of $y$, in terms of the dimensionless quantities defined in Eq. (\ref{eq10A.5}) and in the first row in Eq. (\ref{eq10A.6}). Consequently linearizing with respect to $y(t)$, one obtains the evolution equation for small perturbations to the antisymmetric mode. 

Additional mathematical manipulations, described in \cite{Perchikov2016}, produce the explicit Hill form of the linear stability equation for the purely antisymmetric mode:
\begin{equation}
\label{eq10B.7}
\tilde{y}''(\hat{\tau})+\frac{\Omega_v^{-2}}{1+\hat{\eta}}\frac{1+(3/2)\hat{\epsilon}[\hat{v}(\hat{\tau})]^2}{1+(3/2)\frac{\hat{\eta}}{1+\hat{\eta}}\hat{\epsilon}[\hat{v}(\hat{\tau})]^2}\tilde{y}(\hat{\tau})=0
\end{equation}
with $\Omega_v$ and $\hat{v}(\hat{\tau})$ given by Eqs. (\ref{eq10A.5}) and (\ref{eq10A.6}), and with the Hill function, $h(\hat{\tau})$, being the expression multiplying $\tilde{y}(\hat{\tau})$ in Eq. (\ref{eq10B.7}).

Next, stability analysis of Eq. (\ref{eq10B.7}) is performed using Hill's method. The details of this procedure, namely, rigorous deconvolution and the actual implementation of Hill's method, for identification of instability tongues characterizing the antisymmetric mode, as well as further study of the nature of the occurring instabilities by use of Poincar\'e sections, are given in \ref{AppendixB}. 

\begin{figure}[H]
\begin{center}
{\includegraphics[scale = 0.535]{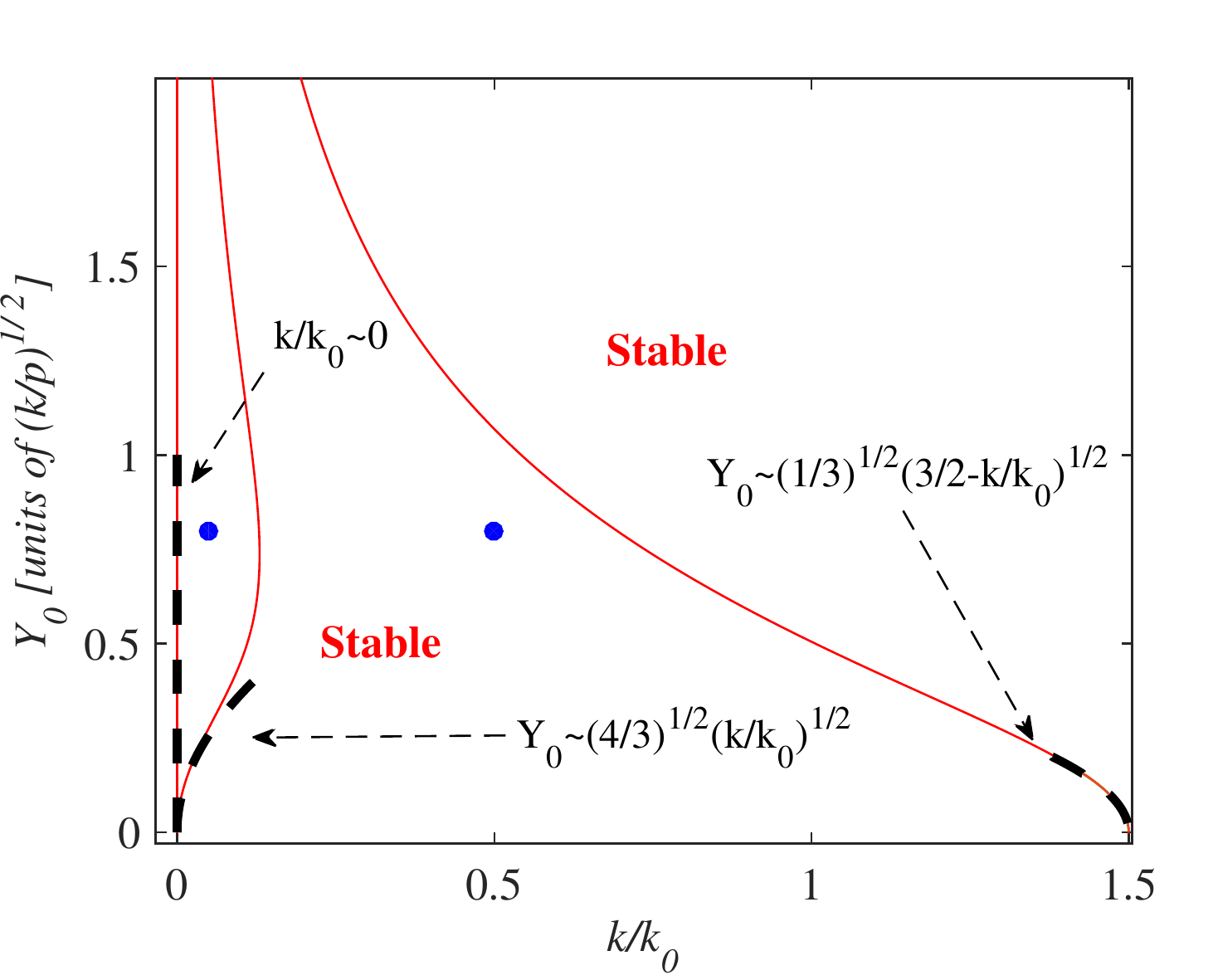}}
\end{center}
\caption{\small Linear stability bounds in solid (red online) with asymptotic estimates in dash black and the two points chosen for numerical integration}
\label{Fig3}
\end{figure}

Of the obtained results, the one mostly relevant for the consequent analysis of a nonlinear chain, namely a stability map in the parameter space (plane) for the representative element, is shown in Fig. \ref{Fig3}.

The single-element analysis presented here is reproduced in part from \cite{Perchikov2016}, for better understanding of the dynamics in the context of a chain. This owes to the fact that for a chain, rigorous stability analysis is not within reach, other than when understanding the single element case as representative of the anti-continuum limit for a chain, taking $k/k_0 \gg 1$. In that limit, as Fig. \ref{Fig3} shows, the compacton is linearly stable (see \cite{Perchikov2016} for details).

\section{Numerical integration for the chain}
\label{Sect2b}

The existence of the compacton in a chain is evident from the fact of compactness itself. Since the compacton corresponds to the antisymmetric mode in the representative element, for which the box displacement is always zero (see Eq. (\ref{eq10A.1})), implying no interaction between boxes, clearly the solution given in Eq. (\ref{eq10A.4}) should also hold for a chain -- it is then simply the difference of the displacements of the two masses in a given box. In fact, several-site compactons can exist just as well. The only arising question would be that of stability. The case addressed in the present work would be, however, that of the stability of a single-site compacton.

The stability of a compacton in the chain shown in Fig. \ref{Fig1a} can be examined by taking parameters inside and outside of the finite-width instability tongue, as calculated for a single element. This embodies the assumption that the stability limits for a chain are somehow correlated with those of a single element. The two points in the stiffness-ratio -- energy plane chosen for stability analysis by direct integration of a chain are shown in Fig. \ref{Fig3}. In order to perform the numerical integration from the perspective of stability of (nonlinear) normal modes (NNMs), first, the equations of motion given in Eq. (\ref{eq0}) for the case of cubic nonlinearity in the internal interaction forces, are rewritten as a dynamic system, in which modal displacements can be integrated directly.

\subsection{Equations of motion}

The equations of motion for a chain of $N$ elements comprised of massless boxes occupied by two identical internal oscillators each, with linear-cubic interaction of each internal particle with the box, are given below. The equations are expressed in terms of the dimensionless quantities $\hat{v}_i$ and $\hat{y}_i$, representing the displacement difference of the internal oscillators and the center-of-mass displacement with respect to the box, for each element. In order to interpret the one-element stability results for the chain, the following relation is set: $k_1\triangleq k_0/2$.
\begin{equation}
\label{SC1}
\begin{split}
\hat{v}_i''=-(1+6\hat{\epsilon}\hat{y}_i^2)\hat{v}_i-\frac{1}{2}\hat{\epsilon}\hat{v}_i^3 \ , \
\hat{y}_i''=-\hat{M}_{ij}\hat{f}_j
\end{split}
\end{equation}
\begin{equation}
\label{SC2}
\begin{split}
\hat{\textbf{M}}\triangleq{\textbf{M}^{-1}} \ , \
\hat{\textbf{B}}\triangleq \hat{\eta} (\textbf{I}-\textbf{B})^{-1} \ , \
\hat{\textbf{Q}}\triangleq \hat{\textbf{B}}\textbf{Q} \\
\hat{\textbf{f}} \triangleq \hat{\textbf{y}}+2\hat{\epsilon} \boldsymbol{\mu} +6\hat{\epsilon}\hat{\textbf{B}}(\boldsymbol{\phi}-\boldsymbol{\lambda})
\end{split}
\end{equation}
\begin{equation}
\label{SC3}
\begin{split}
M_{ij} \triangleq \delta_{ij}+2\hat{B}_{ij}+12\hat{\epsilon}\hat{Q}_{ij} \ , \ I_{ij}=2\delta_{ij} \\
{{B}_{ij}}  \triangleq \delta_{i+1,j}+\delta_{i,j+1} \  \forall \ i,j \in \mathbb{N}< N \\ Q_{ij} \triangleq (\hat{y}_i^2+\hat{v}_i^2/4)\delta_{ij} \ , \ \mu_i \triangleq \hat{y}_i^3+\frac{3}{4}\hat{y}_i\hat{v}_i^2 \\
\phi_i \triangleq  4\hat{y}_i(\hat{y}_i')^2+2\hat{v}_i\hat{y}_i'\hat{v}_i'+\hat{y}_i(\hat{v}_i')^2 \\  {\lambda}_i \triangleq \hat{y}_i\hat{v}_i^2\left[1+\hat{\epsilon}(6\hat{y}_i^2+\hat{v}_i^2/2)\right]
\end{split}
\end{equation}

In order to avoid boundary effects, periodic boundary conditions are applied. In addition, in order to  avoid singularity, the system (described by the matrix given in Eq. \ref{SC3}) is ``statically condensed'' by one degree of freedom (related to rigid-body motion).  Compacton initial conditions are employed.

\subsection{Integration results}

\emph{Unstable case (moderate amplitude).} The first result is obtained by taking the left point in Fig. \ref{Fig3}, the one lying well within the instability tongue for a single element. The initial condition is of a slightly perturbed compacton. A stable integration algorithm with a bounded energy-error \cite{Hosea1996} is employed.

\begin{figure}[H]
\begin{center}
{\includegraphics[scale = 0.47]{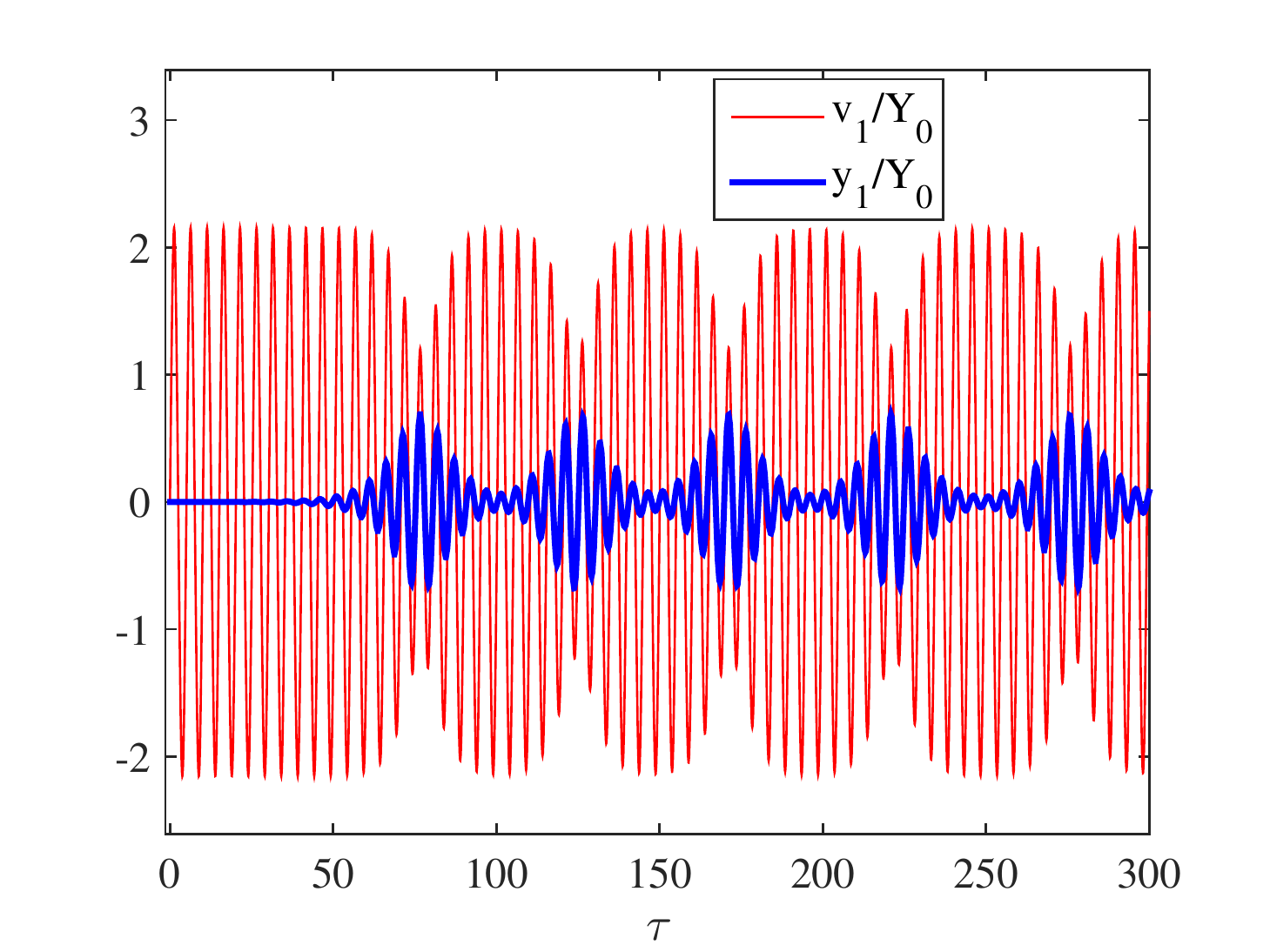}
\includegraphics[scale = 0.47]{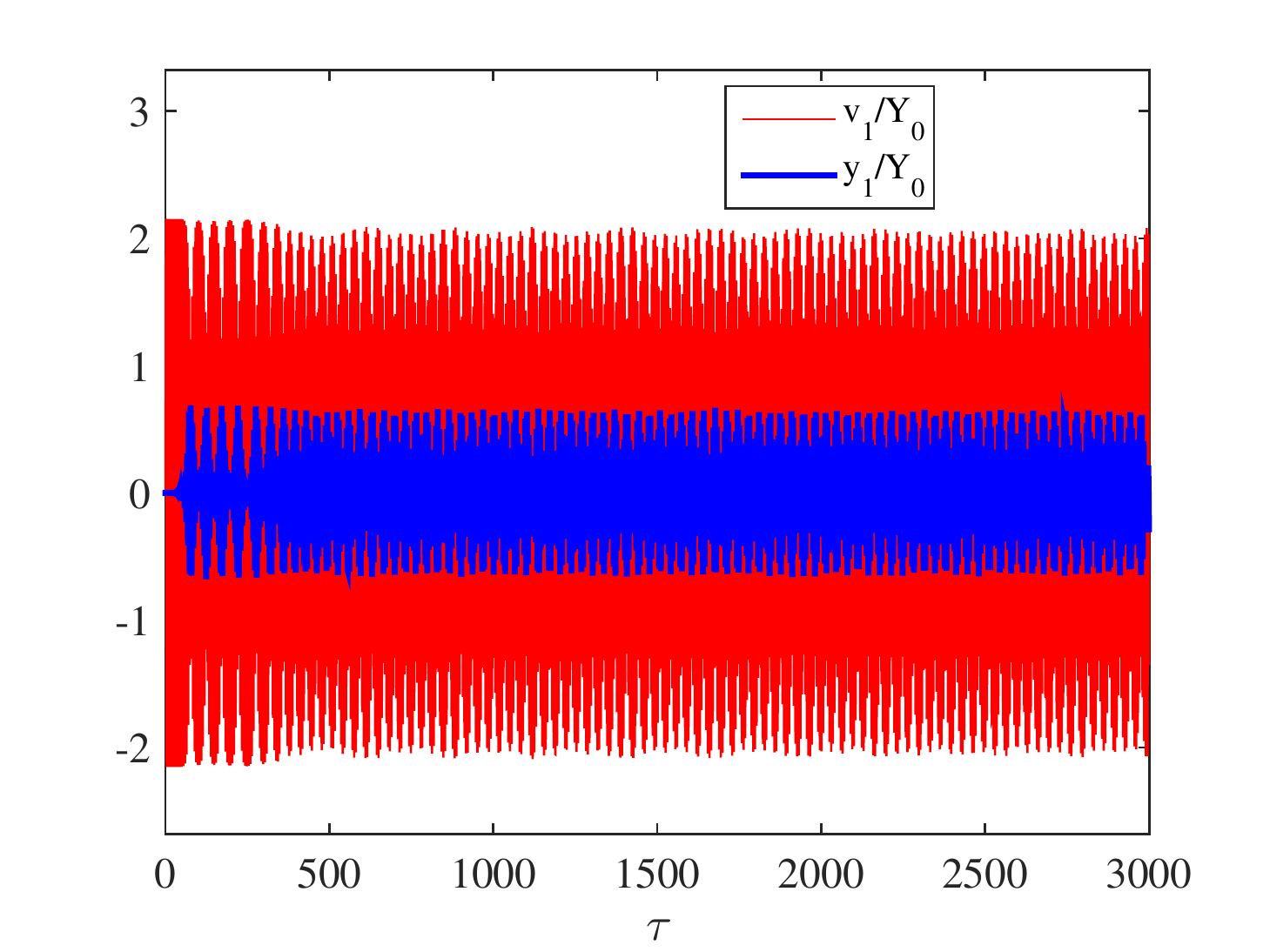}}
\end{center}
\caption{\small Unstable case --  emergence of instability for the principal element in the chain (the compacton site) for short (top) and long (bottom) times, for: $N=10, Y_0=0.8\sqrt{k/p},k/k_0=0.05,y_1(0)=10^{-4}$}
\label{Fig5b}
\end{figure}

It appears, as shown in Fig. \ref{Fig5b}, that for the chosen parameters the stability of the compacton solution is lost, and small, yet finite-amplitude center-of-mass relative (to the boxes) displacements emerge for the ringed chain. The packets-pattern in Fig. \ref{Fig5b} (top) is typical for energy exchange between different NNMs by the mechanism of parametric resonance (at least initially in the process of loss of stability). Fig. \ref{Fig5b} (bottom) shows the long-time history confirming the persistence of the energy-exchange phenomenon after all transients are gone. This energy exchange between modes is similar to what is observed (for other parameter values) on a $y$--$y'$ Poincar\'e section for a \emph{single element} \cite{Perchikov2016}.

\begin{figure}[H]
\begin{center}
{\includegraphics[scale = 0.47]{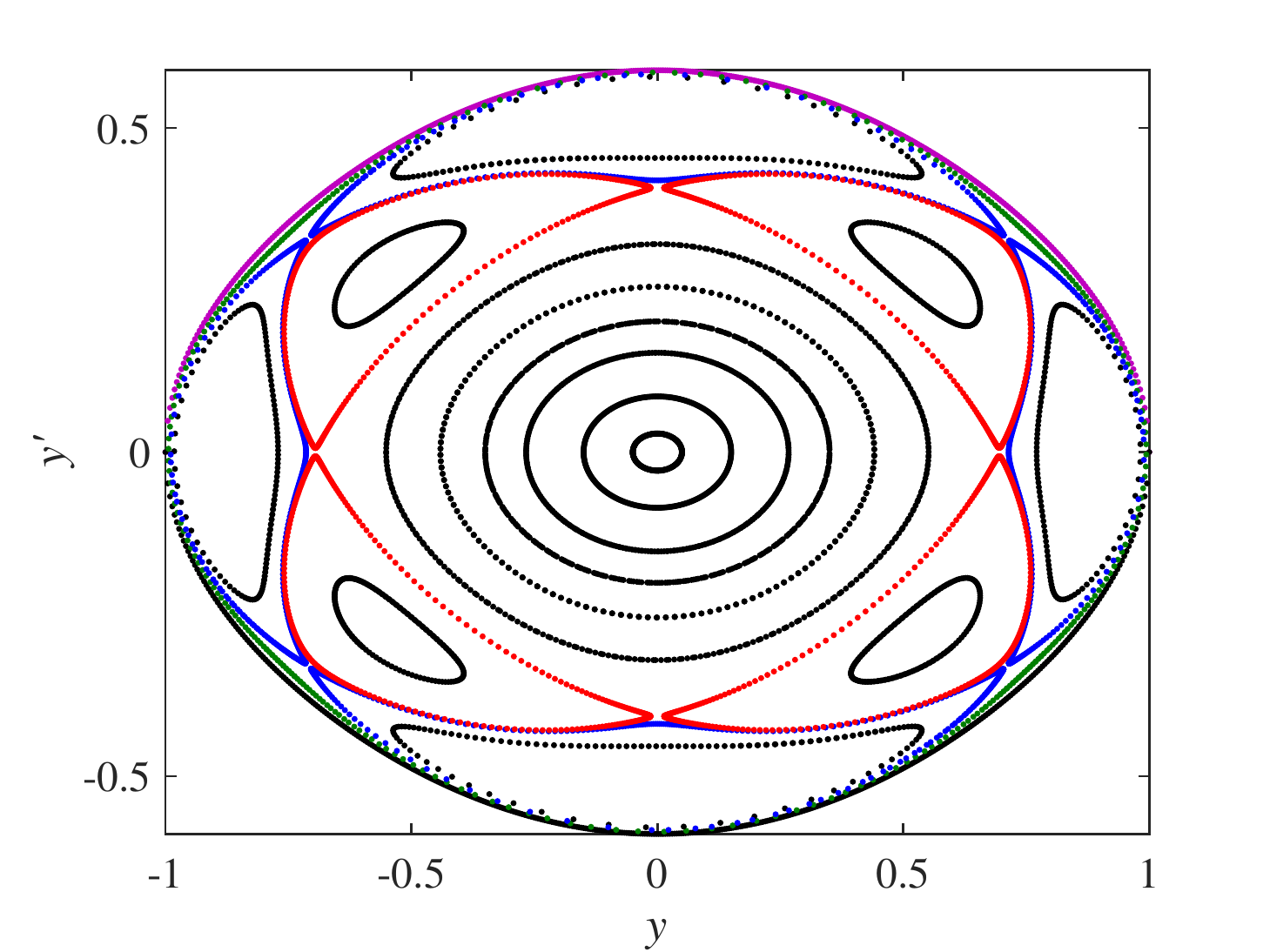}}
\end{center}
\caption{\small  Poincar\'e section: periodic energy exchange between modes (the parameters here correspond to global bifurcation for a single element, with the plot serving as a qualitative illustration for the case of a chain) for $\hat{\epsilon}=0.0645,\hat{\eta}=\frac{5}{2}$ (separatrices in red and blue online)}
\label{Fig5d}
\end{figure}

\begin{figure}[H]
\begin{center}
{\includegraphics[scale = 0.4]{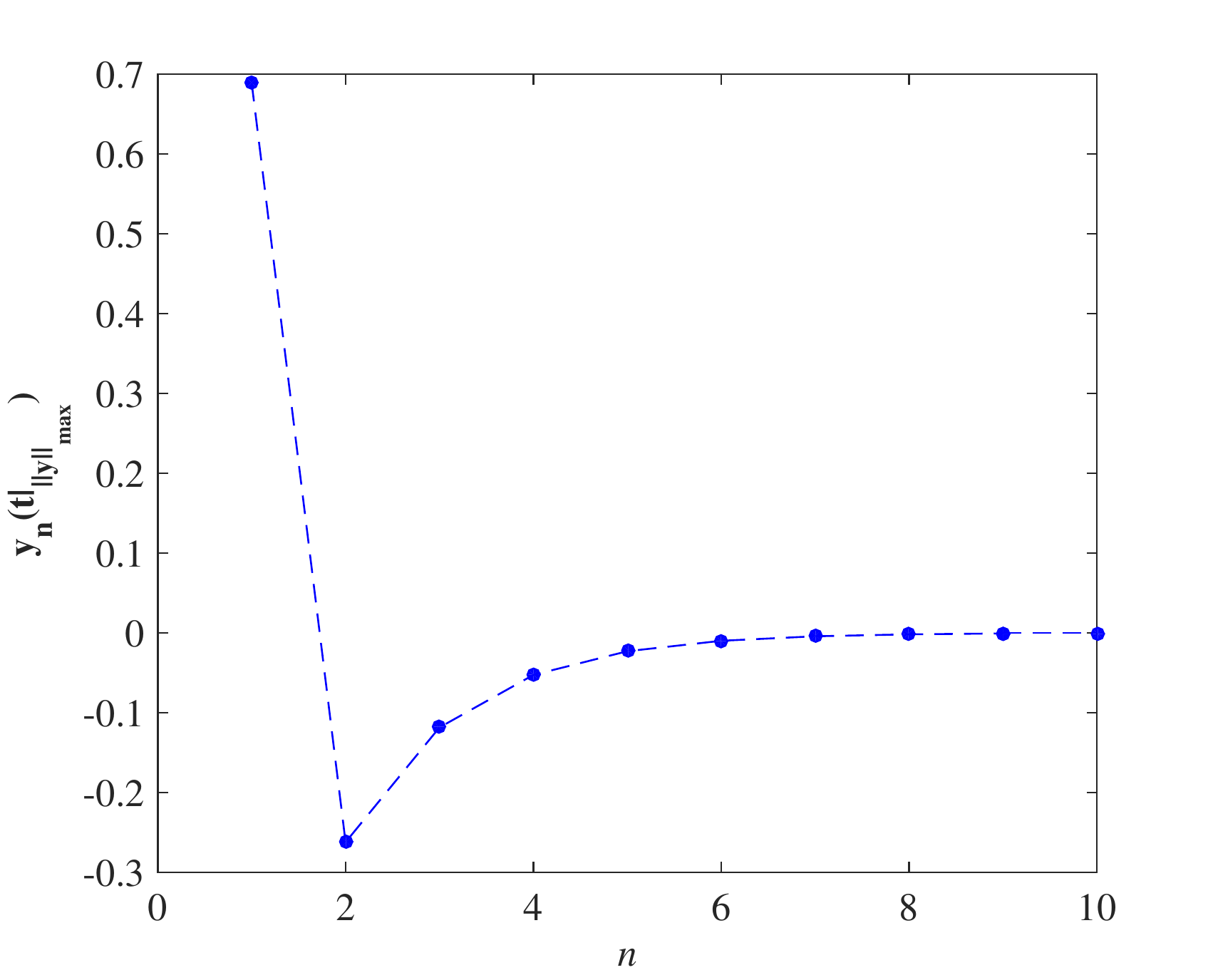}}
\end{center}
\caption{\small Unstable (compacton) case: emerging (from initial perturbation) breather-mode profile for $N=10, Y_0=0.8\sqrt{k/p},
k/k_0=0.05,y_1(0)=10^{-4}$}
\label{Fig5c}
\end{figure}

Figure \ref{Fig5d} shows the flow close to a global bifurcation where KAM (Kolmogorov-Arnold-Moser) islands of the symmetric and antisymmetric modes ``collide'', corresponding to the aforementioned significant (1:2 resonance for both modes) energy exchange (see \cite{Perchikov2016}). Figure \ref{Fig5c} shows the envelope (amplitude) of the symmetric mode displacement for the chain, which has the typical profile of a breather, with partial localization arising from the (cubic) on-site nonlinearity.

\emph{Stable case (moderate amplitude).} The second result is obtained when taking the rightmost point in Fig. \ref{Fig3}, the one lying well inside the stability region for a compacton in a representative element. Numerical integration is performed starting from a slightly perturbed compacton initial-condition.

\begin{figure}[H]
\begin{center}
{\includegraphics[scale = 0.35]{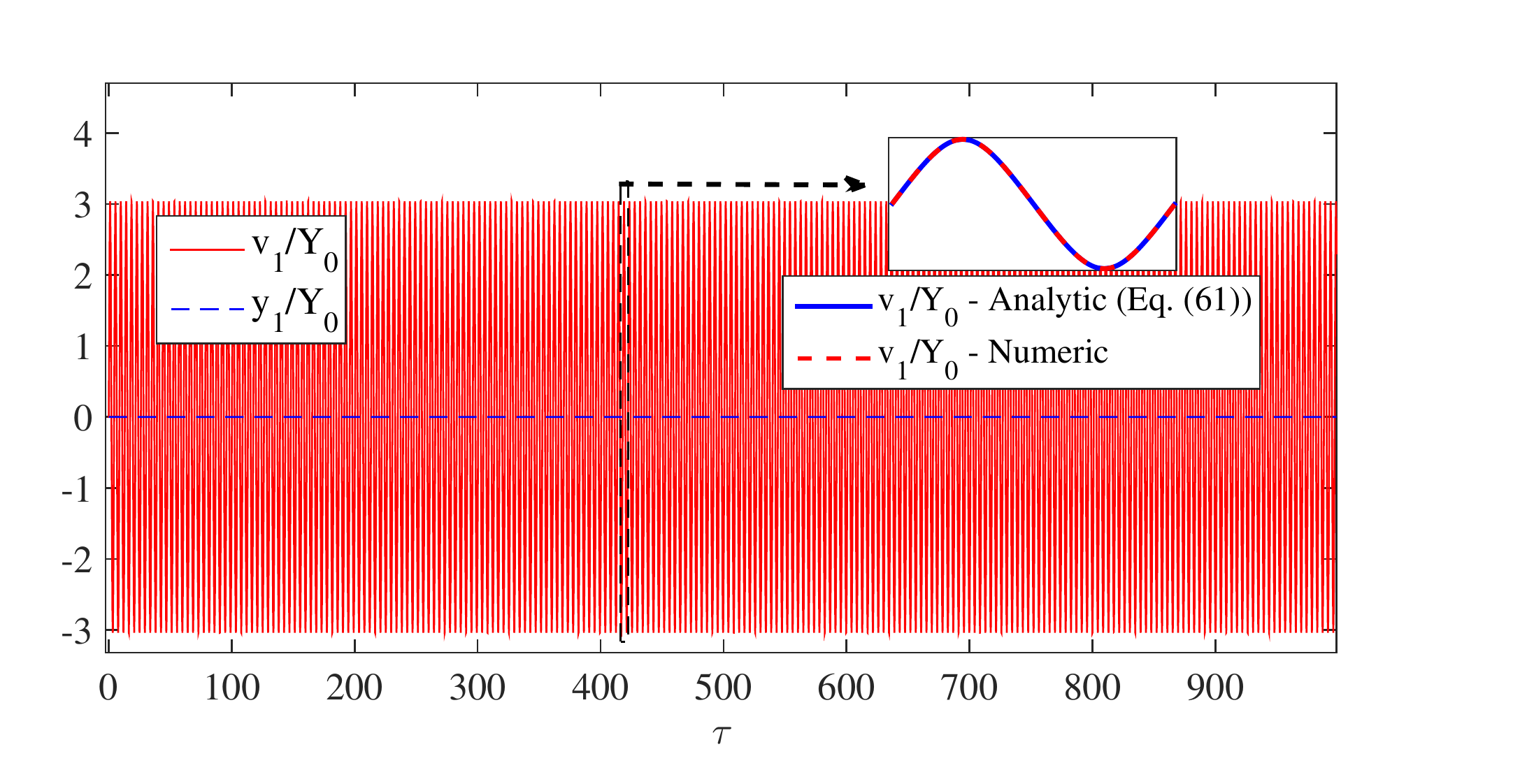}} \\
{\includegraphics[scale = 0.35]{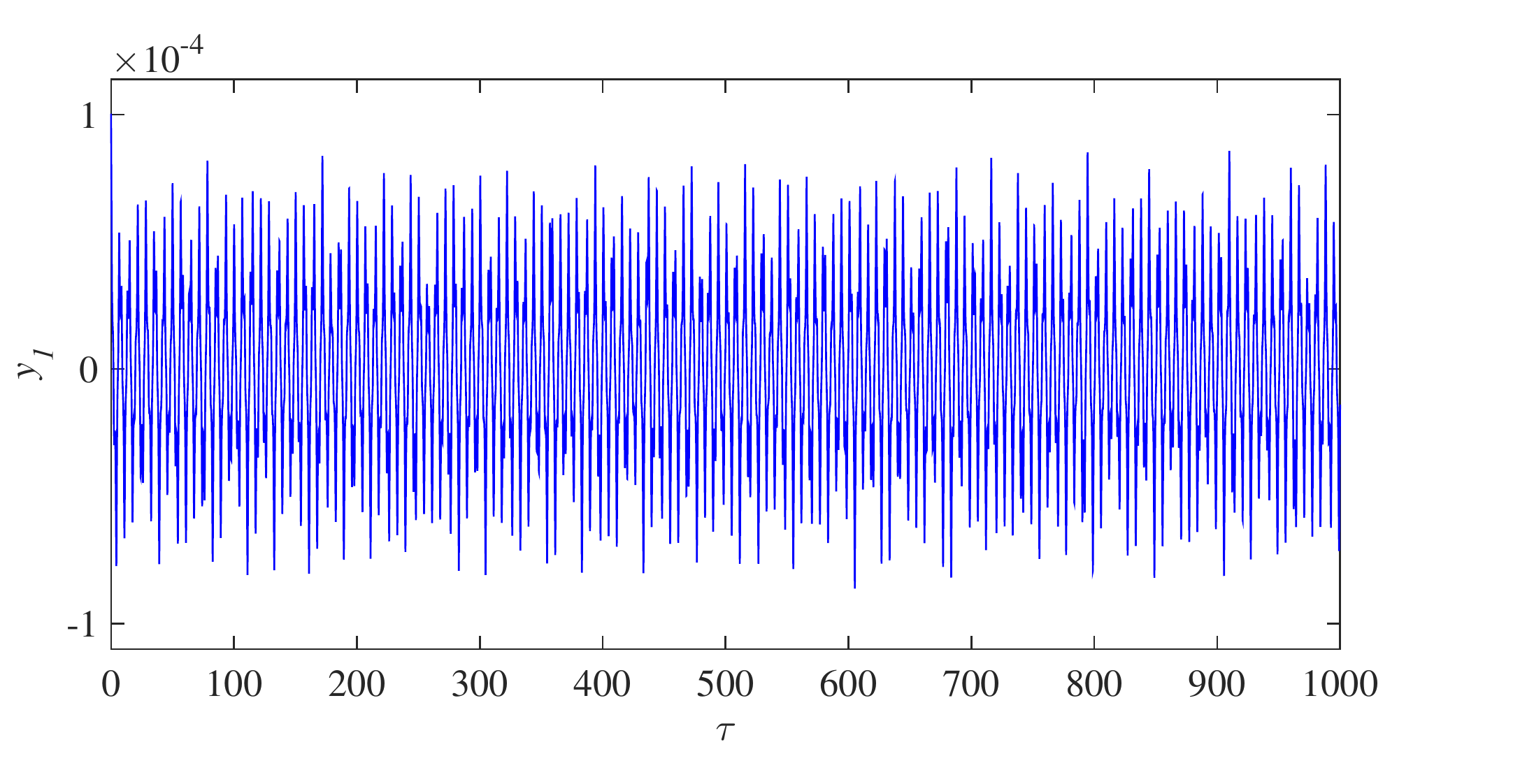}} \\
{\includegraphics[scale = 0.4]{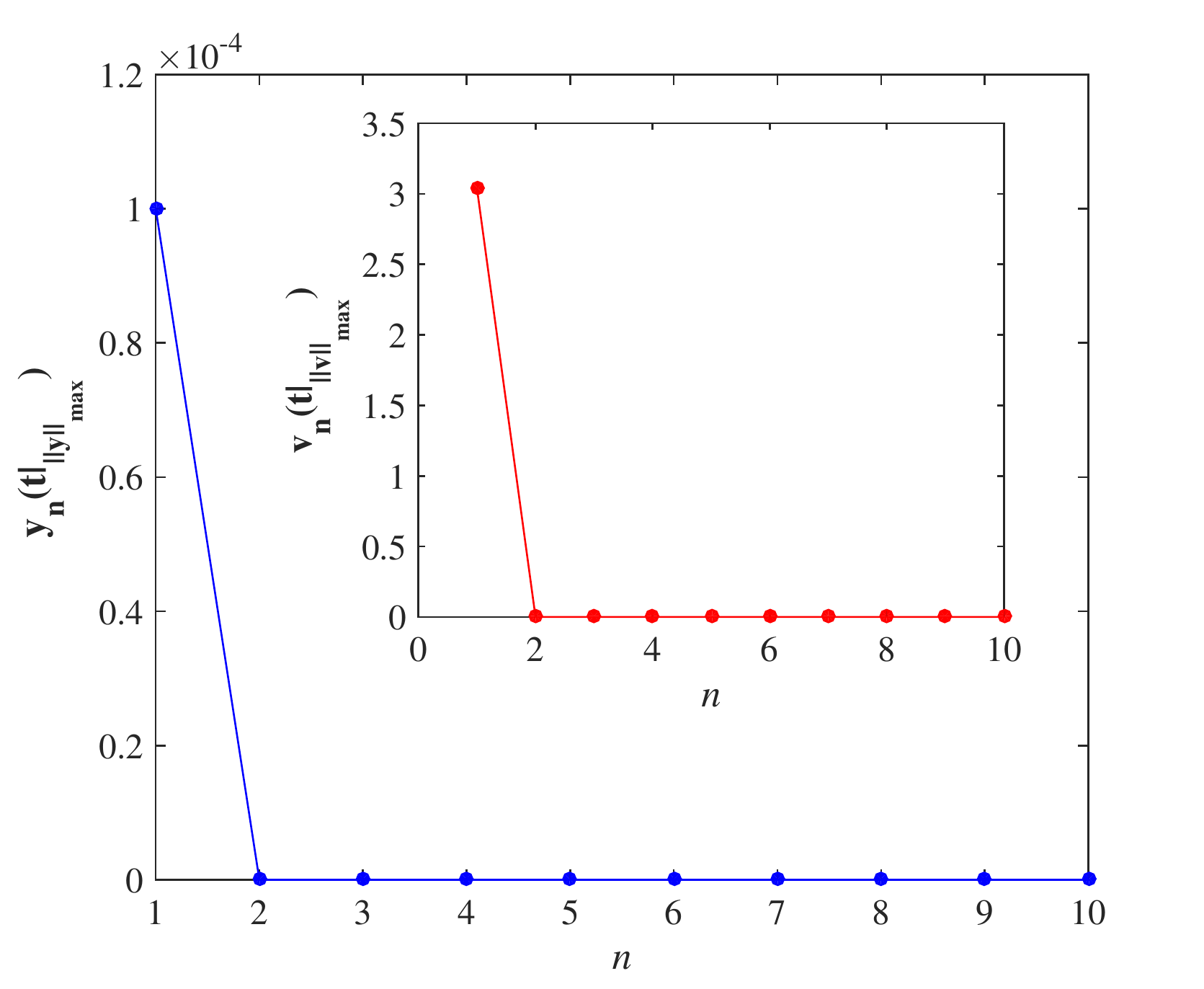}}
\end{center}
\caption{\small Integration results -- stable case (top to bottom): antisymmetric mode displacement history, symmetric mode displacement history, and symmetric (inset: antisymmetric)-mode maximal-displacement-time profile, for $N=10, Y_0=0.8\sqrt{\frac{k}{p}},
\frac{k}{k_0}=0.5,y_1(0)=10^{-4}$}
\label{Fig6}
\end{figure}

Figure \ref{Fig6} shows that when starting from the same perturbation to the compacton as in the unstable-parameters' case, the perturbation stays bounded, i.e. there is no delocalization. This is illustrated by the boundedness of the central-site history and maximal-displacement-time profile of the symmetric mode, and the antisymmetric-mode central-site displacement history, which exhibits stable periodic dynamics. Figures \ref{Fig5b}-\ref{Fig6} show that, as predicted by linear stability analysis for a representative element, the nonlinearity may either retain or destroy the stability of the compacton solution (the sheer existence of the compacton obviously emerges from the configurational symmetry corresponding to the existence of a flat band in the dispersion spectrum of the system).

It also appears that periodic dynamics in a (closed) chain can correspond to a combination of a perfectly localized mode, namely the compacton, emerging due to configurational symmetry, and a partially localized mode, namely a breather mode, corresponding to localization-inducing on-site nonlinearity.

The following subsection extends the analysis to the case of a high-amplitude (high energy) regime, where an additional noteworthy phenomenon is observed.

\subsubsection{High-amplitude regime -- unstable case}

The next result is obtained by increasing the (dimensionless) amplitude one order of magnitude and finding a corresponding stiffness ratio for which the representative element analysis shows the compacton solution to be unstable. This appears to correspond to highly rigid inter-site link potential, and repeating the integration, the obtained results remind those presented in Figs. \ref{Fig5b}-\ref{Fig5c}. There appears to be the additional property that the breather solution possesses a higher level of spatial localization, which can be attributed to the higher energy contained in the system. This matter is elaborated on for the case of the non-smooth system in the analysis in Sec. \ref{sectGD}.

\begin{figure}[H]
\begin{center}
{{\includegraphics[scale = 0.45]{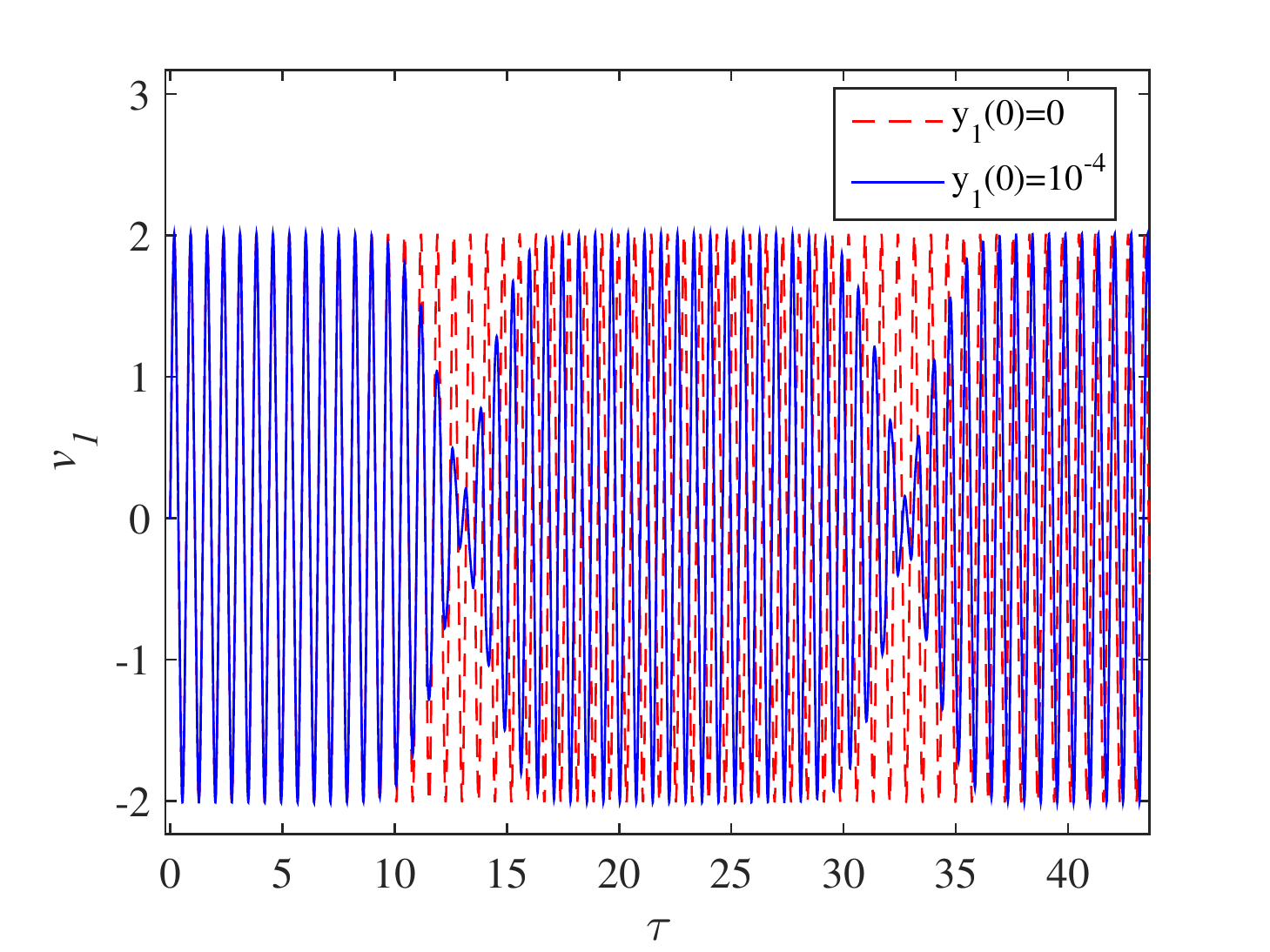}}}
\end{center}
\caption{\small $v_1(\tau)$ -- high-amplitude (unstable) case: $N=10, Y_0=10\sqrt{k/p},k/k_0=0.00005$}
\label{Fig7a}
\end{figure}

\begin{figure}[H]
\begin{center}
{{\includegraphics[scale = 0.45]{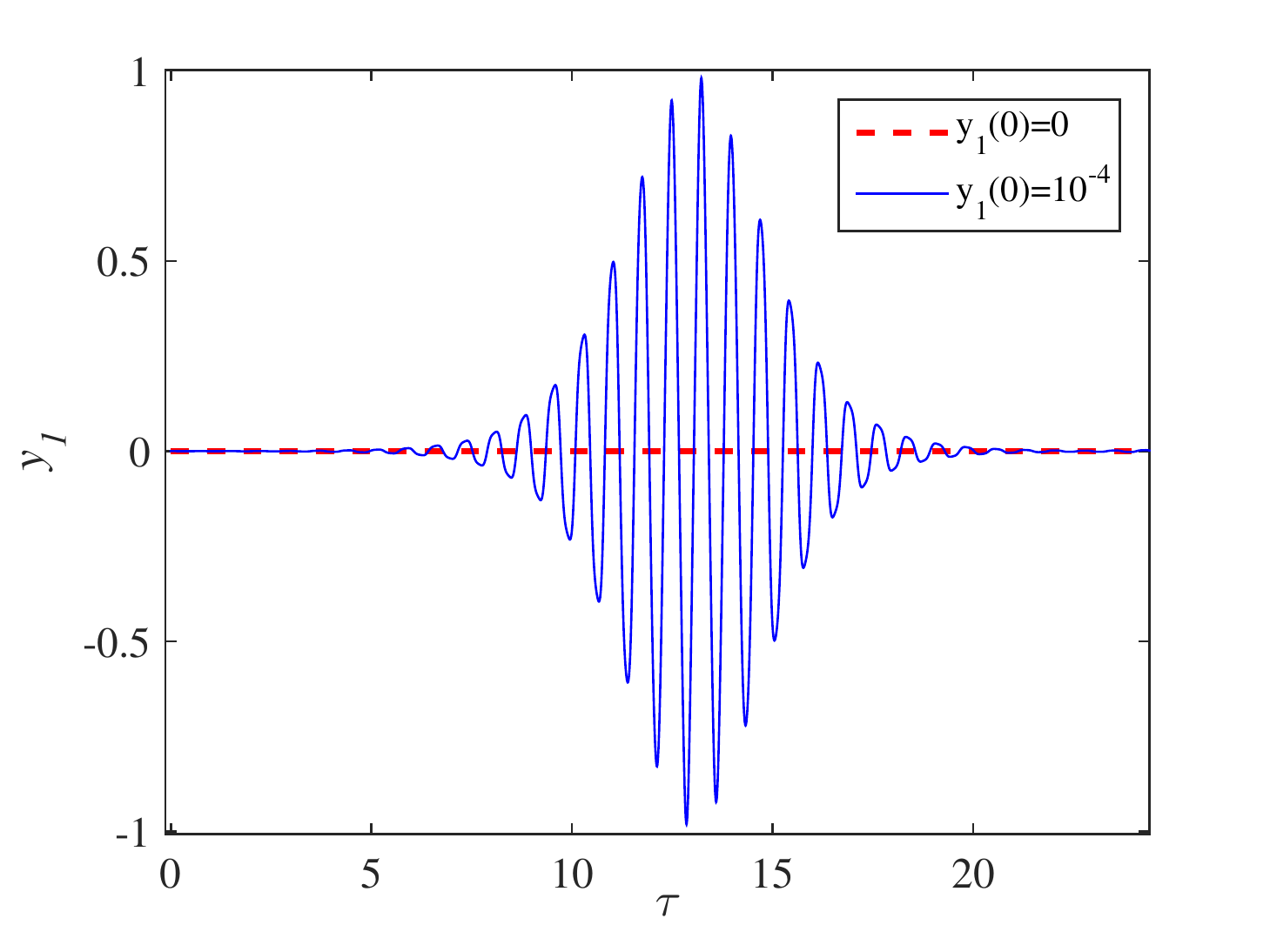}} \\
{\includegraphics[scale = 0.39]{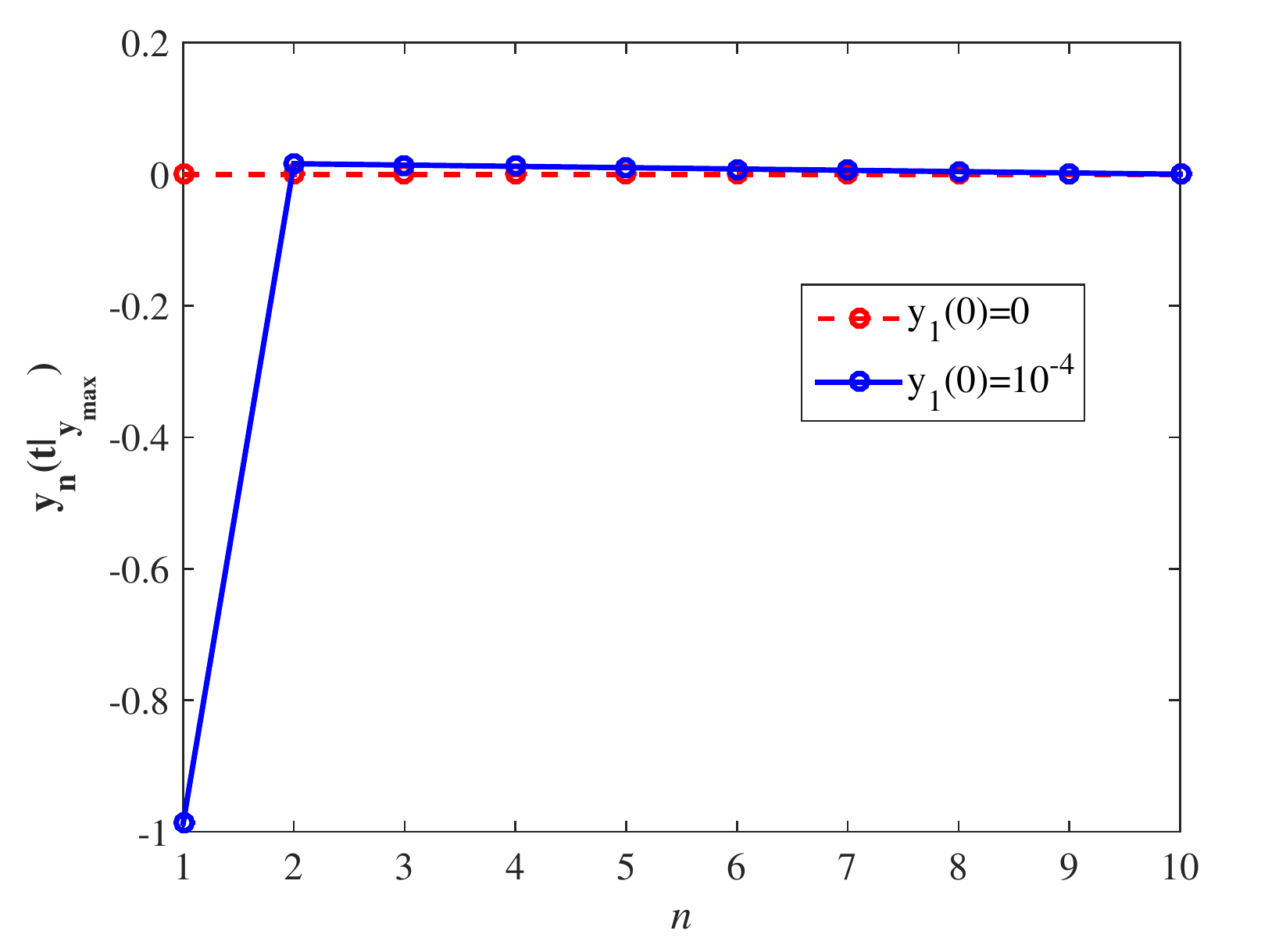}}}
\end{center}
\caption{\small Top to bottom: $y_1(\tau),y_n(t|_{\lVert\textbf{y}\rVert_{\infty}^{max}})$ -- high-amplitude (unstable) case: $N=10, Y_0=10\sqrt{k/p},k/k_0=0.00005$}
\label{Fig7}
\end{figure}

Figure \ref{Fig7a} shows the principal-site antisymmetric-mode displacement history for the perturbed case, on top of the result for the unperturbed case. The result exhibits, once again, the energy exchange feature indicative of resonance between the compacton and the symmetric mode. Figure \ref{Fig7} exhibits the symmetric mode displacement's first energy-packet history, and the symmetric-mode maximal-displacement-time profile (the antisymmetric mode remains perfectly compact). One notes that the extent of energy exchange for the high-energy regime is almost full -- considerably higher than the partial energy exchange observed in the moderate-amplitude regime.

The following subsection exhibits the integration results in the high-amplitude regime corresponding to a stiffness ratio well inside the stability region of the representative-element.

\subsubsection{High-amplitude regime -- stable case}

\begin{figure}[H]
\begin{center}
{{\includegraphics[scale = 0.47]{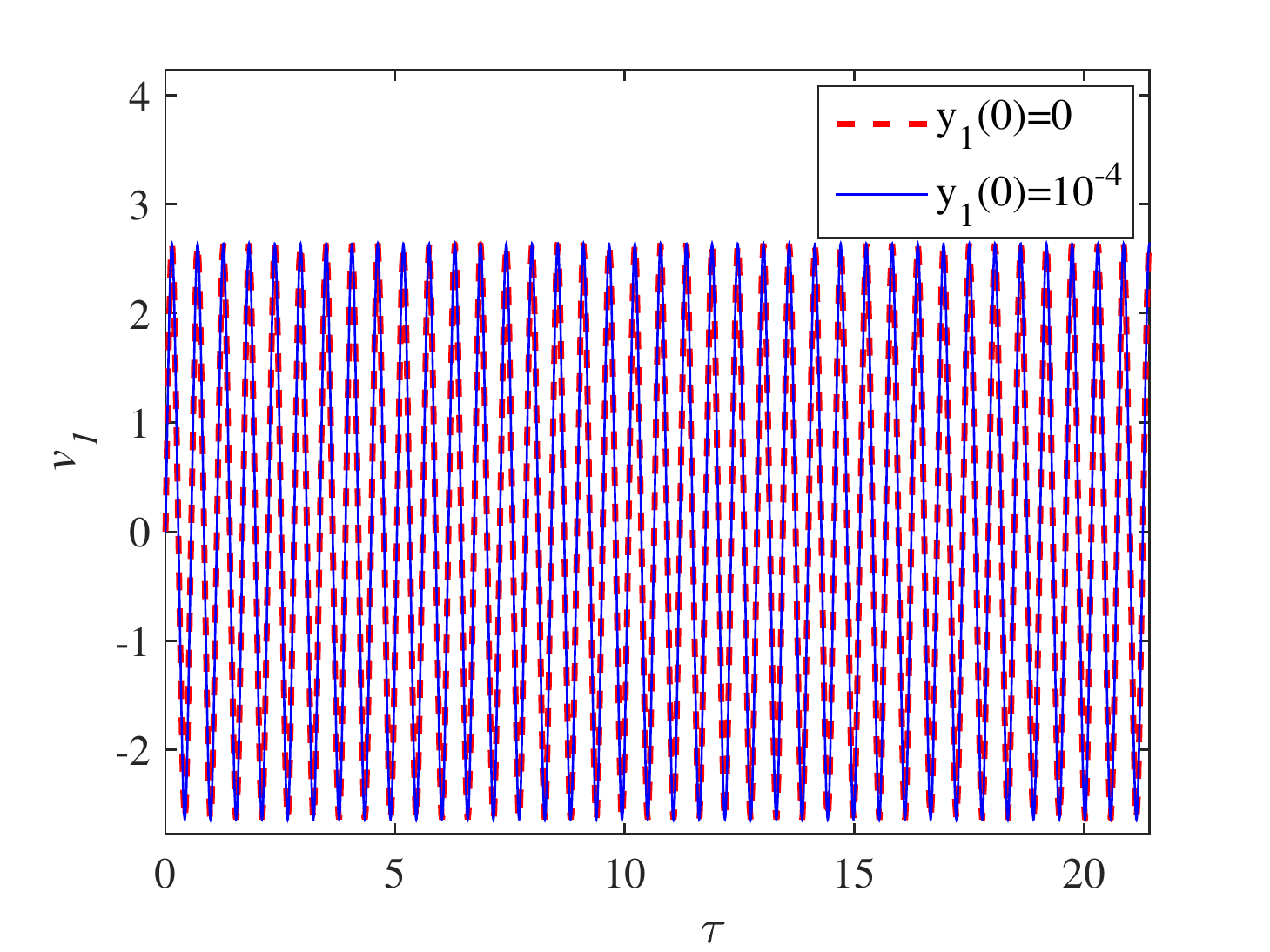}}}
\end{center}
\caption{\small $v_1(\tau)$ -- high-amplitude (stable) case: $N=10, Y_0=10\sqrt{k/p},k/k_0=0.005$}
\label{Fig8}
\end{figure}

\begin{figure}[H]
\begin{center}
{{\includegraphics[scale = 0.467]{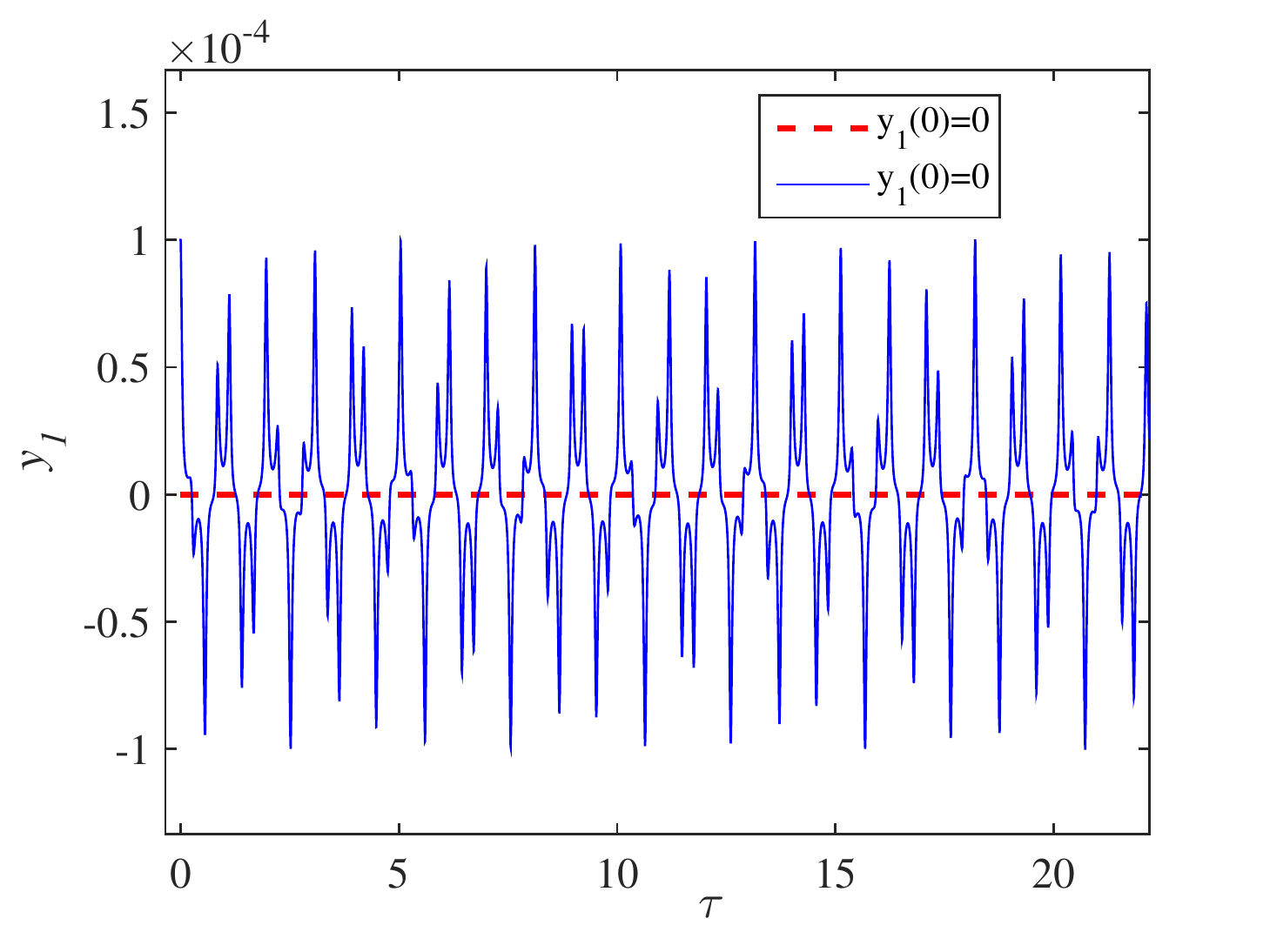}} \\
{\includegraphics[scale = 0.41]{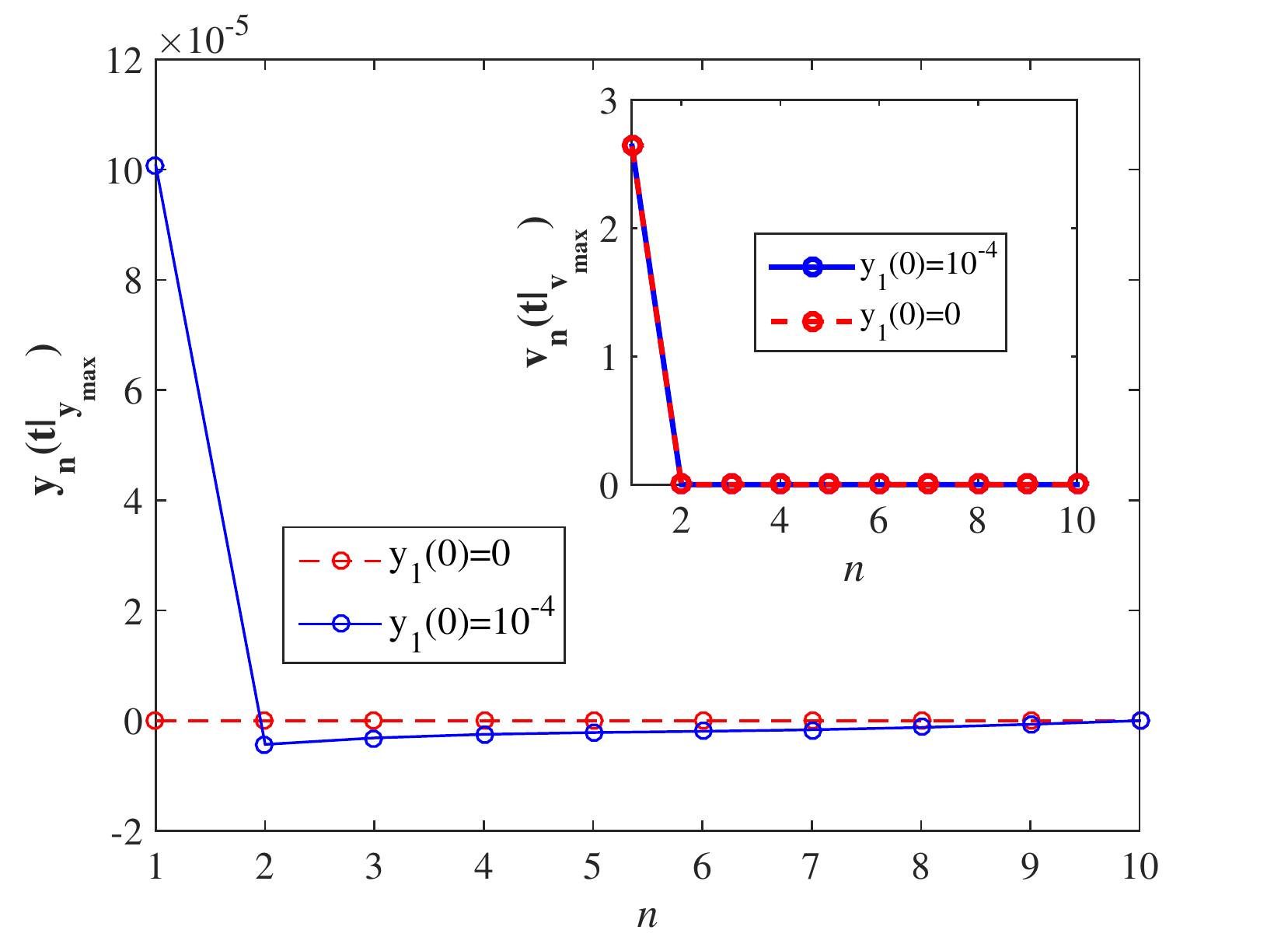}}}
\end{center}
\caption{\small Top to bottom: $y_1(\tau),y_n(t|_{\lVert\textbf{y}\rVert_{\infty}^{max}})$ -- high-amplitude (stable) case: $N=10, Y_0=10\sqrt{k/p},k/k_0=0.005$}
\label{Fig8b}
\end{figure}

Figures \ref{Fig8} and \ref{Fig8b} show the integration results for the slightly perturbed compacton conditions (in solid blue online) on top of the unperturbed ones (in dashed red online). 

The stability of the compacton is evident.

\section{Concluding remarks}
\label{Sect6}

This paper demonstrates that, as one would expect, flat bands and compactons can be realized in purely mechanical lattices. However, mechanical lattices can possess elements that are hardly viable in optical lattices, namely, the impact constraints. This type of nonlinearity allows substantial simplification of the stability analysis; the latter can be accomplished within the framework of linear algebra. Consequently, it becomes possible to distinguish between the different physical reasons for compacton instability. The first one is related to the internal instability of the antisymmetric mode in the unit cell of the lattice. The other is related to resonance with the propagation spectrum. For the local mechanism, it was possible to obtain an analytic evaluation of the local instability threshold as a function of the system parameters. Moreover, it was observed that an unstable compacton can exchange energy with the exponentially localized discrete breather. 

Similar considerations help to understand the stability characteristics of the lattice comprised of the massless boxes with the internal smooth nonlinear oscillators. There, unlike for the non-smooth system, the strict results entailing a complete stability picture are limited to the anti-continuum limit. One can argue that the advantage of the proposed framework beyond the aforementioned asymptotic limits is that once the monodromy matrix is constructed analytically, as it is done for the non-smooth system, the dependence of its eigenvalues on the parameters is proven to be regular, since only linear algebra is involved. This is in contrast with fully numerical Floquet analysis required for a general smoothly nonlinear system, which essentially involves numerical integration of a system formally not known to be integrable.

All lattices having flat bands inevitably possess complex unit-cells. Thus, one can expect that, generically, the stability of compactons will be affected by these two physical mechanisms -- local instability in the cell, and a global one, due to interaction with propagating linear or weakly nonlinear waves. In the former case, it seems that one can, at least asymptotically, rely on the (much more simple) local stability analysis presented herein.

In addition to analytic stability threshold estimates, closed-form expressions were given here for the compacton solutions, for both systems.

\section*{Acknowledgments}
The authors are grateful to the Israel Science Foundation (Grant No. 838/13) for financial support.

\renewcommand{\theequation}{A.\arabic{equation}}
\setcounter{equation}{0}  

\appendix
\section{Derivation of the saltation matrix for asynchronous two-impact jumps}  
\label{AppendixA}

\emph{First impact.}
For the first impact, the following relations hold, assuming small impact time (which owes to impact idealization) and small solution-perturbation (pertinent to linear stability analysis):
\begin{equation}
\label{Taylor1}
\begin{split}
\bar{\textbf{x}}(\bar{T}_{-})=\textbf{x}(\bar{T}_{-})+\delta\textbf{x}_{-} \\
\bar{\textbf{x}}({T}_{-})=\bar{\textbf{x}}(\bar{T}_{+})+\textbf{A}\bar{\textbf{x}}(\bar{T}_{+})\delta t \\
\bar{\textbf{x}}(\bar{T}_{+})=\textbf{g}_1[\bar{\textbf{x}}(\bar{T}_{-}),\bar{T}_{-}] \\
\bar{\textbf{x}}({T}_{+})={\textbf{x}}({T}_{+})+\delta{\textbf{x}}_{+} \\
{\textbf{x}}({T}_{-})={\textbf{x}}(\bar{T}_{+})+\textbf{A}{\textbf{x}}(\bar{T}_{+})\delta t \\
{\textbf{x}}({T}_{+})=\textbf{g}[{\textbf{x}}({T}_{-}),T_{-}] \\
\bar{\textbf{x}}({T}_{-})=\bar{\textbf{x}}({T}_{+}) \ , \\
{\textbf{x}}(\bar{T}_{-})={\textbf{x}}(\bar{T}_{+})
\end{split}
\end{equation}
where the smooth dynamics is linear and is represented by: $\dot{\textbf{x}}=\textbf{A}\textbf{x}$, the jump corresponding to a single mass impact is represented by the mapping $\textbf{g}_1(\textbf{x},t)$, the jump corresponding to synchronous impact of two masses at a two-mass site is represented by the mapping $\textbf{g}(\textbf{x},t)$, the periodic solution is $\textbf{x}$, the perturbed solution is $\bar{\textbf{x}}$, the perturbation in the solution is $\delta\textbf{x}$, the impact instance of the unperturbed solution is $T$, the \emph{first}-impact instance of the perturbed solution is $\bar{T}$, and it is assumed (with no loss of generality) that one has $T>\bar{T}$, and the subscripts `$+$' and `$-$' denote limit-values before and after impacts.

Combining the eight equations in Eq. (\ref{Taylor1}), and expanding the mappings $\textbf{g}_1$ and $\textbf{g}$ as first-order multiple-variable Taylor series (which is justified for the rather general case of a smooth mapping representing impact), and retaining only first-order terms, the following relation is obtained:
\begin{equation}
\label{Taylor2}
\begin{split}
\delta\textbf{x}_{+}=\textbf{g}_1[\textbf{x}(T_{-}),T_{-}]-\textbf{g}[\textbf{x}(T_{-}),T_{-}]+\\ + \lbrace\nabla\textbf{g}_1^{\top}[\textbf{x}(T_{-}),T_{-}]\rbrace^{\top}\delta\textbf{x}_{-} +\left ( \vphantom{ \frac{1}{1}} \textbf{A}\textbf{g}_1[\textbf{x}({T}_{-}),T_{-}] \right. \\ -\lbrace\nabla\textbf{g}_1^{\top}[\textbf{x}(T_{-}),T_{-}]\rbrace^{\top}\textbf{A}\textbf{x}({T}_{-}) +\left.\frac{\partial{\textbf{g}_1[\textbf{x}({T}_{-}),T_{-}]}}{\partial{t}}  \right )\delta{t}
\end{split}
\end{equation}
where $\delta{t}=T-\bar{T}$ is obtained as a function of $\delta{\textbf{x}}_{-}$ by use of the jump conditions, as follows. The jump conditions for the periodic and perturbed solutions as related to the first impact are given by:
\begin{equation}
\label{Taylor3}
\begin{split}
h_1[\textbf{x}(T_{-}),T_{-}]=0 \\
h_1[\bar{\textbf{x}}(\bar{T}_{-}),\bar{T}_{-}]=0
\end{split}
\end{equation}

Substituting lines 5, 8 and 1 of Eqs. (\ref{Taylor1}) into line 2 of  Eq. (\ref{Taylor3}), performing a first-order Taylor series expansion, and employing line 1 of Eq. (\ref{Taylor3}), one obtains an equation relating $\delta{t}$ and $\delta{\textbf{x}}_{-}$, which reads:
\begin{equation}
\label{Taylor4}
\begin{split}
\delta{t}=\frac{\lbrace\nabla^{\top} h_1[\textbf{x}(T_{-}),T_{-}]\rbrace\delta{\textbf{x}_{-}}}{\lbrace\nabla^{\top}  h_1[\textbf{x}(T_{-}),T_{-}]\rbrace\textbf{A}{\textbf{x}(T_{-})-\frac{\partial{h_1[\textbf{x}(T_{-}),T_{-}]}}{\partial{t}}}}
\end{split}
\end{equation}

Substituting Eq. (\ref{Taylor4}) into Eq. (\ref{Taylor2}), one obtains the following relations:
\begin{equation}
\label{Taylor5}
\delta{\textbf{x}}_{+}=\textbf{g}_1[\textbf{x}(T_{-}),T_{-}]-\textbf{g}[\textbf{x}(T_{-}),T_{-}]+\textbf{S}_1\delta{\textbf{x}}_{-}
\end{equation}
\begin{equation}
\label{Taylor6}
\begin{split}
\textbf{S}_1 \triangleq \lbrace\nabla\textbf{g}_1^{\top}[\textbf{x}(T_{-}),T_{-}]\rbrace^{\top} +\left ( \vphantom{ \frac{1}{1}} \textbf{A}\textbf{g}_1[\textbf{x}({T}_{-}),T_{-}] \right. \\ -\lbrace\nabla\textbf{g}_1^{\top}[\textbf{x}(T_{-}),T_{-}]\rbrace^{\top}\textbf{A}\textbf{x}({T}_{-}) +\left.\frac{\partial{\textbf{g}_1[\textbf{x}({T}_{-}),T_{-}]}}{\partial{t}}  \right )  \\ \times
\frac{\nabla^{\top} h_1[\textbf{x}(T_{-}),T_{-}]}{\lbrace\nabla^{\top}  h_1[\textbf{x}(T_{-}),T_{-}]\rbrace\textbf{A}{\textbf{x}(T_{-})-\frac{\partial{h_1[\textbf{x}(T_{-}),T_{-}]}}{\partial{t}}}}
\end{split}
\end{equation}

\emph{Second impact.}
For the second impact, the following relations hold:
\begin{equation}
\label{TaylorS1}
\begin{split}
\bar{\textbf{x}}({T}_{+})=\textbf{x}({T}_{+})+\delta\textbf{x}_{+}
 \\
\bar{\textbf{x}}(\hat{T}_{-})=\bar{\textbf{x}}({T}_{+})+\textbf{A}\bar{\textbf{x}}({T}_{+})\delta \bar t \\
\bar{\textbf{x}}(\hat{T}_{+})=\textbf{g}_2[\bar{\textbf{x}}(\hat{T}_{-}),\hat{T}_{-}] \\
\bar{\textbf{x}}(\hat{T}_{+})={\textbf{x}}(\hat{T}_{+})+\delta{\textbf{x}}_{++} \\
{\textbf{x}}(\hat{T}_{-})={\textbf{x}}({T}_{+})+\textbf{A}{\textbf{x}}({T}_{+})\delta \bar t \\
{\textbf{x}}(\hat{T}_{+})={\textbf{x}}(\hat{T}_{-})
\end{split}
\end{equation}
where the jump corresponding to the second mass impact (for the perturbed periodic solution) is represented by the mapping $\textbf{g}_2(\textbf{x},t)$, the perturbation in the solution after both impacts occurred in both the perturbed and unperturbed solutions is $\delta\textbf{x}_{++}$, the \emph{second}-impact instance of the perturbed solution is $\hat{T}$, and one has $\delta\bar{t}=\hat{T}-T>0$.

Combining the six equations in Eqs. (\ref{TaylorS1}), expanding the mapping $\textbf{g}_2$ as a first order multiple-variable Taylor series, retaining only first-order terms, and recalling line 6 of Eqs. (\ref{Taylor1}), and Eq. (\ref{Taylor5}), one obtains the following relation:
\begin{equation}
\label{TaylorS2}
\begin{split}
\delta\textbf{x}_{++}=\textbf{g}_2\lbrace\textbf{g}_1[\textbf{x}(T_{-}),T_{-}]\rbrace-\textbf{g}[\textbf{x}(T_{-}),T_{-}]+\\ +\left \lbrace\left.\nabla\textbf{g}_2^{\top}(\textbf{x},t)\right|_{\textbf{x}=\textbf{g}_1[\textbf{x}(T_{-}),T_{-}],t=T_{-}}\right\rbrace^{\top}\textbf{S}_1\delta\textbf{x}_{-} \\+ \left (\vphantom{\frac{\partial{t}}{\partial{t}}}\left\lbrace\left.\nabla\textbf{g}_2^{\top}(\textbf{x},t)\right|_{\textbf{x}=\textbf{g}_1[\textbf{x}(T_{-}),T_{-}],t=T_{-}}\right\rbrace^{\top}\textbf{A}\textbf{g}_1[\textbf{x}(T_{-}),T_{-}] \right. \\
 -  \textbf{A}\textbf{g}[\textbf{x}(T_{-}),T_{-}]+\left.\frac{\partial \left.\textbf{g}_2(\textbf{x},t)\right|_{\textbf{x}=\textbf{g}_1[\textbf{x}(T_{-}),T_{-}],t=T_{-}}}{\partial{t}}  \right )\delta{\bar{t}}
\end{split}
\end{equation}
where $\delta{\bar{t}}$ is obtained as a function of $\delta{\textbf{x}}_{-}$ by use of the jump conditions, as follows. The jump conditions for the periodic and perturbed solutions as related to the second impact are given by:
\begin{equation}
\label{TaylorS3}
\begin{split}
h_2[\textbf{x}(T_{+}),T_{+}]=0 \\
h_2[\bar{\textbf{x}}(\hat{T}_{-}),\hat{T}_{-}]=0
\end{split}
\end{equation}

Substituting lines 1, 2 and 6 of Eqs. (\ref{TaylorS1}), and Eq. (\ref{Taylor5}) into line 2 of  Eq. (\ref{TaylorS3}), performing a first-order Taylor series expansion, and employing line 1 of  Eq. (\ref{TaylorS3}), one obtains the equation relating $\delta{\bar{t}}$ and $\delta{\textbf{x}}_{-}$, which reads:
\begin{equation}
\label{TaylorS4}
\begin{split}
\delta{\bar{t}}=\\-\frac{\lbrace\nabla^{\top} h_2[\textbf{x}(T_{-}),T_{-}]\rbrace\textbf{S}_1\delta{\textbf{x}_{-}}}{\lbrace\nabla^{\top}  h_2[\textbf{x}(T_{-}),T_{-}]\rbrace\textbf{A}\textbf{g}_1[\textbf{x}(T_{-}),T_{-}]+\frac{\partial{h_2[\textbf{x}(T_{-}),T_{-}]}}{\partial{t}}}
\end{split}
\end{equation}

Substituting Eq. (\ref{TaylorS4}) into Eq. (\ref{TaylorS2}), one obtains the following relations:
\begin{equation}
\label{TaylorS5}
\begin{split}
\delta{\textbf{x}}_{++}=\textbf{g}_2\lbrace\textbf{g}_1[\textbf{x}(T_{-}),T_{-}]\rbrace-\textbf{g}[\textbf{x}(T_{-}),T_{-}]\\+\textbf{S}_2\textbf{S}_1\delta{\textbf{x}}_{-}
\end{split}
\end{equation}
\begin{equation}
\label{TaylorS6}
\begin{split}
\textbf{S}_2 \triangleq \\ \left\lbrace\nabla\textbf{g}_2^{\top}(\textbf{x},t)\left\lvert_{\underset{t=T_{-}}{\textbf{x}=\textbf{g}_1[\textbf{x}(T_{-}),T_{-}],}}\right.\right\rbrace^{\top}+  \left (  \textbf{A}\textbf{g}[\textbf{x}(T_{-}),T_{-}]\vphantom{\frac{\partial \left.\textbf{g}_2(\textbf{x},t)\right|_{{\textbf{x}=\textbf{g}_1[\textbf{x}(T_{-}),T_{-}],t=T_{-}}}}{\partial{t}}}\right.\\ \left. -\left\lbrace\nabla\textbf{g}_2^{\top}(\textbf{x},t)\left\lvert_{\underset{t=T_{-}}{\textbf{x}=\textbf{g}_1[\textbf{x}(T_{-}),T_{-}],}}\right.\right\rbrace^{\top}\textbf{A}\textbf{g}_1[\textbf{x}(T_{-}),T_{-}] \right. \\
 -\left.\frac{\partial \textbf{g}_2(\textbf{x},t)\left\lvert _{\textbf{x}=\textbf{g}_1[\textbf{x}(T_{-}),T_{-}],t=T_{-}}\right.}{\partial{t}}  \right ) \times \\
\frac{\nabla^{\top} h_2[\textbf{x}(T_{-}),T_{-}]}{\lbrace\nabla^{\top}  h_2[\textbf{x}(T_{-}),T_{-}]\rbrace\textbf{A}\textbf{g}_1[\textbf{x}(T_{-}),T_{-}]+\frac{\partial{h_2[\textbf{x}(T_{-}),T_{-}]}}{\partial{t}}}
\end{split}
\end{equation}

The crux of the present derivation is embodied in Eq. (\ref{TaylorS5}). The linear stability analysis performed here assumes small perturbation to the solution prior to impact, and linear stability can only correspond to the perturbation staying small after the sequence of impacts has occurred. Consequently, a necessary condition for both the possibility of linear stability and the self-consistency of the linear stability analysis as a procedure is that the first two terms in the right-hand side of Eq. (\ref{TaylorS5}) cancel out (formally they can result in an infinitesimal quantity, yet they contain no parameter taken to the limit zero in the context of the linear stability analysis and thus can only result in a finite value that thus has to be an exact zero), namely:
\begin{equation}
\label{TaylorS7}
\begin{split}
\textbf{g}_2\lbrace\textbf{g}_1[\textbf{x}(T_{-}),T_{-}]\rbrace=\textbf{g}[\textbf{x}(T_{-}),T_{-}]
\end{split}
\end{equation}

This consistency requirement can be termed ``impact-integrability'', since it implies that a series of consecutive impacts produces the same result as a single synchronous impact. The collision between two particles and a fixed wall is always impact-integrable. However, in the case of a simultaneous many-body impact, integrability may hold only given some restriction on the problem parameters, solely under which slight asynchronicity will not produce finite change in the outcome. This intuitive observation is represented by a formal linear stability analysis as shown above.

After having been revealed in essence for the two-impacts case, the requirement can be straight-forwardly generalized to $n$ consecutive impacts, the induction procedure being trivial here. For a series of $n$ impacts, the ``impact-integrability'' requirement, or the necessary condition for stability, is:
\begin{equation}
\label{TaylorS8}
\begin{split}
\textbf{g}_n[...(\textbf{g}_2\lbrace\textbf{g}_1[\textbf{x}(T_{-}),T_{-}]\rbrace=\textbf{g}[\textbf{x}(T_{-}),T_{-}])]
\end{split}
\end{equation}

If indeed the system in question is impact-integrable, then it holds that by substituting Eq. (\ref{TaylorS7}) into the first term in the parentheses in the expression for $\textbf{S}_2$ in Eq. (\ref{TaylorS6}), it becomes apparent that the matrix $\textbf{S}_2$ is instantaneous and only depends on the value of $\textbf{x}$ \emph{after} the first impact. Then it becomes clear that $\textbf{S}_2$ is a legitimate saltation matrix and one has:
\begin{equation}
\label{TaylorS9}
\begin{split}
\delta{\textbf{x}}_{++}=\textbf{S}\delta{\textbf{x}}_{-},
\textbf{S}=\textbf{S}_2\textbf{S}_1
\end{split}
\end{equation}
for the two-impacts case, and:
\begin{equation}
\label{TaylorS10}
\begin{split}
\delta{\textbf{x}}_{+(n)}=\textbf{S}\delta{\textbf{x}}_{-},
\textbf{S}=\textbf{S}_n\textbf{S}_{n-1}...\textbf{S}_2\textbf{S}_1
\end{split}
\end{equation}
for the $n$-impacts case.

After establishing the impact-integrability requirement as the necessary condition for stability, it is left to note that in the setting assumed in Secs. \ref{Sect3}-\ref{Sect5}, there are only impacts between a mass and a fixed wall, albeit multiple ones. Hence the aforementioned condition is satisfied automatically, and Eqs. (\ref{TaylorS6}), (\ref{TaylorS7}) and (\ref{TaylorS9}) apply.

Also noteworthy is the following. The partial temporal derivatives of the jump conditions and the jump mappings appearing in Eqs. (\ref{Taylor6}) and (\ref{TaylorS6}) can generally be non-vanishing, as, say, for the case of a periodically externally moved wall and a periodically externally controlled restitution coefficient (as in magnetorheological composites). Hence the corresponding terms (with the associated signs) may be of relevance. However, in the case considered in Secs. \ref{Sect3}-\ref{Sect5}, the walls are fixed and perfect restitution is assumed, for which the jump mapping is realized as a multiplication by a matrix. Consequently, Eqs. (\ref{FS1})-(\ref{FS3}) hold.

\renewcommand{\theequation}{B.\arabic{equation}}
\setcounter{equation}{0}  

\section{Details of the stability analysis of Eq. (\ref{eq10B.7})}  
\label{AppendixB}

Here the procedure of the stability analysis of Eq. (\ref{eq10B.7}) is described. Asymptotic analysis of Eq. (\ref{eq10B.7}), resulting in the identification of the number of emerging instability tongues, derivation of starting points at the zero-amplitude limit for numerical instability-tongue boundaries calculation, as well as the derivation of leading-order asymptotic expansions for the boundaries of the first two instability tongues, using both Mathieu and higher-order approximations, can be found in \cite{Perchikov2016}.

\subsection*{B.1. Fourier decomposition by  numerical deconvolution}
\label{sect10D}
In order to perform stability analysis by Hill's determinants method, the Hill function in Eq. (\ref{eq10B.7}) has to be expressed as a Fourier series.
To this end, first, the numerator and the denominator are decomposed, separately. Both the numerator and the denominator depend linearly on the square of the normalized displacement in the antisymmetric mode, $\hat{v}(\hat{\tau})$, which was already expressed in a sine series form in Eq. (\ref{eq10A.5}). Thus, the first thing one has to do is to perform convolution, since the Fourier series decomposition of the product of two functions (a function and itself, in the present case) equals the (discrete) convolution of the Fourier series decomposition of each of the functions. Multiplying, rearranging and adjusting the indices, one obtains the following expression:
\begin{equation}
\label{eq10D.1}
\begin{split}
[\hat{v}(\hat{\tau})]^2=\frac{V_0^{(2)}}{2}+\sum\limits_{k=1}^{\infty}V_k^{(2)}\cos{\left(2k\hat{\tau}\right)}, \\
V_k^{(2)} \triangleq \left [\sum\limits_{m=1,3,5}^{\infty} V_m V_{2k+m}-\frac{1}{2}V_k^2 \delta_{k,2\mathbb{N}-1} \right. \\ \left. -(1-\delta_{k,1}) \sum\limits_{m=1}^{\infty} V_{k-m} V_{k+m}\right ]\delta_{k,\mathbb{N}-1}
\end{split}
\end{equation}

Next the definition of the Hill function, as arising from Eq. (\ref{eq10B.7}), is rearranged to a product form, $f_1(\hat{\tau})h(\hat{\tau})=f_2(\hat{\tau})$, and then $h(\hat{\tau})$  is formally expanded into a Fourier cosine series (as the ratio of two even functions, $h(\hat{\tau})$  is always even):
\begin{equation}
\label{eq10D.4}
h(\hat{\tau})=\frac{H_0}{2}+\sum\limits_{k=1}^{\infty}H_k\cos{\left(2k\hat{\tau}\right)}
\end{equation}

Substituting Eqs. (\ref{eq10D.1}) and (\ref{eq10D.4}) into the aforementioned product-form equation for $h(\hat{\tau})$ that emerges from Eq. (\ref{eq10B.7}), expanding the resulting products of sums, using trigonometric identities to turn cosine products to single index-shifted cosines and performing the convolution, one finally obtains the product-form equation for $h(\hat{\tau})$ as a single Fourier cosine series that should be equal to zero at all times, namely,

\begin{equation}
\label{eqGFourier}
g(\hat{\tau})=G_0+\sum\limits_{n=1}^{\infty}G_n\cos{(2n\hat{\tau})}\equiv 0
\end{equation}

Fulfillment of this condition corresponds to the vanishing of the coefficients of this Fourier series. The equations representing the vanishing of these $total$ Fourier cosine series coefficients, formulated as equations for the Fourier cosine coefficients of $h(\hat{\tau})$, take the form of linear algebraic equations.

First, there is the condition $G_0=0$, producing the following explicit equation for $H_0$ in terms of $H_{n \geq 1}$:
\begin{equation}
\label{eq10D.5}
\begin{split}
H_0 = \frac{\Omega_v^{-2}(4+3\hat{\epsilon}V_0^{(2)})}{2(1+\hat{\eta})+(3/2)\hat{\eta}\hat{\epsilon}V_0^{(2)}} \\
-\frac{3\hat{\eta}\hat{\epsilon}}{2(1+\hat{\eta})+(3/2)\hat{\eta}\hat{\epsilon}V_0^{(2)}}\sum\limits_{n=1}^{\infty}V_n^{(2)}H_n
\end{split}
\end{equation}

Second, taking the set of conditions $G_{n \geq 1}=0$, which contain both $H_0$ and $H_{n \geq 1}$, and eliminating $H_0$ from these equations by isolating it from Eq. (\ref{eq10D.5}), one obtains an infinite set of linear inhomogeneous equations for $H_{n \geq 1}$, which takes the following form in indicial notation:

\begin{equation}
\label{eq10D.6}
\sum\limits_{m=1}^{\infty}M_{nm} H_m = \mu_n \ \ ; \  \forall \ \  n,m \in \mathbb{N}
\end{equation}
where

\begin{equation}
\label{eq10D.6s}
\begin{split}
\mu_n = \frac{3\hat{\epsilon}\Omega_v^{-2}\delta_{n,\mathbb{N}}V_n^{(2)}}{2+2\hat{\eta}+(3/2)\hat{\eta}\hat{\epsilon}V_0^{(2)}} \\
M_{nm} =  \left(1+\hat{\eta}+\frac{3}{4}\hat{\eta}\hat{\epsilon}V_0^{(2)} \right)\delta_{n,m} \\
+ \frac{3}{4}\hat{\eta}\hat{\epsilon}V_{n+m}^{(2)} \delta_{n+m,\mathbb{N}}+ \frac{3}{4}\hat{\eta}\hat{\epsilon}V_{n-m}^{(2)} \delta_{n-2m,\mathbb{N}}\\
+\frac{3}{4}\hat{\eta}\hat{\epsilon}V_{m-n}^{(2)} \delta_{m-n,\mathbb{N}}+\frac{3}{4}\hat{\eta}\hat{\epsilon}V_m^{(2)} \delta_{n,2m}
\\-\frac{3}{4}\hat{\eta}\hat{\epsilon}\frac{3\hat{\eta}\hat{\epsilon}V_n^{(2)}V_m^{(2)}}{2+2\hat{\eta}+(3/2)\hat{\eta}\hat{\epsilon}V_0^{(2)}}
\end{split}
\end{equation}

Finally, Eqs. (\ref{eq10D.6}) are truncated, which is achieved by exchanging $\mathbb{N}$ with $\mathbb{N}\leq\hat{N}$ everywhere in these equations (this is the reason for the presence of Kronecker's delta in the first term in the third line in Eq. (\ref{eq10D.6s})), and deconvolution is performed by solving the truncated system numerically, by a standard algorithm for the solution of linear systems, thus obtaining $H_{n \geq 1}$. The coefficient $H_0$ is then derived using Eq. (\ref{eq10D.5}).

The issue of convergence for this procedure is discussed in \cite{Perchikov2016}. 

Having obtained the Fourier cosine coefficients of the Hill function, one performs the stability analysis by Hill's (truncated) infinite determinants method. The result is shown in Fig. \ref{Fig3}.

\subsection*{B.2. Asymptotic stability-limit estimates}
\renewcommand{\thefigure}{\arabic{figure}}

The small-amplitude analytic estimates for the stability boundaries (given for energy in amplitude units, following Eq. (\ref{eq7.5})), are (see \cite{Perchikov2016} for derivation):
\begin{equation}
\label{cubest}
\begin{split}
\left.\frac{k}{k_0}\right|^{(1)}_{-}\underset{Y_0\to 0}\to 0 \ ,\\ \left.Y_0\right|^{(1)}_{+}\underset{k/k_0\to 0}\to\sqrt{\frac{4}{3}}\sqrt{\frac{k}{k_0}} \ , \\
Y_0\left.{\lvert^{(2)}_{\pm}}\right.\underset{k/k_0\to 0}\to\sqrt{\frac{1}{3}}\sqrt{\frac{3}{2}-\frac{k}{k_0}}
\end{split}
\end{equation}

Plots of the first (finite-width) instability tongue boundaries for the antisymmetric mode, and the second tongue, which is (at least to second order) a collapsed, zero-width one, are presented in Fig. \ref{Fig3}. Asymptotic approximations for the boundaries, as given in Eq. (\ref{cubest}), are shown therein in dashes.

\subsection*{B.3.  Poincar\'e sections}
\renewcommand{\thefigure}{\arabic{figure}}

\begin{figure}[H]
\begin{center}
{\includegraphics[scale = 0.316,trim={1.35cm 0 1.32cm 0},clip]{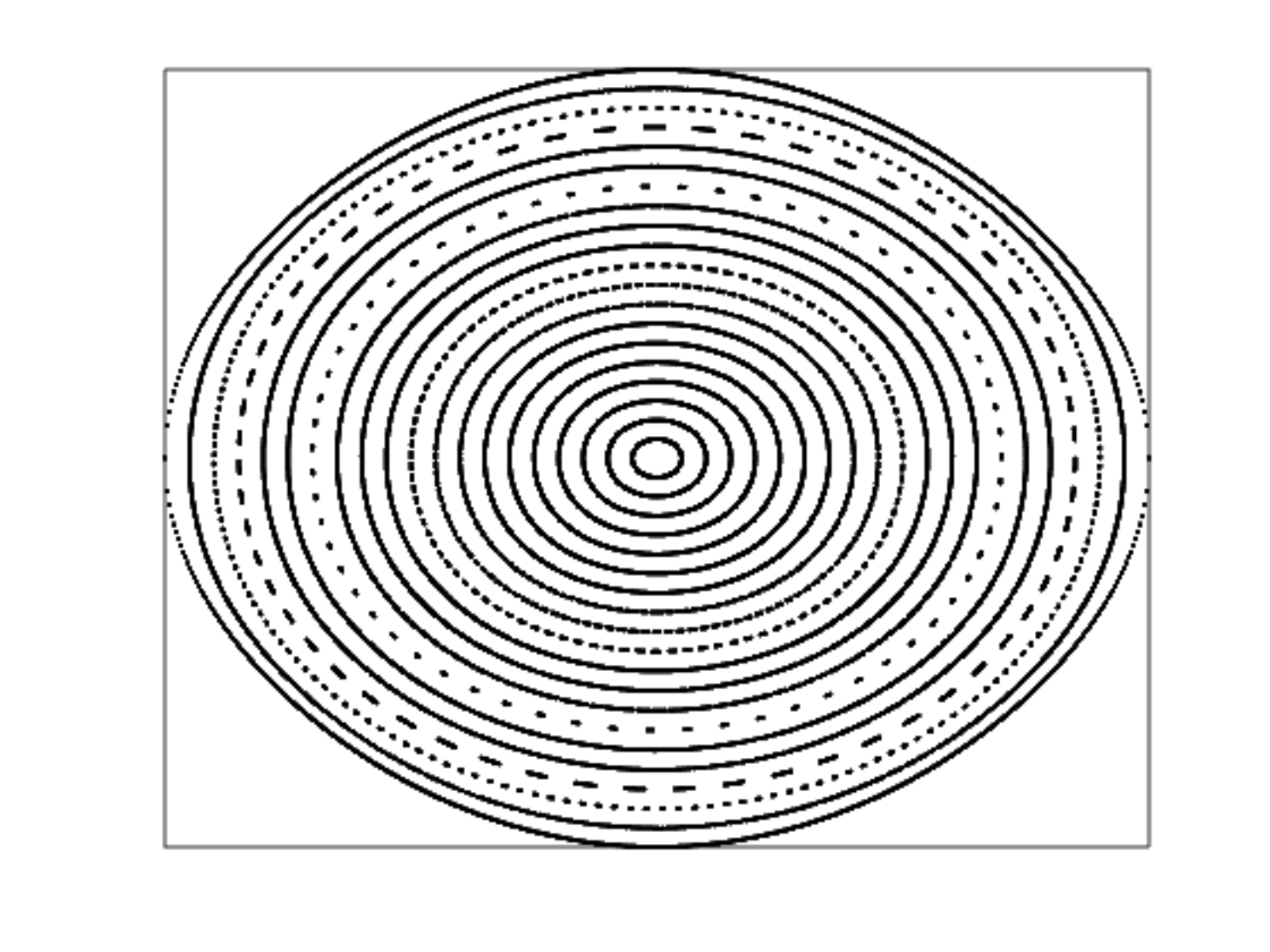}}
{\includegraphics[scale = 0.316,trim={1.35cm 0 1.32cm 0},clip]{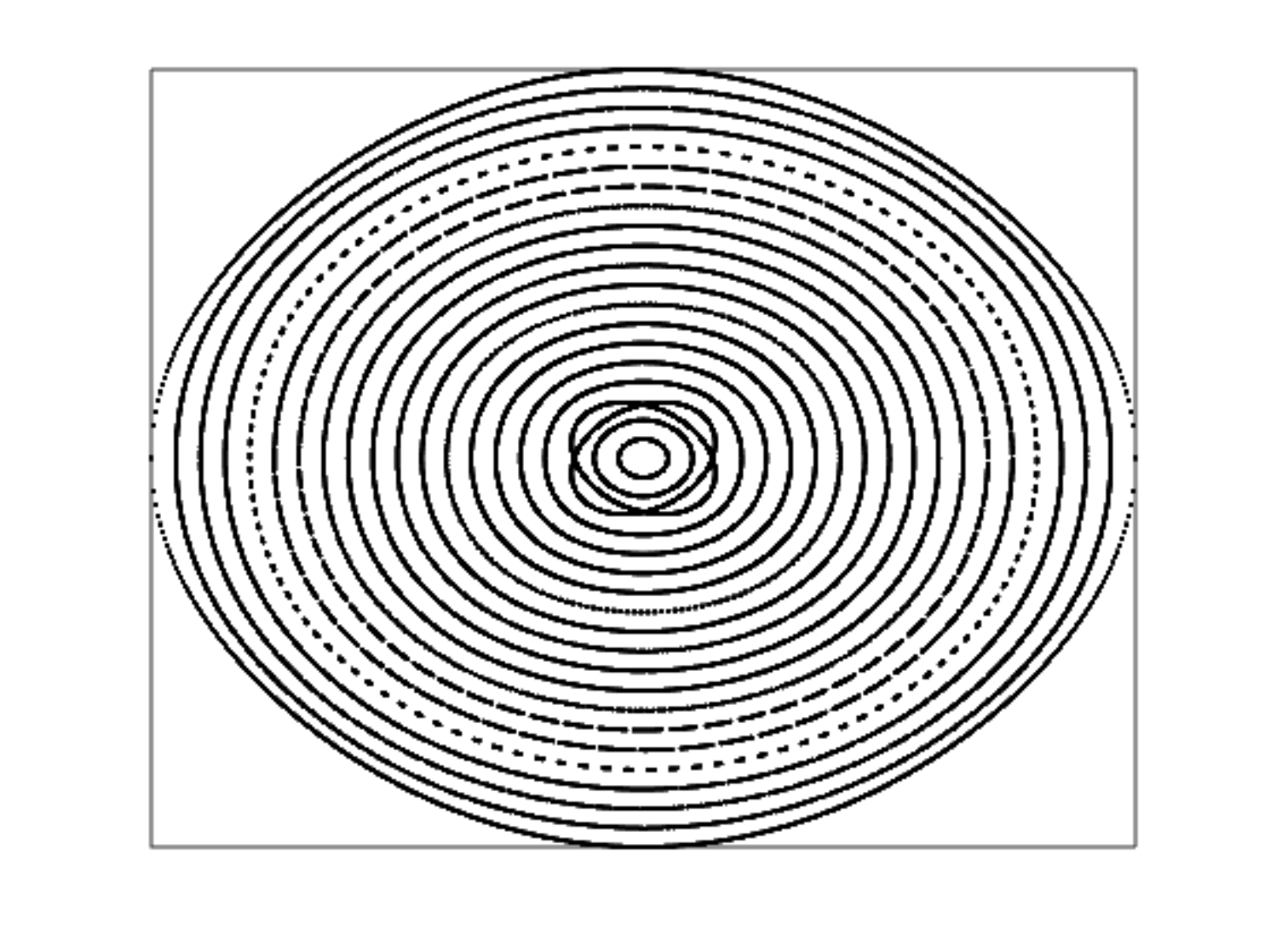}}
\end{center}
\caption{\small Poincar\'e sections in (normalized) symmetric-mode displacement and velocity, for energies just below and above the degenerate instability tongue (for $\hat{\eta}=5/2$), specified by: $\hat{\epsilon}=0.048$ (left) and $\hat{\epsilon}=0.049$ (right)}
\label{Fig4a}
\end{figure}

Figure \ref{Fig4a} presents Poincar\'e sections of the flow in the vicinity of the second (degenerate) instability tongue. The emergence of KAM islands represents the effect of the loss of stability of the compact mode through parametric resonance. 

The coordinates in the plots are the symmetric mode displacement and its (dimensionless) time derivative. Hence, energy localized entirely in the compact, antisymmetric mode is represented as a point in the center (or an oval around the center, in the perturbed case). As stability is lost, the oval around the center turns into four KAM islands, corresponding to the 1:2 resonance of the second instability tongue.

The first, finite-width instability tongue is standard manifestation of instability through parametric resonance -- hence the derivation of the stability range boundaries seems sufficient. The second tongue appears to be a collapsed one, and therefore further investigation is called for, in order to understand the finite-amount topological implications (for parameter values at finite distance from the zero-width tongue) of the appearance of such a feature. 

To this end Poincar\'e section analysis was employed, as a means to reveal the existence of instability through higher-order resonance manifested in the appearance of KAM tori (of course, the first instability tongue has its own second validation --  by direct numerical integration, albeit for a chain, as performed in Sec. \ref{Sect2b}).

\begin{figure}[H]
\begin{center}
{\includegraphics[scale = 0.303,trim={1.35cm 0.1cm 1.33cm 0.6cm},clip]{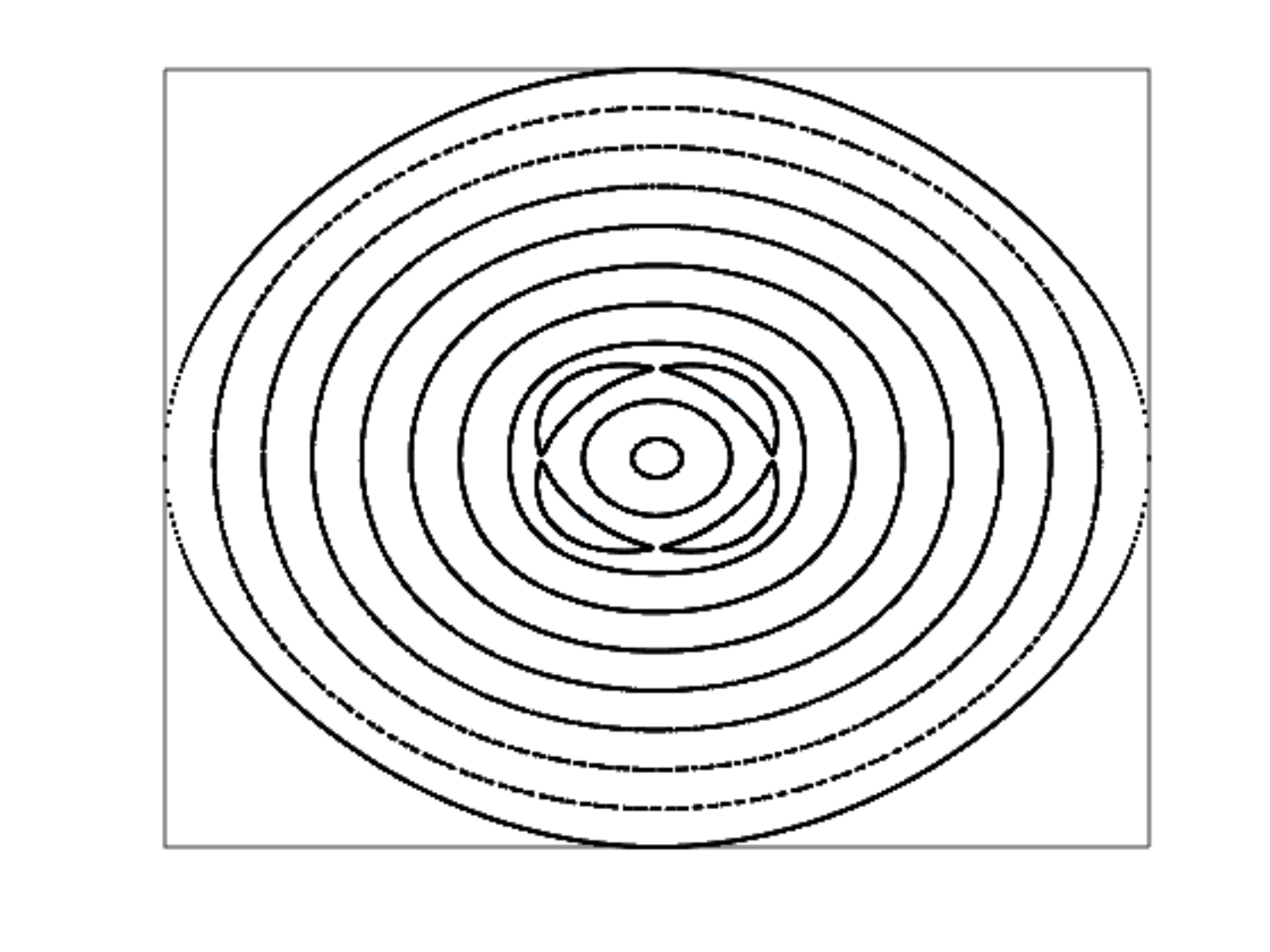}}
{\includegraphics[scale = 0.31,trim={1.5cm 0.3cm 1cm 0.5cm},clip]{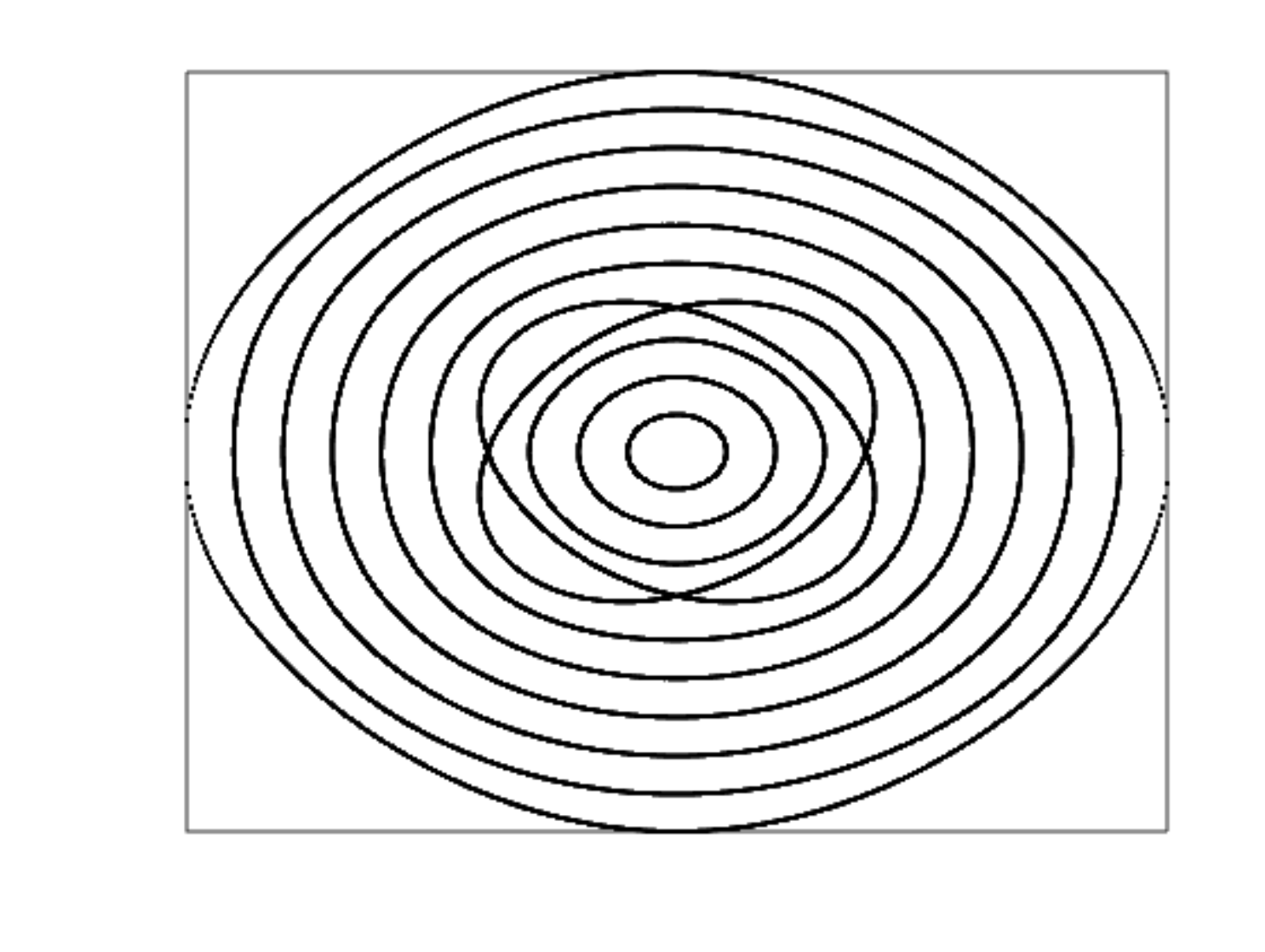}}
\end{center}
\caption{\small Poincar\'e sections, illustrating the effect of the degenerate instability tongue on the topology, for energies further above the critical line --  KAM islands emanate from the compact mode as energy is increased ($\hat{\eta}=5/2$ and $\hat{\epsilon}=0.05$ -- left and $\hat{\epsilon}=0.0528$ -- right)}
\label{Fig4}
\end{figure}

Figure \ref{Fig4}, much like Fig. \ref{Fig5d}, shows resonance-originating KAM islands (increasing with energy), confirming the existence of periodic dynamics of energy exchange between the different modes, for higher energies, for the representative element. The same interpretation can be attributed to the non-divergent dynamics of energy exchange in the case of a chain, as observed in the histories of the principal-box (initial localization site) masses, as shown in Fig. \ref{Fig5b}.

The choice of a massless box in the conservative case allows, at least for a representative element, to describe regular (non-chaotic) dynamics completely, using Poincar\'e sections. 

The emergence of KAM islands spanning from the center to the envelope implies the instability of the compact solution and corresponding significant energy exchange with the symmetric mode (this scenario corresponds to the high-energy points inside the first instability tongue). Chain dynamics can then be inferred -- approximately for arbitrary link-stiffness, with the help of direct numerical integration, and exactly in the anti-continuum limit.


\end{document}